\documentclass[a4paper,11pt]{article}
\usepackage{jheppub}
\usepackage{graphicx}
\usepackage{epsfig}
\usepackage{rotating}
\usepackage{amssymb}
\usepackage{subfigure}
\usepackage{dsfont}
\usepackage{psfrag}
\usepackage{amsmath,euscript,array,mathrsfs}
\usepackage{slashed}
\usepackage{array}
\usepackage{youngtab}
\usepackage{color}
\usepackage{bbold}

\newcommand{\beq}{\begin{equation}}
\newcommand{\eeq}{\end{equation}}
\newcommand{\beqs}{\begin{eqnarray}}
\newcommand{\eeqs}{\end{eqnarray}}
\newcommand{\Tr}{{\rm Tr}}

\newcommand{\dd}{\mbox{d}}
\newcommand{\arctanh}{{\rm arctanh}}

\renewcommand{\arraystretch}{1.9}
\newcommand{\be}{\begin{equation}}
\newcommand{\ee}{\end{equation}}
\newcommand{\ba}{\begin{array}}
\newcommand{\ea}{\end{array}}
\newcommand{\eq}[1]{Eq.~(\ref{#1})}

\newcommand\figwidth{15.2cm}
\newcommand\figwidthsmall{15.2cm}
\newcommand\ecap{-0.25cm}

\begin{document}

\title{Holographic models of composite Higgs in the Veneziano limit. Part I. Bosonic sector}

\author[a]{Daniel Elander,}
\author[a]{Michele Frigerio,}
\author[b]{Marc Knecht,}
\author[a]{Jean-Lo\"ic Kneur}

\affiliation[a]{Laboratoire Charles Coulomb (L2C), University of Montpellier, CNRS, Montpellier, France}
\affiliation[b]{Centre de Physique Th\'eorique, CNRS/Aix-Marseille Univ./Univ. de Toulon (UMR 7332), CNRS-Luminy Case 907, 13288 Marseille Cedex 9, France}

\date{\today}

\abstract{
We study strongly-coupled, approximately scale-invariant gauge theories, which develop a mass gap in the infrared. We argue that a large number of fermion flavours is most suitable to provide an ultraviolet completion for the composite Higgs scenario. The holographic approach allows to describe the qualitative features of the non-perturbative dynamics in the Veneziano limit. We introduce new bottom-up holographic models, which incorporate the backreaction of flavour on the geometry, and show that this can correlate the mass gap to the scale of flavour-symmetry breaking. We compute the mass spectrum for the various composite bosonic states, 
and study its dependence on the scaling dimension of the symmetry-breaking operators, as well as on the number of flavours. The different regions with a light dilaton are critically surveyed. We carefully assess the domain of validity of the holographic approach, and compare it with lattice simulations and the Nambu--Jona-Lasinio model.
}

\maketitle


\section{Introduction}

A strongly-coupled extension of the Standard Model (SM) has the potential to protect the hierarchy between the electroweak scale, $v\simeq 246$ GeV, and large new physics scales. A new, strong interaction can dynamically induce a mass gap in the multi-TeV range, and the Higgs boson can emerge as a composite, pseudo Nambu-Goldstone boson (NGB) of such a strongly-coupled sector \cite{Kaplan:1983fs}. In this scenario, reviewed e.g.~in 
\cite{Contino:2010rs,Panico:2015jxa}, not only should the Higgs properties depart from the SM predictions by order-$(v/f)^2$ corrections, with $f$ the Goldstone decay constant, but also a number of additional composite states should appear in the spectrum. While their typical mass should be of the order of the mass gap or higher, $m\sim 4\pi f$, it is plausible that some states are significantly lighter, providing a target for experimental searches.

In order to characterise the composite spectrum, one needs to specify the ultraviolet (UV) completion of the composite-Higgs scenario. We will introduce a new gauge theory, hypercolour (HC), with no fundamental scalars, and fermions in specific representations of the HC gauge group. To ensure that such a scenario respects electroweak precision tests, and to have sizeable Yukawa couplings to the SM fermions via partial compositeness \cite{Kaplan:1991dc}, the HC fermions must carry non-trivial SM charges, and the HC chiral symmetries must include the SM accidental global symmetries (custodial, baryon, and lepton number), which otherwise would be badly broken by the strong HC dynamics. In addition, one wishes to keep the HC sector in a strongly-coupled and quasi scale-invariant (walking) regime \cite{Holdom:1984sk,Yamawaki:1985zg,Appelquist:1986an} for a large range of energies above the mass gap, while avoiding Landau poles in 
the SM gauge couplings. In the infrared (IR), the HC sector should confine, resulting in a plethora of HC-invariant composite states appearing in the spectrum.

We will show that all these requirements could be satisfied if one introduces a large number, $N_F \ge 5$, of Dirac fermions in the fundamental representation of HC. This is in contrast with most HC models of composite Higgs,
usefully classified in \cite{Ferretti:2013kya},
where fermions in different HC representations also carry SM charges.
Some noteworthy exceptions, closer to our approach, are motivated and analysed in \cite{Vecchi:2015fma,Gertov:2019yqo}. In order to exploit large-$N_C$ arguments (where $N_C$ is the number of hypercolours), and because of the comparatively large number of flavours, we are lead to consider the Veneziano limit, with the ratio $x_F \equiv N_F/N_C$ of order one \cite{Veneziano:1976wm}.

A main challenge to the study of strongly-coupled dynamics comes from calculability. In addition to lattice simulations and non-perturbative field theory methods, gauge-gravity duality~\cite{Maldacena:1997re,Gubser:1998bc,Witten:1998qj} provides a useful tool, relating certain strongly-coupled field theories in the large-$N_C$ limit to weakly-coupled gravity theories in higher dimensions (for a review, see~\cite{Aharony:1999ti}). Using the formalism of holographic renormalization~\cite{Bianchi:2001kw,Bianchi:2001de,Skenderis:2002wp}, it is known how to systematically compute 
field-theory correlators, from which the spectrum of composite states can be extracted. Since its original inception in~\cite{Maldacena:1997re}, gauge-gravity duality has been extended to include examples describing non-conformal, confining dynamics on the field theory side, see e.g.~\cite{Witten:1998zw,Klebanov:2000hb,Chamseddine:1997nm,Maldacena:2000yy,Butti:2004pk}. Moreover, it is possible to incorporate matter in the fundamental representation of the gauge group by including flavour branes~\cite{Karch:2002sh,Kruczenski:2003be} (see also the review \cite{Erdmenger:2007cm}). As long as $N_F \ll N_C$, such flavour branes can be treated as probes on top of given gravity backgrounds. However, in the Veneziano limit, when $N_F \sim N_C$, their backreaction on the geometry has to be taken into account (for a review, see~\cite{Nunez:2010sf}). This is analogous to going beyond the quenched approximation in field theory.

There exists by now a large literature on holographic models of composite Higgs, following the original papers of~\cite{Contino:2003ve,Agashe:2004rs}. Most studies so far exist within the bottom-up approach to holography,
that is, phenomenologically motivated models where the gravity description is not required to have a fundamental origin in string theory. 
The simplest, commonly used setting, in order to describe non-conformal IR physics, is to consider an Anti-de Sitter (AdS) background geometry in five dimensions (5D), and introduce a hard-wall that ends the geometry in the IR. More realistically, one expects some dynamics to induce a departure from AdS in the IR region. Such models, sometimes referred to as soft-wall models, have been studied both in the context of holographic QCD \cite{Karch:2006pv,Csaki:2006ji,Gursoy:2007er,Batell:2008zm,Colangelo:2008us} and of the stabilisation of the electroweak scale \cite{Falkowski:2008fz,Batell:2008me,Cabrer:2009we,Cabrer:2011fb}. In some of these papers the deformation of the geometry was determined by the backreaction of a bulk scalar field, with a profile chosen ad hoc \cite{Csaki:2006ji,Gursoy:2007er,Batell:2008zm}\cite{Batell:2008me,Cabrer:2009we}. A recent holographic analysis of composite-Higgs models, inspired by top-down constructions can be found in~\cite{Erdmenger:2020lvq,Erdmenger:2020flu}. Although bottom-up studies of holographic QCD addressing the Veneziano limit exist \cite{Jarvinen:2011qe}, such limit does not seem to have been considered within the context of holographic composite Higgs.

Since the composite Higgs models we study are required to exhibit walking dynamics, the question arises whether the spectrum contains a light dilaton~\cite{Leung:1985sn,Bardeen:1985sm,Yamawaki:1985zg}, the pseudo-Goldstone boson associated with spontaneous breaking of approximate scale invariance. It has been suggested that such dynamics may arise close to the conformal window, hence raising the question of the dependence of the dilaton mass on the number of flavours. Within the rigorous context of top-down holography, there is evidence of dilaton dynamics in theories for which the dual supergravity background is related to the conifold~\cite{Elander:2017cle,Elander:2017hyr} (see also~\cite{Nunez:2008wi,Elander:2009pk,Elander:2012yh,Elander:2014ola}). Additional examples are provided by the existence of a light dilaton on a metastable branch of solutions in both Romans supergravity~\cite{Elander:2020ial} as well as in a model~\cite{Elander:2020fmv} related to the one proposed by Witten as a description of confinement~\cite{Witten:1998zw}. Yet, none of these studies rely on incorporating flavour. In a bottom-up context, the phase transition at the lower edge of the conformal window has been discussed in~\cite{Kaplan:2009kr} (see also \cite{Kutasov:2012uq,Goykhman:2012az,Evans:2013vca,Arean:2012mq}), and it was argued in \cite{Pomarol:2019aae} that, even though a singlet scalar is typically the lightest resonance, it cannot become parametrically light without fine tuning.

We will propose three models, built within the bottom-up approach to holography, aimed at capturing the non-perturbative dynamics of the aforementioned HC gauge theory. Our models are able to incorporate both non-conformal dynamics, leading to a dynamically generated mass gap, as well as the backreaction of flavour on the geometry, as required by the Veneziano limit. Since global symmetries on the field theory side correspond to gauge symmetries in the bulk, we consider a bulk scalar field, charged under a gauged $SU(2N_F)$ symmetry, that is dual to a scalar-meson operator in the field theory. In this model, which we refer to as Model I, the scalar field acquires a non-trivial profile along the extra radial dimension, which can be used to describe the spontaneous breaking of flavour symmetry on the field theory side, due to the dual meson operator acquiring a vacuum expectation value (VEV). Moreover, the backreaction of the scalar on the metric causes the geometry to depart from AdS, eventually yielding
an end of the space in the deep IR, which in turn induces a mass gap in the spectrum. Since these two phenomena are described by the dynamics of a single bulk scalar field, on the field theory side the mass gap becomes dynamically linked to the scale $f$ of flavour symmetry breaking.

The spectrum contains spin-0 and spin-1 composite states in various flavour representations, as well as spin-2 states. Our Model I is engineered, by a special choice of the scalar profile, to realise a purely spontaneous breaking of the flavour symmetry, with the associated massless NGBs. At the same time, such scalar profile breaks spontaneously scale invariance, leading to a massless dilaton. 
In a more realistic scenario we expect additional sources to break explicitly scale invariance, and thus lift the mass of the dilaton, while the flavour NGBs could remain massless. To this end, we introduce a second bulk scalar field that is a flavour singlet, and whose dynamics encodes the explicit breaking of scale invariance. We will discuss two varieties of such scenario, referred to as Model IIA and IIB, which differ in the choice of scalar potential, allowing us to assess the generality of our results. Our framework turns out to be flexible enough to describe all possible combinations of spontaneous and explicit symmetry breaking, of the flavour symmetry and of scale invariance, and to study correspondingly the various limits where the Goldstone bosons and the dilaton become parametrically light.

While our models are constructed within the bottom-up approach to holography, the bulk scalar potentials that we consider can be thought of as modifications of the one of the GPPZ model~\cite{Girardello:1999bd}, originally proposed as the gravity dual of $\mathcal  N = 1^*$ super Yang-Mills (SYM), a confining four-dimensional gauge theory obtained as a mass deformation of $\mathcal  N = 4$ SYM. We choose this scalar potential because the corresponding backreaction of the scalar field on the geometry dynamically generates an end of space and a mass gap. Our modification is such that the amount of backreaction is related to $x_F$. When $x_F$ is small {\it and} the explicit breaking of conformal invariance is small, our models can be made to resemble the ones of \cite{Contino:2003ve,Agashe:2004rs} for which the background geometry is a slice of AdS. On the other hand, when $x_F$ is large, the backreaction of the flavour sector plays a dominant role in determining the geometry. Although our models incorporate the backreaction on the geometry, as required by the Veneziano limit, another subtlety arises due to the large number (order $N_F^2$) of degrees of freedom in the bulk, which implies potentially large quantum corrections on the gravity side~\cite{Pomarol:2019aae}.\footnote{Similar comments apply to top-down models, where in order to realize a $U(N_F)$ global symmetry, one would need to consider the non-Abelian DBI action of $N_F$ number of (coincident) flavour branes, which becomes strongly coupled in the Veneziano limit. This problem is to some extent overcome by the procedure of smearing the flavour branes~\cite{Nunez:2010sf,Bigazzi:2005md,Casero:2006pt,Bigazzi:2008zt,HoyosBadajoz:2008fw}, which however breaks explicitly the flavour symmetry to $U(1)^{N_F}$.} Since we treat our holographic models in the classical approximation, these effects will not be accounted for. As a consequence, when $x_F$ becomes large, our predictions for the spectrum are not expected to be accurate. Nevertheless, we will give arguments for why the classical approximation may still capture the spectrum of light states, even as $x_F \sim 1$.

The predictions of our holographic analysis can be confronted with other non-perturbative techniques applied, ideally, to the same HC gauge theory. These include lattice simulations, as well as phenomenological models that provide some analytic approximation of the strong dynamics, such as Nambu--Jona-Lasinio (NJL) \cite{NJL}. The latter approach was pursued by some of the authors in \cite{Bizot:2016zyu}, where general results on HC theories were also derived. We attempt a comparison of the bosonic spectrum from holography, with some of the existing lattice and NJL literature, underlining the similarities and differences in the specific gauge theory that is considered, and in the regime of parameters where each technique is more reliable. 

While in this paper we provide a comprehensive description of the composite spectrum in the bosonic sector, we postpone to a companion paper \cite{part2} the analysis of the fermionic spectrum in this scenario, especially relevant to implement fermion partial compositeness. While the fermions will be sensitive to the symmetry breaking pattern---through the background geometry and scalar profiles of the 5D gravity dual---which we define in this paper, their addition will not modify any of the results presented here.

This paper is organised as follows. In section~\ref{model} (see also Appendices~\ref{BL} and~\ref{beta}), we introduce the new gauge theory, proposed as the UV completion of the composite-Higgs scenario. In particular, in section~\ref{QFTops}, we present the most relevant operators that interpolate bosonic composite states. A full classification of the bosonic operators is provided in appendix~\ref{zoo}. In section~\ref{sec:warmup}, we introduce some minimal holographic models (with no flavour symmetry), to pedagogically illustrate the holographic dictionary, as well as the dependence of the results on the choice of the bulk scalar potential. This also allows us to demonstrate the formalism used to compute spectra, detailed in appendices~\ref{sec:sigmamodel} and~\ref{sec:AVandPS}. The holographic models describing the flavour sector are introduced in section~\ref{sec:models}, while the computation of their spectra is detailed in section~\ref{sec:bosonicspectrum}, where our results are illustrated as a function of the number of flavours $x_F$, the 5D gauge coupling, and the scaling dimension of the scalar operator(s). Section~\ref{sec:latticeNJL} contains the comparison of the spectrum to lattice simulations and the NJL model. Finally, we summarise and discuss our results in section~\ref{sec:Conclusions}.

\section{Requirements on the hypercolour gauge theory}\label{model}

We aim to describe electroweak symmetry breaking by a strongly-coupled sector, with the observed Higgs boson emerging as a composite, pseudo NGB.
In this section we characterise the gauge theory we will adopt as the ultraviolet (UV) completion of this scenario. Let us start from general considerations on the suitable class of gauge theories, before motivating our specific choice. Our analysis presents many similarities and some crucial differences with respect to the classification in \cite{Ferretti:2013kya}.

We restrict ourselves to gauge theories of fermions, with no fundamental scalar fields.
The latter would reintroduce a hierarchy problem, unless they are protected by supersymmetry, an avenue we do not pursue here.
We consider a HC gauge group $G_{HC}$ that is simple, with gauge coupling $g_{HC}$. We assume the theory is asymptotically free at very high energies, then enters a strongly-coupled, walking (approximately scale-invariant) regime at some UV scale $\Lambda_{UV}$, and eventually develops a mass gap at some IR scale $m_*$. 
A large walking region, that is, a hierarchy $m_* \ll \Lambda_{UV}$, is required to induce the SM Yukawa couplings and to suppress flavour violation at the same time.
The holographic models that we present in section~\ref{sec:models} are meant to describe the theory for scales $\mu < \Lambda_{UV}$, by providing a dual gravitational description in five dimensions; the geometries that we consider are asymptotically AdS in the UV, which on the field theory side corresponds to a UV fixed point, and end in the IR at some finite value of the radial coordinate, which produces a mass gap.

A crucial issue is the choice of the global symmetry of the theory. A generic gauge theory with (massless) fermions in $k$ different representations of $G_{HC}$ has flavour symmetry 
\begin{equation}\label{GF}
G_F= SU(N_1)\times \dots \times SU(N_k)\times U(1)^{k-1}~,
\end{equation}
 where $N_i$ is the number of Weyl fermions in the representation $R_i$. Note that the fermion kinetic term of each sector has symmetry 
 $G_F^i=U(N_i)=SU(N_i)\times U(1)$ at the classical level, 
but only $k-1$ linear combinations of the $U(1)$ symmetries have no anomaly with respect to $G_{HC}$.
We restrict ourselves to vector-like gauge theories, with complex representations in conjugate pairs, and/or real representations, and/or an even number of pseudoreal representations, that are free from local and global anomalies. In each sector the strong dynamics may either preserve $G_F^i$, or break it spontaneously to its vector subgroup $H_F^i$ \cite{Vafa:1983tf,tHooft:1979rat},
e.g. via the vacuum expectation value of a fermion-bilinear operator. The vector subgroup depends on the type of representation \cite{Peskin:1980gc},
\begin{equation}\ba{l}
\psi^1,\dots,\psi^N\sim R_i {~\rm real~}:~~ H_F^i = SO(N)~, \\
\psi^1,\dots,\psi^{2N}\sim R_i {~\rm pseudoreal~} :~~ H_F^i = Sp(2N) ~, \\
(\psi^1,\bar\psi_1),\dots,(\psi^N,\bar\psi_N) \sim (R_i,\bar{R}_i) {~\rm complex~} :~~ H_F^i = SU(N)_V \times U(1)_V~. 
\ea\label{RPC}\end{equation}
This stage of spontaneous symmetry breaking (SSB) occurs at some scale of order $m_*$. 

At least some of the constituent fermions should transform non-trivially under some of the SM symmetries,
providing several requirements on the coset $G_F\to H_F$,  as detailed in the following.
The electroweak group $SU(2)_L\times U(1)_Y$ must be embedded in $H_F$ in order to realise the SM Higgs 
as a composite NGB, $h\sim 2_{1/2}$. Generic strong dynamics would break the SM custodial symmetry, leading to a strong lower bound on $m_*$ from 
the electroweak precision tests, e.g. the $T$-parameter, which introduces a significant hierarchy problem.
This minimal scenario cannot be excluded, in the absence of a new physics signal at colliders. To insist on naturalness, we rather assume the 
strong dynamics to preserve the custodial symmetry, $H_F\supset SU(2)_L\times SU(2)_R$, with a NGB Higgs $h\sim (2_L,2_R)$.
The minimal realisation \cite{Galloway:2010bp,Barnard:2013zea,Cacciapaglia:2014uja}
  satisfying these requirements is provided by four Weyl fermions in the fundamental, 
pseudoreal representation of $G_{HC}=Sp(2N_C)$, $\psi^a$ with $a=1,2,3,4$ the flavour index.
The associated flavour symmetry is $G_F=SU(4)\to H_F=Sp(4)$, the only $U(1)$ factor being anomalous.

In addition to providing the composite NGB Higgs, the HC sector should couple to the SM fermions in order to induce their Yukawa couplings.
A linear mixing between the SM fermions and composite operators at the scale $\Lambda_{UV}$ is preferable, in order to generate sizeable Yukawa couplings
while suppressing flavour and CP violation. In contrast, the couplings of SM fermion-bilinears to
the composite sector can hardly satisfy all constraints \cite{Contino:2010rs,Panico:2015jxa},
in particular for the case of the large top-quark Yukawa.
This means that the HC sector has to contain operators with the quantum numbers of the top-quark doublet $q=(t~b)^T$ and singlet $t^c$,
in order to linearly mix with them.
Therefore, some of the constituent fermions should also carry colour and hypercharge, and the coset should satisfy
$H_F\supset SU(3)_c \times SU(2)_L \times SU(2)_R \times U(1)_X$.
Here the $U(1)_X$ factor is required to properly embed hypercharge, according to $Y=T^R_3+X$.

Besides the SM gauge symmetries and the approximate custodial symmetry, the SM has additional accidental symmetries, the baryon number $U(1)_B$
and the lepton number $U(1)_L$. As violations of these symmetries are severely constrained experimentally, the HC sector 
(as any new physics close to the electroweak scale) is bound to preserve them to a high degree of accuracy. 
As proved in appendix \ref{BL},
the (only) safe possibility is to require $U(1)_{B,L}$ to be part of the unbroken global symmetry $H_F$,
\begin{equation}
H_F\supset SU(3)_c \times SU(2)_L \times SU(2)_R  \times U(1)_B  \times U(1)_L~.
\label{HFparent}\end{equation}
This would imply ${\rm rank}(H_F) \ge 6$,
but for simplicity we will drop the $U(1)_L$ factor and restrict our analysis to $U(1)_B$. The latter is sufficient to realise quark partial compositeness, thus allowing for a minimal ${\rm rank}(H_F) = 5$.
In \eq{HFparent} we dropped the $U(1)_X$ factor introduced previously, because we will see below that, in the minimal model, $U(1)_B$ is sufficient to embed hypercharge satisfactorily.
The additional $U(1)_L$ can be introduced along the same lines, as outlined at the end of appendix \ref{BL}.

The required large rank of the global symmetry group calls for many flavours of constituent fermions. Here we argue that is preferable to
ascribe the large rank of $H_F$ to sufficiently many copies of the fundamental representation of $G_{HC}$.
In order to build all desired hypercolour-singlet operators, such as the top-quark partners,
it is sufficient to add one single fermion in a larger representation of $G_{HC}$, with no SM charges.
This class of models presents both practical and conceptual motivations:
\begin{itemize}
\item{Asymptotic freedom can be preserved in the large $N_C$ limit, despite a large number of flavours $N_F$.}
\item{The contribution of the HC sector to the running of the SM gauge coupling is minimised, in order to postpone Landau poles well above the scale $m_*$.}
\item{By embedding the SM gauge group into a {\it simple} subgroup of $G_F$, one can actually control gauge coupling unification, despite a
strongly-coupled composite sector \cite{Agashe:2005vg,Frigerio:2011zg}.}
\item{A ratio $x_F \equiv N_F/N_C$ of order one is typically required to sit close to the conformal window, and thus realise the desired approximately scale-invariant, walking behaviour over a large range of scales, $m_*\ll\Lambda_{UV}$.}
\end{itemize}
The analysis of the beta-function of HC is presented in detail in appendix \ref{beta}, where we also comment on the (perturbative) conditions for an IR fixed-point, and on the SM gauge coupling evolution.
The gauge-gravity duality in the Veneziano limit, large $N_C$ and large $N_F$, was not extensively explored in the composite-Higgs literature. In this regime the flavour sector backreacts significantly on the metric, while most studies treat the flavour sector as a probe on a fixed background. While such study is technically challenging, we will see in the coming sections that it may lead to physically novel effects.

To explicitly realise the above requirements on the gauge theory, we will focus for definiteness on the case
\begin{equation}
G_F = SU(2N_F) \times U(1) \to H_F = Sp(2N_F)~,
\label{SSB}\end{equation}
corresponding to $2N_F$ Weyl fermions in the fundamental of $G_{HC}=Sp(2N_C)$, $\psi^a\sim {\Yvcentermath1\tiny\yng(1)}$\,, with $a=1,\dots,2N_F$.
The extra $U(1)$ factor originates from the presence of one additional Weyl fermion in a two-index 
representation of $Sp(2N_C)$, see the discussion following \eq{GF}.
We will consider both possible options: $\chi\sim{\Yvcentermath1 \tiny \yng(1,1)}$ 
in the two-index antisymmetric, traceless representation, or $\chi'\sim{\Yvcentermath1 \tiny \yng(2)}$
 in the two-index symmetric representation.
Note that, while the fundamental is pseudoreal, both two-index representations are real, so such single Weyl fermion does not introduce a gauge anomaly.
A two-index fermion is needed \cite{Barnard:2013zea,Bizot:2016zyu} to build hypercolour-singlet trilinear operators 
$(\psi\psi \chi)$ or $(\psi\psi\chi')$, that correspond to fermionic composite states such as top-quark partners. The composite fermion spectrum will be studied in detail in a companion paper \cite{part2}.

The minimal model has $N_F=5$, with
\be
H_F = Sp(10)\supset SU(3)_c \times SU(2)_L \times SU(2)_R \times U(1)_B~.
\label{3221}
\ee
The detailed decomposition of the relevant $Sp(10)$ representations is provided in \eq{decompo}.
The hypercharge can be embedded in two ways, according to $Y=\pm T^R_3+ B/2$. One can check that $(\psi^a\psi^b)$ contains a component with
the Higgs quantum numbers, $h\sim (1,2,2)_0$, and that $(\psi^a\psi^b\chi)$ contains components with the quark quantum numbers,
\be
q_L
\in (3,2,1)_{1/3}~, \qquad t_R^{\ c}\in (\overline 3, 1,2)_{-1/3}~, \qquad b_R^{\ c}\in (\overline 3, 1,2)_{-1/3} {\rm~or~} (\overline 3, 1,1)_{2/3}~.
\label{topP}
\ee

A large number of flavours in the HC fundamental representation implies large multiplets of composite states. 
Indeed, the bilinear $\psi^a \psi^b$ carries two $Sp(2N_F)$ indexes, which amount to $(2N_F)^2$ components. This is not a problem for heavy multiplets, whose mass is of the order of the mass gap. However one may worry about the Goldstone multiplet, that includes several states beside the Higgs. 
They eventually receive non-zero masses from various sources of $SU(2N_F)$ explicit symmetry breaking: SM couplings and/or HC fermion masses. The NGB masses induced by the SM gauge couplings and from a
mass term $m_{ab}\psi^a\psi^b$ are detailed in section II.E of \cite{Bizot:2016zyu}. For example, the coloured NGBs, subject to the strongest experimental constraints, 
receive a sizeable mass from gluon one-loop contributions. If needed, they can be further lifted by giving a mass $m_a$ only to those flavours $\psi^a$ which carry colour. For the phenomenology of composite Goldstone bosons see e.g. \cite{Belyaev:2016ftv,Cacciapaglia:2019bqz,Cacciapaglia:2020vyf}. We will see that the spectrum can also include a light SM-singlet associated with SSB of scale invariance, the dilaton: for its phenomenology see e.g. \cite{Serra:2013kga,Abu-Ajamieh:2017khi,Ahmed:2019csf}.

We remark that our scenario, both for the gauge theory and for the dual gravity description, could be easily generalised to other classes of theories. By replacing $Sp(2N_C)$ by $SO(N_C)$, whose fundamental representation is real, the analysis would remain analogous in most respects. 
In this case, for $N_F$ fundamental fermions $\psi^a$ and one two-index fermion $\chi$, 
the associated flavour coset becomes $SU(N_F)\times U(1)\to SO(N_F)$. The case $G_{HC}=SU(N_C)$, with a complex fundamental representation, could be treated similarly as well, by introducing conjugate pairs
$(\psi,\psi^c)^a$.

\subsection{Significant operators}\label{QFTops}
\label{sec:sigop}

Before developing the gravity dual description, let us identify the set of field theory operators that we want to study.
Analogously to QCD, there is a long (infinite) series of HC-invariant operators, in all sorts of Lorentz and flavour representations, 
and an associated plethora of composite states of various spins organised in flavour multiplets. 
In this paper we focus on bosonic operators only, while an analogous analysis of fermionic operators will be presented in \cite{part2}.
While a systematic classification is presented in appendix \ref{zoo}, here we restrict ourselves to a significant subset of bosonic operators,
which (i) are especially relevant to characterise the low energy spectrum of the theory, and (ii) admit a simple gravity-dual description.

Let us begin by recalling the definition of the energy-momentum tensor. In terms of the action $\mathcal S = \int \dd^4x \sqrt{-g} \mathcal L$, where ${\cal L}$ is the Lagrangian density of matter fields and $g_{\mu\nu}$ is the four-dimensional metric, it is given by
\be
T_{\mu\nu} \equiv - \frac{2}{\sqrt{-g}}\frac{\delta {\mathcal S}}{\delta g^{\mu\nu}} =
\left(T_{\mu\nu} - \dfrac 14 T_{~\rho}^\rho \, g_{\mu\nu}\right) + \dfrac 14 T_{~\rho}^\rho \, g_{\mu\nu}~,\qquad
T_{\mu\nu}=T_{\nu\mu}~,\qquad \nabla^\mu T_{\mu\nu}=0~. 
\label{tmunu}\ee
As $T_{\mu\nu}$ is symmetric and conserved, it contains six independent degrees of freedom: the trace $T_{~\mu}^{\mu}$ 
transforms as a Lorentz scalar, while the traceless part, in brackets in \eq{tmunu}, transforms in the Lorentz representation $(1,1)$. In the gravity description, the UV-boundary value of the 5D metric will be the source for  $T_{\mu\nu}$.
 
Operators involving only the HC gauge fields describe the glueball sector of the spectrum. Let us consider the gauge kinetic term, ${\cal L}^{(g)}=-(1/4)F^A_{\mu\nu}F^{A\mu\nu}=-(1/2)\Tr(F_{\mu\nu}F^{\mu\nu})$,
where we defined $F_{\mu\nu}\equiv F^A_{\mu\nu}T^A$, with $T^A$ the $Sp(2N_C)$ generators. Here and elsewhere we normalise the generators 
by $2\,\Tr (T^A T^B) = \delta^{AB}$.
The corresponding part of the energy-momentum tensor reads
\be
T_{\mu\nu}^{(g)}\equiv - \frac{2}{\sqrt{-g}}\frac{\delta {\mathcal S}^{(g)}}{\delta g^{\mu\nu}} 
= 2 \Tr (F_{\mu\rho} F_{\nu}^{\ \rho}) - \frac{1}{2} g_{\mu\nu} \Tr (F_{\rho\sigma} F^{\rho\sigma}) ~.
\label{tmunug}\ee
When the model includes fermion flavours, glueballs and mesons mix. In the absence of fermions, the trace operator $\Tr (F_{\rho\sigma} F^{\rho\sigma})$ describes spin-0 glueballs, while the traceless part of $T^{(g)}_{\mu\nu}$ describes spin-2 glueballs. In appendix \ref{glue} we present a more general classification of the operators that can be built with $F_{\mu\nu}$.

Coming to HC fermions, $\psi^a$ and $\chi$, the simplest HC-invariant operators are fermion bilinears (their structure remains the same if $\chi$ is replaced by  $\chi'$). Let us begin from the Lorentz scalars,
\be
S^{ab}\equiv (\psi^a\psi^b)
~,\qquad s\equiv (\chi\chi)~,
\label{eq:LorentzScalarOps}\ee
which will be sourced by bulk scalar fields $\Phi^{ab}$ and $Z$, respectively.
The antisymmetry $S^{ab}=-S^{ba}$ is a combined consequence of the anticommutation of the two spinors, and the contraction of Lorentz and HC indexes, which are both antisymmetric.
To minimise notations, we relegate the analysis of index contractions to appendix \ref{Sbili}, see in particular table \ref{bili}.
The SSB of \eq{SSB} can be realised by  the VEV of $S^{ab}$, that is proportional to a real, antisymmetric, dimensionless matrix $\Sigma^{ab}$, normalised such that $\Sigma^2 = -\mathds 1$. The VEV of $s$ also contributes to the spontaneous breaking of $U(1)$.
The complex operator $S^{ab}$ can be decomposed in scalar and pseudoscalar parts, as well as in trace ($S^{ab}\Sigma_{ab}$) and traceless parts.
The Goldstone bosons $\pi^{ab}$ are associated to the pseudoscalar, traceless part.
An additional, singlet Goldstone $\eta$ is associated to a linear combination of the pseudoscalar parts of $S^{ab}\Sigma_{ab}$ and $s$. 
The spectrum includes also a pseudoscalar $\eta'$, whose mass is generated by  the HC anomaly.

Among the fermion-bilinear, Lorentz-vector operators we have the
currents
of the flavour symmetry,
\be
\label{eq:current}
\mathcal J^{~~~b}_{\mu a} \equiv (\overline{\psi}_a
\overline{\sigma}_\mu \psi^b)~,\qquad
\mathcal J_\mu\equiv (\overline{\chi}\,\overline{\sigma}_\mu\chi)~.
\ee
The traceless part of $\mathcal J^{~~~b}_{\mu a}$ transforms in the
adjoint of $SU(2N_F)$.
The trace $\mathcal J^{~~~a}_{\mu a}$ as well as $\mathcal J_\mu$ are
$SU(2N_F)$ singlets, associated to the fermion-number symmetries
$U(1)_\psi\times U(1)_\chi$.
Only one linear combination of these is anomaly-free with respect to
HC, and therefore a true $U(1)$ global symmetry of the model.
These currents are associated to vector and axial-vector resonances,
according to the
symmetry-breaking pattern in \eq{SSB}: the vector generators
$T^A$ of the global $SU(2N_F)$ symmetry group are provided by those of
the unbroken $Sp(2N_F)$ subgroup, whereas the axial generators $T^{\hat
A}$ are those that span the symmetric coset space $SU(2N_F)/Sp(2N_F)$.
They are conveniently characterized by the identities
\beq
\label{eq:genid}
        T^A \Sigma + \Sigma (T^A)^T = 0 \,, \hspace{1cm}
        T^{\hat A} \Sigma - \Sigma (T^{\hat A})^T = 0 \,,
\eeq
which will prove useful in the following. Explicitly, these vector
and axial
currents read, respectively,
\be
\label{eq:vector_and_axial_currents}
J_\mu^A = \mathcal J^{~~~b}_{\mu a} (T^A)^a_{~~b} \,, \hspace{1cm}
J_\mu^{\hat A} = \mathcal J^{~~~b}_{\mu a} (T^{\hat A})^a_{~~b} \,.
\ee
The $SU(2N_F)$ currents will be sourced by bulk gauge fields $\mathcal
A^{~~~b}_{Ma}$ in the adjoint,
with $M$ a 5D spacetime index. The anomaly-free $U(1)$ current can be sourced by a bulk gauge field $a_M$.
The Goldstone decay constant $F_G$ can be extracted from the residue at $q^2 = 0$ of the two-point function of the axial current $J^{\hat A}_\mu$,
according to\footnote{\label{FourierConvention}
Our conventions for the transverse projector and the Fourier transforms are
\beq
	P^{\mu\nu} = \eta^{\mu\nu}-\frac{q^\mu q^\nu}{q^2}~, \qquad
	\phi(q) = \int \frac{\dd^4x}{(2\pi)^2} e^{-i q_\mu x^\mu} \phi(x) ~, \qquad
	\delta^4(q) = \int \frac{\dd^4x}{(2\pi)^4} e^{i q_\mu x^\mu} ~. \nonumber
\eeq
with Minkowski metric $\eta^{\mu\nu}={\rm diag}(-1,+1,+1,+1)$.
} 
\beqs
        q^2 \Pi_A(q^2) P^{\mu\nu} \delta^{{\hat A}{\hat B}} &\equiv & i \, \int \dd^4
x \, e^{-i q_\sigma x^\sigma} \langle J^{{\hat A}\mu}(x) J^{{\hat
B}\nu}(0) \rangle = i \, \int \dd^4 p \langle J^{{\hat A}\mu}(q)
J^{{\hat B}\nu}(p) \rangle \,, \\
        F_G^2 &\equiv& \lim\limits_{q^2 \rightarrow 0} \left\{ - q^2 \Pi_A(q^2)
\right\} \,.\label{FG}
\eeqs
We note that $F_G$ is related to the decay constant $f$ commonly used in the composite-Higgs literature according to 
\beq
f\equiv \sqrt{2} F_G~.
\label{fF}\eeq

\section{Symmetry breaking in holography with backreaction}
\label{sec:warmup}

As a warm-up exercise, let us first study two examples of holographic models that exhibit explicit and/or spontaneous breaking of a global $U(1)$ symmetry. In these models, a bulk scalar field charged under a $U(1)$ gauge symmetry acquires a non-trivial profile as a function of the radial coordinate. We are interested in the case when backreaction on the metric is important, and hence we consider that symmetry breaking occurs in a sector that in the bulk is described by an action that is order $N_C^2$. This can be the case when the operator dual to the bulk scalar is built up of fields in a two-index representation of the HC group. We find these simple examples to be useful in illustrating the formalism as well as building intuition for the case that we ultimately have in mind, namely field theories with fermions in the fundamental representation of hypercolour, for which backreaction in the bulk description becomes important at large number of flavours $N_F \sim N_C$.

To this end, consider a model consisting of gravity, two gauge fields $V_M$ and $A_M$ corresponding to a $U(1)_V \times U(1)_A$ gauge symmetry, and a complex scalar $\mathcal X = \frac{\sigma}{\sqrt{2}} e^{i \pi}$ charged under $U(1)_A$. The scalar $\mathcal X$ is uncharged under the $U(1)_V$, which remains unbroken, but which we include in view of the larger cosets that we will consider in the next section. The action is given by
\beqs
\label{eq:simpleaction}
	\mathcal S &=& \frac{1}{4\pi G_5} \int \dd^5x \sqrt{-g} \, \bigg\{ \hspace{-0.1cm}
	\frac{R}{4} - |D_M \mathcal X|^2 - \mathcal V(|\mathcal X|) - \frac{1}{4} \sum_{i=V,A} F_{MN}^{(i) \, 2} \bigg\} \\ \nonumber
	&=& N_C^2 L^{-3} \int \dd^5x \sqrt{-g} \, \bigg\{ \hspace{-0.1cm}
	\frac{R}{4} - \frac{1}{2} (\partial_M \sigma)^2 - \mathcal V(\sigma) - \frac{1}{4} \sum_{i=V,A} F_{MN}^{(i) \, 2} 
	- \frac{1}{2} \sigma^2 (\partial_M \pi + g_5 A_M)^2
	 \bigg\} \,,
\eeqs
where $D_M \mathcal X = (\partial_M + i g_5 A_M) \mathcal X$, and our conventions are such that the metric has mostly plus signature, five-dimensional indices take values $M,N = 0,1,2,3,5$, and we denote four-dimensional indices by Greek letters $\mu, \nu = 0,1,2,3$. The bulk fields $\mathcal X$, $A_M$, and $V_M$ are normalized to be dimensionless, while the five-dimensional gauge coupling $g_5$ has dimensions of inverse length.

We will assume the potential $\mathcal V$ has a maximum at $\mathcal X = 0$ with $\mathcal V(0) = - \frac{3}{L^2}$, so that the action admits an AdS solution with scale $L$. In the following, we fix units such that $L = 1$, and hence the 5D Newton's constant is given by $G_5 = \frac{1}{4\pi N_C^2}$. Note that the prefactor $N_C^2$ of the action implies that correlators of the operators dual to the bulk fields scale as $N_C^2$. The precise overall normalization can only be fixed in a fully top-down model. We have also made the simplifying assumption that the prefactors of the kinetic terms are constant, rather than functions of the scalar fields of the model as is generically the case in supergravities with dual field theory descriptions.

Let us consider background solutions in which only the scalar $\sigma$ and the metric are turned on. In order to preserve four-dimensional Poincare invariance, we furthermore assume that the background fields only depend on the radial coordinate, and that the metric has the domain wall form
\beq
\label{eq:domainwall}
	\dd s^2 = g_{MN} \dd x^M \dd x^N = \dd r^2 + e^{2A} \dd x_{1,3}^2 \,,
\eeq
where $A(r)$ is the warp factor, and the Minkowski part $\dd x_{1,3}^2$ has the metric $\eta_{\mu\nu} = {\rm diag} (-1,1,1,1)$. The background equations of motion are then given by
\beqs
\label{eq:eomsAB}
	\partial_r^2 \sigma + 4 \partial_r A \partial_r \sigma - \frac{\partial \mathcal V}{\partial \sigma} &=& 0 \,, \\ \nonumber
	6 (\partial_r A)^2 - (\partial_r \sigma)^2 + 2 \mathcal V &=& 0 \,.
\eeqs

The solutions we will consider flow to the UV fixed point at $\sigma = 0$ as $r \rightarrow +\infty$, and are asymptotically AdS corresponding to $A \simeq r$. Under this assumption, the general solution of $\sigma$ has the UV expansion given by\footnote{For the special case $\Delta_\pm = \Delta = 2$, the two modes behave as $e^{-2r}$ and $r e^{-2r}$.}
\beq
\label{eq:sigmaUV}
	\sigma = \sigma_- (e^{-\Delta_- r} + \cdots) + \sigma_+ (e^{-\Delta_+ r} + \cdots) \,, \hspace{1cm} \Delta_\pm \equiv 2 \pm \sqrt{4 + M_\sigma^2} \,,
\eeq
where $M_\sigma^2$ is the bulk mass (squared) of $\sigma$ obtained from expanding the potential around the UV fixed point, i.e. $\mathcal V(\sigma) = -3 + \frac{1}{2} M_\sigma^2 \sigma^2 + \cdots$. As usual in gauge-gravity duality, $\sigma_-$ corresponds to turning on a source for the operator $\mathcal O_\sigma$ (dual to the bulk field $\sigma$), while the presence of $\sigma_+$ indicates that the same operator acquires a VEV. The scaling dimension of $\sigma_\pm$ is $[\sigma_\pm] = \Delta_\pm$, while for the dual operator $[\mathcal O_\sigma] = \Delta_+$.

In a geometry given by the metric of Eq.~\eqref{eq:domainwall} with a warp factor $A$, the field theory energy scale $\Lambda$ corresponding to the bulk radial coordinate $r$ can be estimated to be \cite{Csaki:2000cx}\footnote{For AdS, $A(r) = r$, and one obtains the familiar result $\Lambda(r) = e^{r} \equiv z^{-1}$.}
\beq
\label{eq:Lambdar}
	\Lambda(r)^{-1} = \int_r^{\infty} \dd \tilde r \, e^{-A(\tilde r)} \,.
\eeq
The solutions we will consider have an end of space in the IR of the geometry at some value of the radial coordinate $r = r_o$, either dynamically generated or imposed by hand. The IR scale, typically related to the mass gap of the theory, can be estimated to be $\Lambda_{\rm IR} \equiv \lim\limits_{r \rightarrow r_o} \Lambda(r)$.

It is convenient to pick the scalar potential $\mathcal V$ so that it can be written in terms of a function $\mathcal W$, commonly called superpotential,
\beq
	\mathcal V(\sigma) = \frac{1}{2} (\partial_\sigma \mathcal W)^2 - \frac{4}{3} \mathcal W^2 \,.
\label{specialV}\eeq
Even though this form is inspired by supergravity, we do not assume any supersymmetry here.
It is then straightforward to verify that solutions to the equations of motion given in Eq.~\eqref{eq:eomsAB} can be found by solving 
the first-order equations
\be
\label{eq:eomWsimp}
	\partial_r \sigma = \partial_\sigma \mathcal W \,, \qquad\qquad
	\partial_r A =  - \frac{2}{3} \mathcal W \,.
\ee
This conveniently allows one to generate a subset of the general class of solutions of the system.

We will soon consider two explicit forms of the superpotential, dubbed Examples A and B below, leading to different kinds of dynamics (the scalar sector in both cases was studied in \cite{Elander:2010wd}). Our motivation is to gain intuition to build holographic models suitable to describe composite-Higgs models, that we will construct later in section~\ref{sec:models}. In Example~A, the superpotential is quadratic in $\sigma$ and, depending on the coefficient of the quadratic term, the background solution may describe either the spontaneous or explicit breaking of the $U(1)_A$ symmetry. This simple setting illustrates the formalism for computing the spectrum and the Goldstone decay constant. However, we are ultimately interested in models for which the bulk scalar dynamics is responsible for the end of the space in the IR. This requires knowing also higher order terms in the superpotential, which affect the IR dynamics of the dual field theory. Such terms can only be determined in a top-down framework, where the holographic model has a rigorous origin in supergravity. Our models will be constructed within the bottom-up approach to holography. Nonetheless, in Example B we take inspiration from the top-down  GPPZ model  \cite{Girardello:1999bd}, slightly generalising its superpotential. The GPPZ model was originally introduced to describe $\mathcal N = 1^*$ SYM, a four-dimensional confining gauge theory obtained as a mass deformation of $\mathcal N = 4$ SYM. Its superpotential is such that the end of space is dynamically generated, leading to a mass gap and a discrete spectrum. These features will allow us to construct a model in which the symmetry breaking scale becomes dynamically linked to the mass gap. Example~B will provide the starting point for the composite-Higgs models in section~\ref{sec:models}, where we will identify which kind of scalar potential is suitable to capture the essential features of the dual field theory.

The field-theory dynamics can be probed by computing the spectrum of composite states and, in the cases when a Goldstone boson is present, its decay constant $F_G$. We summarise here the formalism used to compute these observables, and refer the Reader to appendices~\ref{sec:sigmamodel} and~\ref{sec:AVandPS} for more details. The spectrum can be found by solving the linearised equations of motion for the fluctuations of the bulk fields around a given background and imposing appropriate boundary conditions in the IR and the UV, chosen such that they extract the poles of the two-point functions of the corresponding dual field theory operators. This typically limits the space of solutions to discrete values of the four momenta $q^2$, which in turn give the mass spectrum according to $m^2 = -q^2$. In practice, and in order to make the problem numerically tractable, one introduces IR (UV) cutoffs at $r = r_1$ ($r = r_2$). Setting up the boundary conditions at $r_{1,2}$, one then studies the spectrum taking $r_1 \rightarrow r_o$ in the IR, and $r_2 \rightarrow +\infty$ in the UV, making sure that cutoff effects are negligible. 

The scalar sector of the model consists of a fluctuation $\varphi(q,r)$ of $\sigma(q,r) =  \sigma(r) +\varphi(q,r)$ around the background solution $\sigma(r)$. Such fluctuation in general mixes with the scalar fluctuation $h(q,r)$ of the metric defined in Eq~\eqref{eq:ADM}. As explained in appendix~\ref{sec:sigmamodel}, it is convenient to form a  gauge-invariant (diffeomorphism-invariant) combination~\cite{Bianchi:2003ug,Berg:2005pd,Berg:2006xy,Elander:2009bm,Elander:2010wd} of the two which in the present case reads $\mathfrak a(q,r) = \varphi(q,r) - \frac{\sigma'}{6A'} h(q,r)$ where primes denote derivatives with respect to $r$. The spectrum of scalar composite states can then be obtained by solving the equation of motion given in Eq.~\eqref{eq:fluceoms}
\beq
	\left[ \partial_r^2 + 4 A' \partial_r - \left( \partial_\sigma^2 \mathcal V + \frac{8 \sigma'}{3A'} \partial_\sigma \mathcal V + \frac{16\sigma'^2}{9A'^2} \mathcal V \right) - e^{-2A} q^2 \right] \mathfrak a = 0 \,,
\eeq
imposing boundary conditions given in Eq.~\eqref{eq:flucbcs}
\beq
	\partial_r \mathfrak a \Big|_{r_i} = \frac{3A'}{2\sigma'^2}  \left[ e^{-2A} q^2 - \frac{A'}{2} \partial_r \left( \frac{A''}{A'^2} \right) \right] \mathfrak a \Big|_{r_i} \,.
\eeq
Here and below it is understood that the coefficients appearing in the equations of motion for the fluctuations and in the boundary conditions are evaluated on the background solution.

Similarly, the spectrum of spin-2 resonances can be found by solving the equations of motion for the tensor part of the metric $\mathfrak e^\mu{}_\nu(q,r)$ given in Eq.~\eqref{eq:fluceoms2}
\beq
\label{eq:fluceoms22}
	\Big[\partial_r^2 + 4A' \partial_r - e^{-2A} q^2 \Big] \mathfrak e^\mu{}_\nu = 0 \,,
\eeq
with boundary conditions $\partial_r \mathfrak e^\mu{}_\nu |_{r_i} = 0$.

The transverse part of the vector $V_M(q,r)$ satisfies the equation of motion
\beq
\label{eq:eomV}
	\Big[ \partial_r^2 + 2A' \partial_r - q^2 e^{-2A} \Big] P^{\mu\nu} V_\nu(q,r) = 0 \,,
\eeq
where $P^{\mu\nu}$ is defined in footnote \ref{FourierConvention}. The boundary conditions are given by $P^{\mu\nu} \partial_r V_\nu  |_{r_1} = 0$ and $P^{\mu\nu} V_\nu  |_{r_2} = 0$.

The transverse part of the axial-vector $A_M(q,r)$ satisfies the equation of motion given in Eq.~\eqref{eq:eomA}
\beq
\label{eq:eomA1}
	\Big[ \partial_r^2 + 2A' \partial_r - \left( q^2 e^{-2A} + g_5^2 \sigma^2 \right) \Big] P^{\mu\nu} A_\nu(q,r) = 0 \,,
\eeq
with boundary conditions $P^{\mu\nu} \partial_r A_\nu  |_{r_1} = 0$ and $P^{\mu\nu} A_\nu |_{r_2} = 0$. The pseudoscalar $X(q,r)$ (which is a combination of $\pi$ and the longitudinal part of $A_M$ defined by Eq.~\eqref{eq:piAL}) satisfies the equation of motion given in Eq.~\eqref{eq:eomX}
\beq
\label{eq:eomX2}
	\bigg[ \partial_r^2 - 2 \left( A' + \frac{\sigma' }{\sigma} \right) \partial_r - \left( q^2 e^{-2A} + g_5^2 \sigma^2 \right) \bigg] X(q,r) = 0 \,,
\eeq
with boundary conditions $X \big|_{r=r_1} = 0$ and $\partial_r X \big|_{r=r_2} = 0$. For illustrative purposes, in appendices~\ref{sec:sigmamodel} and~\ref{sec:AVandPS}, we show the explicit calculation of two-point functions in the axial-vector, pseudoscalar sector for Example~B introduced below.

Finally, when the breaking of $U(1)_A$ is purely spontaneous, the Goldstone decay constant $F_G$ can be extracted from \eq{FG},
using the expression for the axial current two-point function $\langle J^{\mu}(q) J^{\nu}(-q) \rangle$ obtained from Eq.~\eqref{eq:JJT}.\footnote{With a slight abuse of notation, we will frequently write
\beq
	\langle \mathcal O(q) \mathcal O(p) \rangle = \delta^4(q+p) \langle \mathcal O(q) \mathcal O(-q) \rangle \,, \nonumber
\eeq
thus extracting the non-trivial part of a two-point function, that does not follow from momentum conservation.
} 
Observing that its longitudinal part vanishes due to the purely spontaneous breaking of $U(1)_A$, we then have
\beq
\label{eq:decayconstant}
	F_G^2 = \lim\limits_{r \rightarrow \infty} \bigg\{ N_C^2 \frac{e^{2A}}{g_5^2} \frac{\partial_r a}{a} \Big|_{q^2 = 0} \bigg\} \,,
\eeq%
where in the derivation we wrote $P^{\mu\nu} A_\nu(q,r) = \tilde A^\mu(q) a(q,r)$, such that $a(q,r)$ satisfies the equation of motion given in Eq.~\eqref{eq:eomA1} with IR boundary condition $\partial_r a |_{r_1} = 0$. Note that $a(q,r)$ is defined up to an overall rescaling that does not affect the result for the decay constant $F_G$. In the class of models of this section, the decay constant scales as $F_G \sim N_C$. Taking out the factor of $N_C$, we define $\tilde F_G \equiv F_G / N_C$.

\subsection{Example A}
\label{sec:ExampleA}

A simple superpotential giving rise to dynamics that deviate from the pure AdS case is given by
\beq
	\mathcal W(\sigma) = - \frac{3}{2} - \frac{\Delta}{2} \sigma^2 \,,
\label{WA}\eeq
where $\Delta$ is a free parameter. By solving the first order equations Eq.~\eqref{eq:eomWsimp} coming from the superpotential, one finds the solution
\beqs
\label{eq:ExampleAsolutions}
	\sigma(r) &=& \tilde \sigma \, e^{-\Delta r} \,, \\ \nonumber
	A(r) &=& r - \frac{\tilde \sigma^2}{6} e^{-2 \Delta r} \,,
\eeqs
where $\tilde \sigma$ is an integration constant, and we fixed another integration constant corresponding to a shift in the warp factor $A$ such that $A \simeq r$ for large values of $r$.\footnote{This can always be accomplished by rescaling the boundary coordinates and hence corresponds to an overall rescaling of energies.} In order to obtain a mass gap, we introduce a hard-wall cutoff in the IR, roughly mimicking confinement, as triggered by the vacuum expectation value of a operator with infinite dimension~\cite{Rattazzi:2000hs}. Without loss of generality, we can choose the IR cutoff to be at $r_o=0$. The parameters of the model are hence $\Delta$, $g_5$, and $\tilde \sigma$.

Let us first consider the meaning of $\Delta$. By comparing the general UV expansion of $\sigma$ given in Eq.~\eqref{eq:sigmaUV} with the solutions picked by the superpotential and given in Eq.~\eqref{eq:ExampleAsolutions}, we see that for $\Delta < 2$ the CFT is deformed by the operator $\mathcal O_\sigma$ with scaling dimension equal to $4 - \Delta$, whereas for $\Delta > 2$, the same operator has scaling dimension $\Delta$ and acquires a VEV. In the latter case, the VEV breaks the global $U(1)_A$ in the dual field theory, and we hence expect a Goldstone boson in the spectrum. In addition, we also expect a massless dilaton to be present, because $\sigma_-$ vanishes identically in \eq{eq:sigmaUV}, and consequently scale invariance is not broken explicitly. Notice that this is an artefact
of choosing background solutions generated by  the first-order equations (\ref{eq:eomWsimp}), with a superpotential given by  \eq{WA}.

The meaning of the bulk gauge coupling $g_5$ could be estimated from considering the large $q^2$ behaviour of the transverse part of the (axial-)vector current two-point function, and matching with the perturbative field theory result along the lines of holographic QCD \cite{Erlich:2005qh}. However, since such a calculation matches a strong coupling result with a perturbative one, it should be taken with caution. We will instead investigate the dependence of the spectrum and decay constant on $g_5$, treating it as a free parameter, with the aim of extracting the generic behaviour. We will return to this point at the end of section~\ref{Form}.

\begin{figure}[t]
\begin{center}
\includegraphics[width=\figwidthsmall]{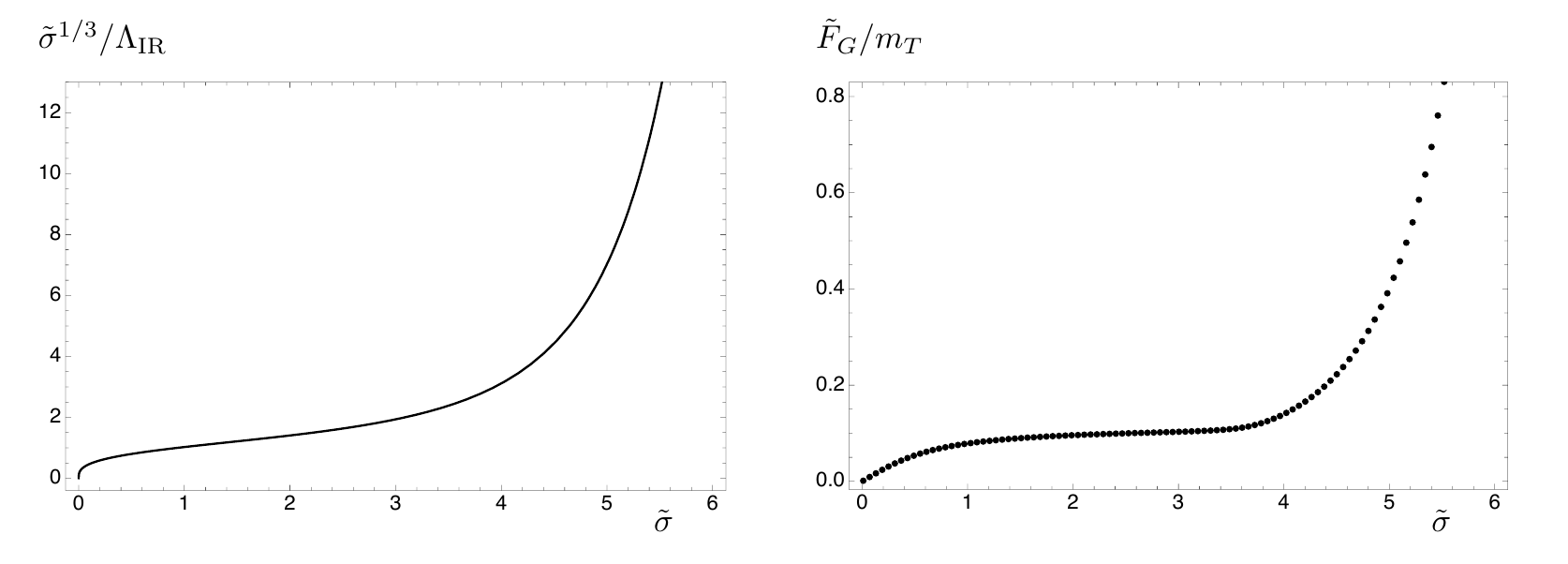}
\caption{Example A. The dimensionless ratio $\tilde \sigma^{1/3} / \Lambda_{\rm IR}$ (left panel) and the decay constant (right panel) as a function of $\tilde \sigma$ for $\Delta = 3$, $g_5 = 5$. We used the UV cutoff $r_2 = 8$ when extracting the decay constant.}
\label{fig:ExampleA2}
\end{center}
\begin{center}
\includegraphics[width=\figwidth]{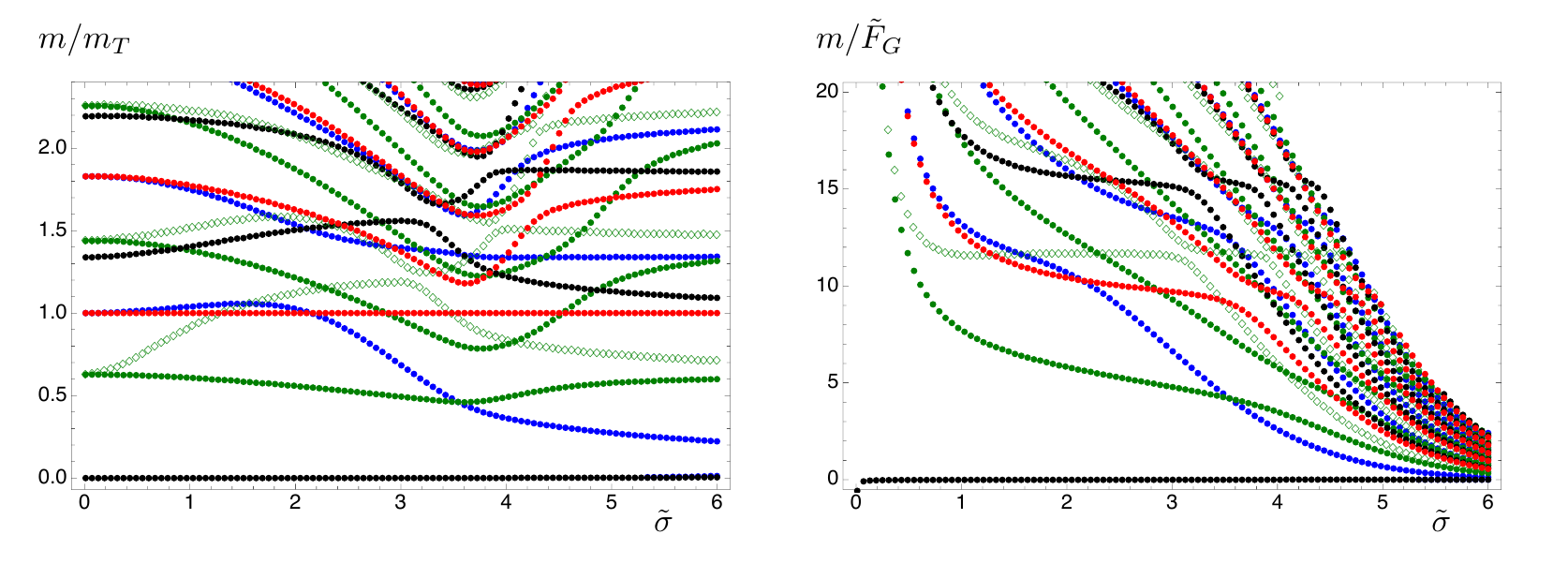}
\caption{Example A. Spectrum as a function of $\tilde \sigma$ for $\Delta = 3$, $g_5 = 5$, $r_1 = 0$, $r_2 = 8$. The left panel is normalized to the mass $m_T$ of the lightest tensor, while the right panel is normalized to the decay constant $\tilde F_G = F_G/N_C$. The colour coding for the spectrum is: scalar (blue), tensor (red), pseudoscalar (black), vector (green), axial-vector (green diamonds). Since $\Delta > 2$, there is also a massless dilaton in the spectrum, as can be seen from Fig.~\ref{fig:ExampleA3}; due to the numerical resolution deployed only the Goldstone is shown in the present two plots. In the right panel, it is understood that heavier states are present in the right-upper corner.}
\label{fig:ExampleA1}
\end{center}
\end{figure}

From Eq.~\eqref{eq:sigmaUV} and Eq.~\eqref{eq:ExampleAsolutions}, we see that $\tilde \sigma$ is related to the size of the source (for $\Delta < 2$) or the VEV (for $\Delta > 2$) of the operator $\mathcal O_\sigma$. On the gravity side, one can independently vary the value of $\tilde \sigma$ and $\Lambda_{\rm IR}$ (through the choice of $r_o$). We may consider three regimes, in which the dimensionless ratio $\tilde \sigma^{1/\Delta} / \Lambda_{\rm IR}$ is either much smaller than one, order one, or much larger than one. On the field theory side, these three regimes are characteristed by different physics. We illustrate this for the case $\Delta = 3$ in Fig.~\ref{fig:ExampleA2}, where we plot the decay constant $F_G$ in units of mass $m_T$ of the lightest tensor (characterising the scale of the mass gap) as a function of the parameter $\tilde \sigma$. As can be seen, when $\tilde \sigma^{1/\Delta} / \Lambda_{\rm IR} \ll 1$ the decay constant can be made arbitrarily small close to $\tilde \sigma = 0$. Conversely, when $\tilde \sigma^{1/\Delta} / \Lambda_{\rm IR} \gg 1$ the decay constant becomes large compared to $m_T$. Finally, when $\tilde \sigma^{1/\Delta} / \Lambda_{\rm IR} \sim 1$, there is a plateau between $1 \lesssim \tilde \sigma \lesssim 4$ in which the decay constant (in units of $m_T$) stays nearly constant. This intermediate regime can be expected to capture the features of a generic strongly coupled field theory with a single characteristic energy scale. In Fig.~\ref{fig:ExampleA1}, we show the spectrum as a function of $\tilde \sigma$ for the case $\Delta = 3$. For all values of $\tilde \sigma$, there is a massless dilaton and Goldstone boson, associated with the spontaneous breaking of scale invariance and the global $U(1)_A$ symmetry, respectively.

\begin{figure}[t]
\begin{center}
\includegraphics[width=\figwidth]{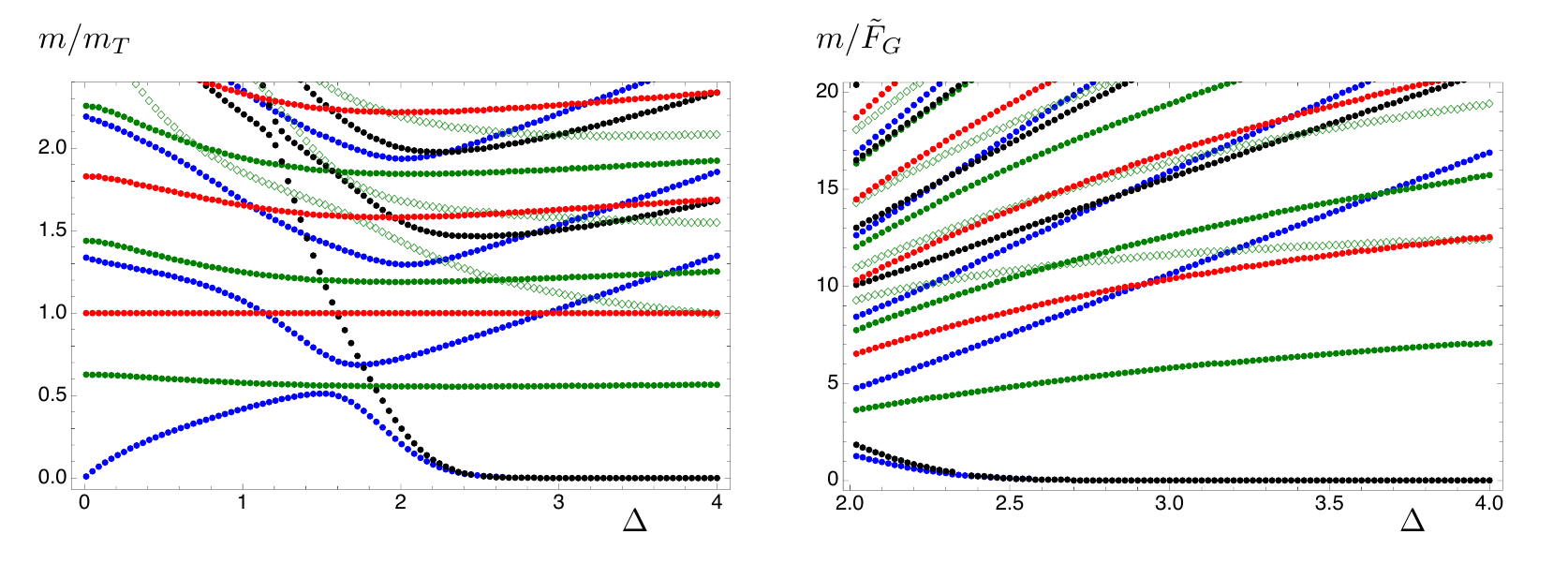}
\caption{Example A. Spectrum as a function of $\Delta$ for $g_5 = 5$, $\tilde \sigma = 2$, $r_1 = 0$, $r_2 = 8$. The left panel is normalized to the mass $m_T$ of the lightest tensor, while the right panel is normalized to the decay constant $\tilde F_G = F_G/N_C$. The colour coding is: scalar (blue), tensor (red), pseudoscalar (black), vector (green), axial-vector (green diamonds).}
\label{fig:ExampleA3}
\end{center}
\begin{center}
\includegraphics[width=\figwidth]{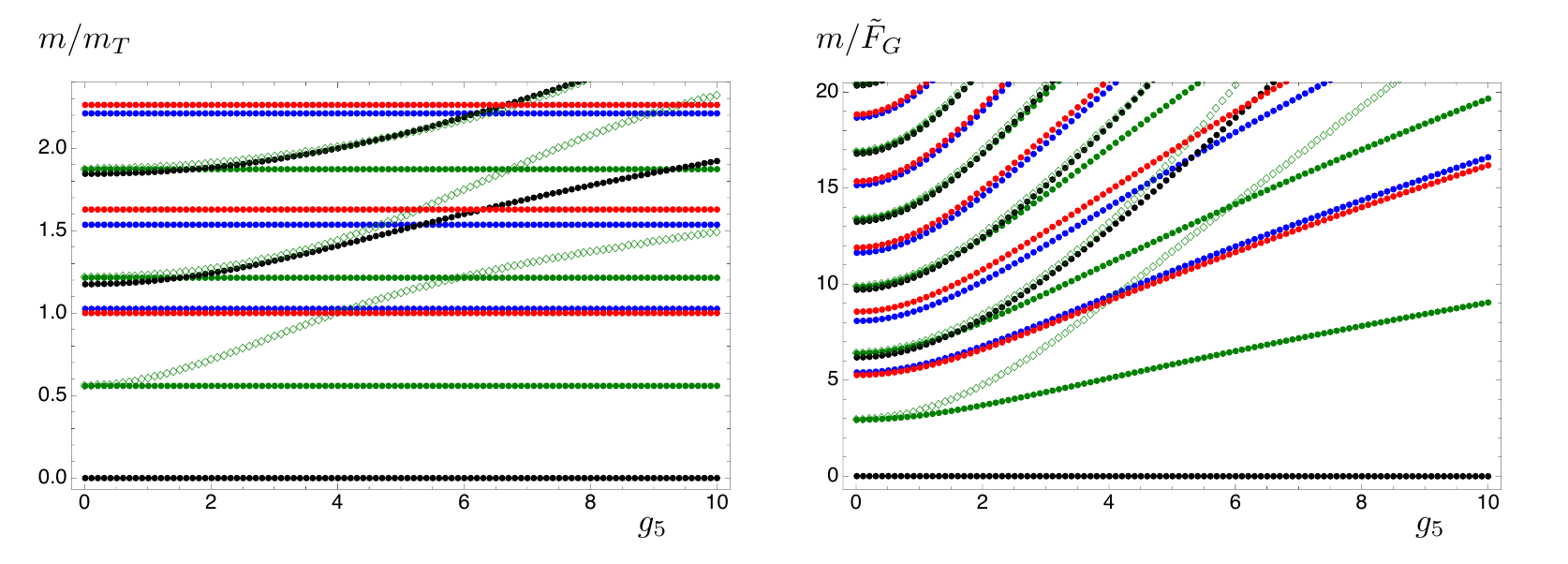}
\caption{Example A. Spectrum as a function of $g_5$ for $\Delta = 3$, $\tilde \sigma = 2$, $r_1 = 0$, $r_2 = 8$. The left panel is normalized to the mass $m_T$ of the lightest tensor, while the right panel is normalized to the decay constant $\tilde F_G = F_G/N_C$. The colour coding is: scalar (blue), tensor (red), pseudoscalar (black), vector (green), axial-vector (green diamonds). Since $\Delta > 2$, there is also a massless dilaton in the spectrum, as can be seen from Fig.~\ref{fig:ExampleA3}; due to the numerical resolution deployed only the Goldstone is shown in the present two plots.}
\label{fig:ExampleA4}
\end{center}
\end{figure}

Fig.~\ref{fig:ExampleA3} shows the dependence of the spectrum on $\Delta$. There is a light dilaton in the spectrum for $\Delta$ close to zero, as well as for $\Delta > 2$. In the former case, this is due to the CFT being deformed by the nearly marginal operator $\mathcal O_\sigma$ (as familiar from Goldberger-Wise \cite{Goldberger:1999uk,Kofman:2004tk}), while in the latter, the breaking of conformal invariance is spontaneous. In addition, for $\Delta > 2$, there is also a Goldstone corresponding to the spontaneous breaking of the $U(1)_A$ symmetry. Note that the cutoff effects due to finite $r_2$ are especially pronounced around $\Delta = 2$, increasing the masses of both the Goldstone and dilaton, which in the limit $r_2 \to + \infty$ remain massless all the way down to $\Delta = 2$.

Finally, in Fig.~\ref{fig:ExampleA4}, we show the spectrum as a function of $g_5$. As seen from Eq.~\eqref{eq:eomV} and Eq.~\eqref{eq:eomA1}, the spectra of the vector and axial-vector resonances coincide in the limit of $g_5$ going to zero. In order to depart from this special and unrealistic case, one should increase the bulk gauge coupling to (at least) $g_5 \sim 5$, thus capturing the more generic behaviour.

\subsection{Example B}
\label{sec:ExampleB}

\begin{figure}[t]
\begin{center}
\includegraphics[width=\figwidth]{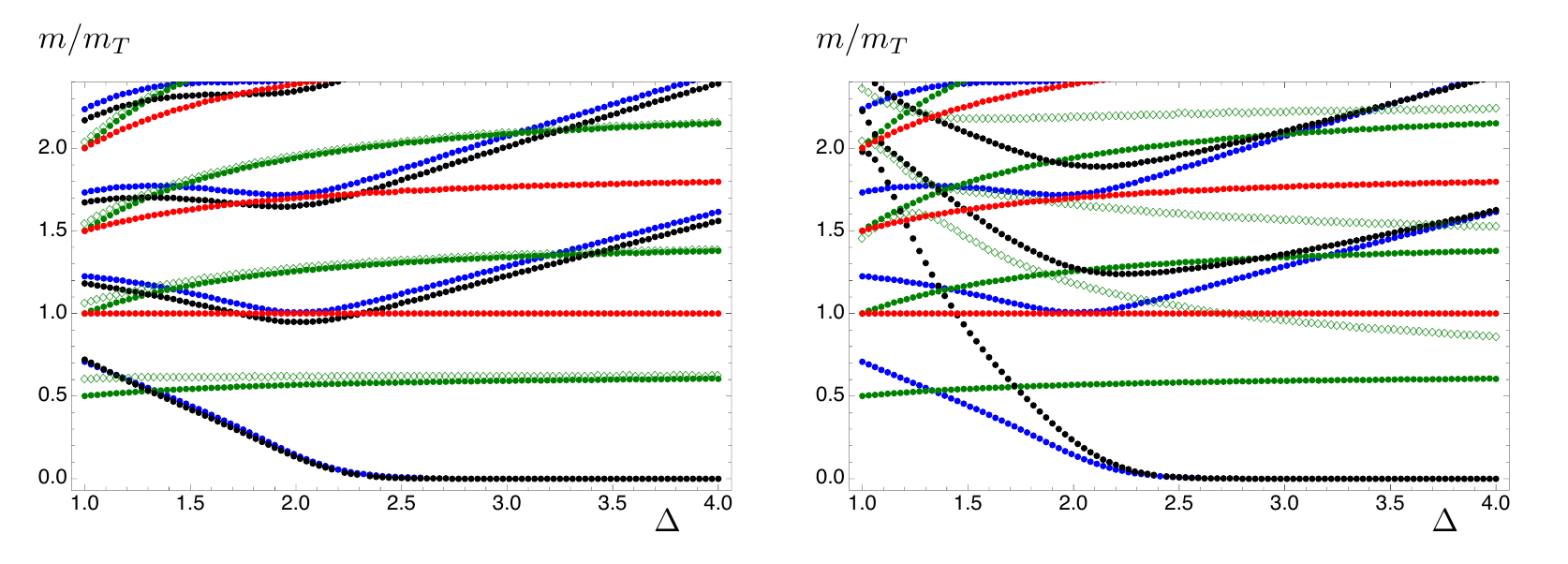}
\caption{Example B. Spectrum as a function of $\Delta$ for $r_1 = 10^{-4}$, $r_2 = 8$, and $g_5 = 1$ (left), $g_5 = 5$ (right), normalized to the mass $m_T$ of the lightest tensor. The colour coding is: scalar (blue), tensor (red), pseudoscalar (black), vector (green), axial-vector (green diamonds).}
\label{fig:ExampleB1}
\end{center}
\end{figure}

\begin{figure}[t]
\begin{center}
\includegraphics[width=\figwidth]{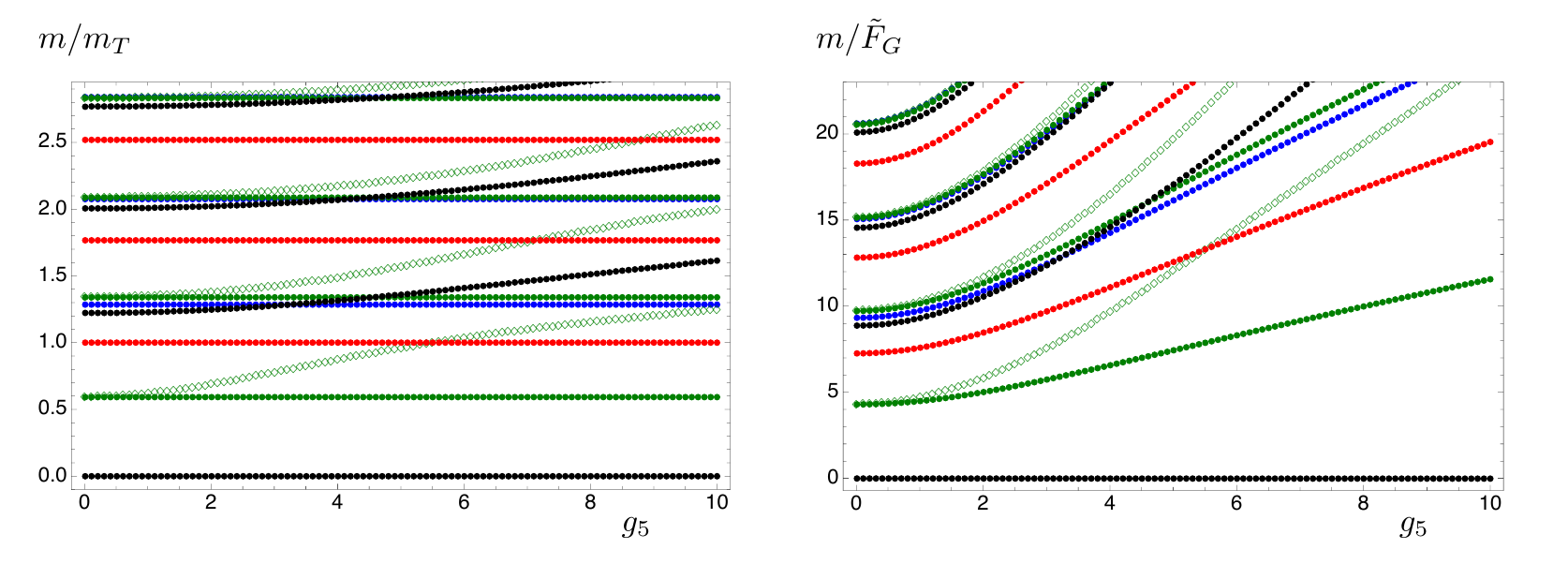}
\caption{Example B. Spectrum as a function of $g_5$ for $\Delta = 3$, $r_1 = 10^{-4}$, $r_2 = 8$. The left panel is normalized to the mass $m_T$ of the lightest tensor, while the right panels are normalized to the decay constant $\tilde F_G = F_G/N_C$. The colour coding is: scalar (blue), tensor (red), pseudoscalar (black), vector (green), axial-vector (green diamonds). Since $\Delta > 2$, there is also a massless dilaton in the spectrum, as can be seen from Fig.~\ref{fig:ExampleB1}; due to the numerical resolution deployed only the Goldstone is shown in the present two plots.}
\label{fig:ExampleB2}
\end{center}
\end{figure}

The second example that we consider has the superpotential
\beq
\label{eq:WExampleB}
	\mathcal W(\sigma) = - \frac{3}{4} \left[ 1+ \cosh\left( 2 \sqrt{\frac{\Delta}{3}} \sigma \right) \right] \,,
\eeq
where again $\Delta$ is a free parameter. This choice of superpotential is inspired by the top-down GPPZ model \cite{Girardello:1999bd}, which admits consistent truncations to the cases $\Delta = 1$ and $\Delta = 3$.

Solving the first order equations~\eqref{eq:eomWsimp} coming from the superpotential leads to
\beqs
\label{eq:ExampleBbackground}
	\sigma(r) &=& \sqrt{\frac{3}{\Delta}} \arctanh \left(e^{-\Delta r} \right) = \frac{1}{2} \sqrt{\frac{3}{\Delta}} \log\left(\coth\left( \frac{\Delta r}{2} \right)\right) \,, \\ \nonumber
	A(r) &=& \frac{1}{2\Delta} \log \left(e^{2\Delta r} - 1 \right) \,,
\eeqs
where we have fixed an integration constant corresponding to a shift in the radial coordinate so that the end-of-space in the IR is at $r_o = 0$, and again required that $A \simeq r$ for large values of $r$. The parameters of the model are hence $\Delta$ and $g_5$. By requiring that the end-of-space coincides with the value of the radial coordinate at which the scalar $\sigma$ becomes singular, so that the mass gap becomes dynamically determined, we have in effect eliminated one of the parameters as compared to Example~A.

For large $r$, we have that
\beq
\label{eq:ExampleBsigmaUV}
	\sigma = \sigma_c e^{-\Delta r} + \cdots \,, \hspace{0.5cm} \sigma_c \equiv \sqrt{\frac{3}{\Delta}} \,,
\eeq
which (as in Example~A) for $\Delta < 2$ has the interpretation that the CFT is deformed by the operator $\mathcal O_\sigma$ with scaling dimension equal to $[\mathcal O_\sigma] = 4 - \Delta$, whereas for $\Delta > 2$, $\mathcal O_\sigma$ has scaling dimension $[\mathcal O_\sigma] = \Delta$ and acquires a VEV. Hence, as before, in the latter case we expect a Goldstone boson associated with the spontaneous breaking of the global $U(1)_A$ symmetry, as well as a massless dilaton associated with the spontaneous breaking of scale invariance.

As previously alluded to, the main difference between Example~A and Example~B is that the latter model has one less free parameter. In Example~B, the mass gap gets generated by the same dynamics that breaks the $U(1)_A$, encoded in the bulk as the radial profile of the scalar $\sigma$. Therefore, the scale of the mass gap cannot be varied independently from that of the decay constant $F_G$. We believe that for this reason, Example~B is a more realistic (and constraining) model than Example~A.

Fig.~\ref{fig:ExampleB1} and Fig.~\ref{fig:ExampleB2} show the spectrum as a function of $\Delta$ and $g_5$. As can be seen, the ratio $F_G/m_T$ strongly decreases as $g_5$ grows, while we have checked that its dependence on $\Delta$ is mild. We also note that when $\Delta < 2$, as $g_5$ grows the lightest pseudoscalar resonance becomes progressively heavier than the lightest scalar resonance. In later sections, we will show how to lift the mass of the dilaton independently from that of the Goldstone bosons by including an additional bulk scalar in the model, that encodes the explicit breaking of conformal invariance due to a deformation of the dual field theory by a relevant operator.

\begin{figure}[t]
\begin{center}
\includegraphics[width=\figwidth]{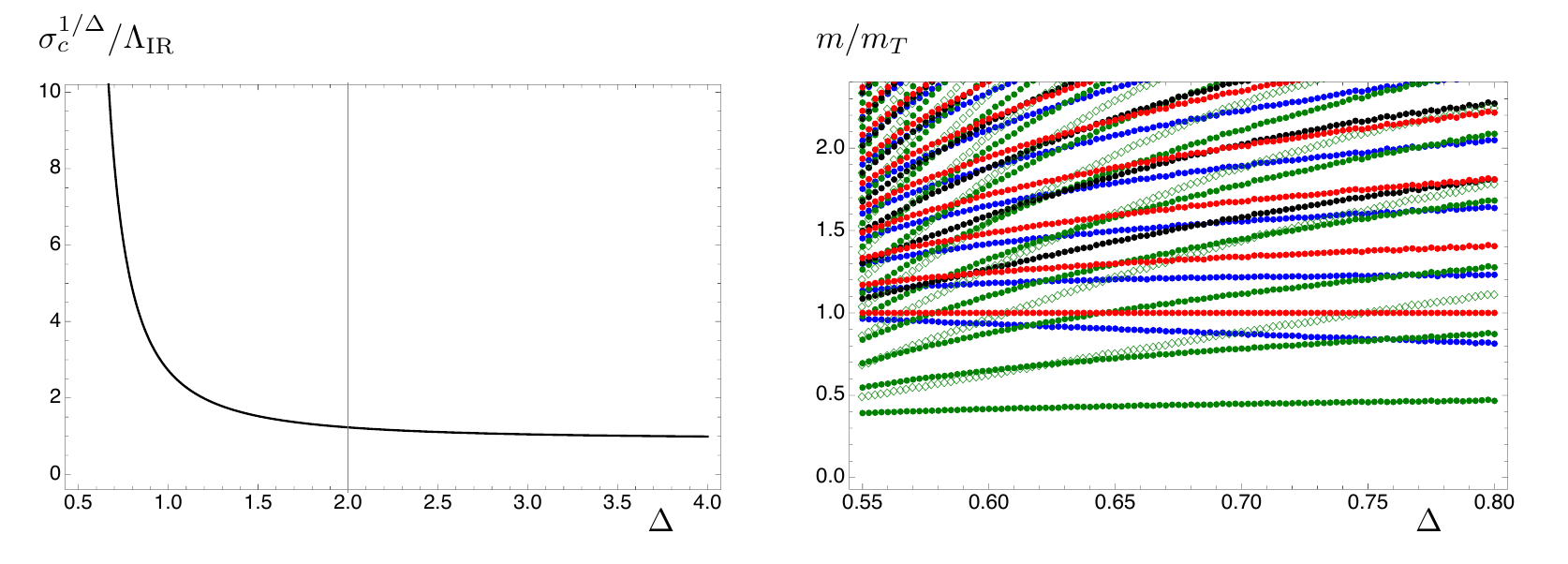}
\caption{Example B. The left panel shows the dimensionless ratio $\sigma_c^{1/\Delta}/\Lambda_{\rm IR}$. The right panel shows the spectrum as a function of $\Delta$ (close to $\Delta \simeq \frac{1}{2}$) for $g_5 = 5$, $r_1 = 10^{-10}$, $r_2 = 8$, normalized to the mass $m_T$ of the lightest tensor.}
\label{fig:ExampleB3}
\end{center}
\end{figure}

Finally, let us comment on the possibility of more than one characteristic energy scale arising in this model. 
From the IR expansion of $A(r)$, we see that as $\Delta \rightarrow \frac{1}{2}$, the typical IR scale defined below Eq.~\eqref{eq:Lambdar} diverges, i.e. $\Lambda_{\rm IR} \rightarrow +\infty$. 
Since, in the same limit, $\sigma_c$ remains finite, following the same reasoning as for Example~A, we are led to the conclusion that in the vicinity of $\Delta \simeq \frac{1}{2}$, the model exhibits multi-scale dynamics. 
We illustrate this in Fig.~\ref{fig:ExampleB3}, where the left panel shows $\sigma_c^{1/\Delta}/\Lambda_{\rm IR}$ as a function of $\Delta$, while the right panel shows the resulting spectrum close to $\Delta \simeq \frac{1}{2}$. 
As can be seen, close to $\Delta \simeq \frac{1}{2}$, the spectrum approaches that of a gapped continuum, and the presence of two scales is apparent in that the mass gap becomes parametrically larger than the typical spacing of the resonances above it. 
While this behaviour is not a generic feature for strongly-coupled field theories, it has been observed in certain special cases, such as close to the CVMN solution \cite{Chamseddine:1997nm,Maldacena:2000yy} on the baryonic branch of the Klebanov-Strassler system~\cite{Elander:2017cle,Elander:2017hyr}, 
as well as on the Coulomb branch of $\mathcal N = 4$ SYM~\cite{Brandhuber:2000fr,Brandhuber:2002rx}. It is also a characteristic feature of linear-dilaton/clockwork models \cite{Antoniadis:2011qw,Cox:2012ee,Giudice:2017fmj,Teresi:2018eai}.

\section{Holographic composite Higgs with many flavours}
\label{sec:models}

In this section, we will construct holographic models built from the bottom-up with the aim of capturing the dynamics of the kinds of field theories described in section~\ref{model}.

In the scalar sector, there are two field theory operators, $\psi^a \psi^b$ and $\chi \chi$, that we will consider. On the bulk side they are dual to a complex antisymmetric scalar $\Phi_{ab}$ and a complex scalar $Z$. Both of these operators have correlators that scale as $N_C^2$ in the Veneziano limit, which implies that in general $\Phi_{ab}$ and $Z$ will backreact on the metric. In order to study the breaking of the global symmetry to $Sp(2N_F)$, it is convenient to parametrize $\Phi$ as\footnote{To show that (at least locally) this parametrizes all the degrees of freedom of $\Phi$, note that to first order in $\Pi$, we have that
\beq
	\Phi = \left[ \left(1 + \frac{i \pi_0}{\sqrt{N_F}} \right) \left(\frac{\sigma}{2} \mathds{1} + S_{\hat A} T^{\hat A} \right) + i \pi_{\hat A} \left( \sigma T^{\hat A} + S_{\hat B} \{ T^{\hat A} , T^{\hat B} \} \right) \right] \Sigma \,, \nonumber
\eeq
where we made use of the identities given in Eq.~\eqref{eq:genid}. Furthermore, since $d^{A \hat B \hat C} \equiv \frac{1}{2} \Tr (T^A \{ T^{\hat B} , T^{\hat C} \}) = 0$, it follows that the expression in the parenthesis before $\Sigma$ is a (complex) linear combination of the broken generators. This is the general form of a complex antisymmetric matrix.}
\beqs
\label{eq:Phidecomp}
	\Phi &=& U(\Pi) \Phi_H \Sigma U(\Pi)^T \, \\ \nonumber
	\Phi_H &=& \frac{\sigma}{2} \mathds{1} + S_{\hat A} T^{\hat A} \,.
\eeqs
where $\Sigma$ is the anti-symmetric matrix introduced below Eq.~\eqref{eq:LorentzScalarOps}, $\Phi_H$ is a Hermitian matrix ($\sigma$ and $S$ are real), $T^{\hat A}$ are the broken generators of $SU(2N_F)$, and $U(\Pi)$ is a $U(2N_F)$-transformation along the broken directions:
\beqs
\label{eq:Phidecomp2}
	U(\Pi) &=& \exp (i \Pi) \,, \\ \nonumber
	\Pi &=& \frac{\pi_0}{2\sqrt{N_F}} \mathds{1} + \pi_{\hat A} T^{\hat A} \,.
\eeqs
As we shall see, a non-trivial radial profile of $\sigma$ in the bulk will encode the breaking of the global symmetry on the field theory side. We also decompose the complex scalar $Z$ into its absolute value and a phase, as $Z = |Z| \, e^{i\theta}$.

Now let us comment on the gauge symmetry on the gravity side, related to the global symmetry of the dual field theory. As explained in section~\ref{model}, there is a linear combination of $U(1)_\psi$ and $U(1)_\chi$ transformations (associated with shifting the phases $\pi_0$ and $\theta$) that is a global symmetry, while the remaining factor is anomalous. This latter fact results in $\eta'$ receiving a mass, which (contrary to e.g. QCD) is of order $\mathcal O(1)$ in the large-$N_C$ limit due to $\chi$ transforming in a two-index representation of the HC group. Furthermore, in the Veneziano limit we have that $N_F \sim N_C$, and the contribution of the fundamental fermions $\psi^a$ to $m_{\eta'}^2$ is of order $N_F / N_C = \mathcal O(1)$ \cite{Witten:1979vv,Veneziano:1979ec}. In order to capture these effects in a holographic model, one would have to gauge both $U(1)_\psi$ and $U(1)_\chi$ symmetries. One would also have to account for the operator mixing with $\Tr (\tilde F_{\mu\nu} F^{\mu\nu})$, related to pseudoscalar glueballs. This is a non-trivial, but interesting, exercise that we leave for a future study (for the treatment of the $U(1)_A$ anomaly in the context of holographic QCD, see for example \cite{Sakai:2004cn,Katz:2007tf,Arean:2016hcs}). In this work, we hence only gauge the remaining $SU(2N_F)$ symmetry in the bulk, and put $\pi_0 = \theta = 0$, thus neglecting the aforementioned sector in our study. Although this results in one fewer Goldstone boson in the spectrum, we stress that including this extra sector would not affect our results for the observables that we do compute. 
We parametrise the $SU(2N_F)$ gauge field $\mathcal A_M$ as
\beq
\label{eq:Adecomp}
	 \mathcal A_M = A_{\hat A M} T^{\hat A} + V_{A M} T^{A} \,,
\eeq
where $T^{\hat A}$ ($T^A$) are the broken (unbroken) generators of $SU(2N_F)$.

We will present three models. The first one, which we refer to as Model I, is simpler in that it does not contain the complex scalar $Z$. The complex scalar $\Phi_{ab}$ is in the antisymmetric representation of $SU(2N_F)$, and its radial profile will be responsible for the breaking of the global symmetry $SU(2N_F) \rightarrow Sp(2N_F)$. In addition, the model contains the metric $g_{MN}$ and the $SU(2N_F)$ gauge field $\mathcal A_M$. As usual, the former is dual to the stress-energy tensor $T^{\mu\nu}$ of the field theory, while the latter is dual to the current $\mathcal J^\mu{}_a{}^b$ defined in Eq.~\eqref{eq:current}. We will see that Model I is too simple to break explicitly conformal invariance without explicit breaking of the flavour symmetry. To overcome this drawback, the second and third models, referred to as Models IIA and IIB, additionally contain a bulk field $\phi$, whose radial profile 
will correspond to adding a relevant deformation, thus lifting the mass of the dilaton. One possibility is that $\phi$ is the modulus of $Z = \phi \, e^{i \theta}$, in which case the relevant deformation in question is a mass term for $\chi$. This would then describe a situation in which the vanishing of said mass term leads to the field theory flowing to an IR fixed point, such as is the case for example when $x_F = N_F/N_C$ is within the conformal window. Note that the presence of the mass term for $\chi$ not only would lift the mass of the dilaton, but also that of the extra NGB, alluded to in the previous paragraph, although we do not include it in our present study of the spectrum. This being said, our analysis does not rely on which operator is dual to $\phi$, only that it encodes the explicit breaking of scale invariance.

Before presenting the models described above, let us make some general comments about the Veneziano limit. In holographic models, the large-$N_C$ limit usually implies that the gravitational description becomes weakly coupled. This then allows to describe non-perturbative physics on the field theory side by translating to a classical computation on the bulk side. In the Veneziano limit, the number of degrees of freedom in the bulk grows as $N_F^2$. This implies that loops on the gravity side are no longer suppressed, and hence one should strictly speaking consider an expansion in $x_F = N_F / N_C$. On the field theory side, these issues are reflected by the fact that, in the Veneziano limit, multi-resonance states contribute to, for instance, correlators of fermion bilinears, contrary to what is the case in the usual large-$N_C$ limit. As a consequence, the poles of two-point functions may acquire an imaginary part, leading to resonances with a  non-zero decay width. This effect clearly cannot be described by the classical approximation on the gravity side, and since we do not have anything to add in this respect we will not include it, assuming that the single-resonance exchanges provide a sufficiently correct description of the spectrum. Nevertheless, one may expect that such approximation captures reasonably well the masses of the lightest states even when $x_F \sim 1$, as these have fewer decay channels available, leading to a narrow width, and in turn suggesting that the impact of multi-resonance exchange on the masses of the lightest states is small.

\subsection{Model I}
\label{sec:Model1}

The first model that we consider consists of gravity, an antisymmetric complex scalar $\Phi$, and an $SU(2N_F)$ gauge field $\mathcal A_M$. $\Phi$ is parametrized as in Eqs.~\eqref{eq:Phidecomp} and~\eqref{eq:Phidecomp2} (recall that $\pi_0 = 0$), while $\mathcal A_M$ is parametrized as in Eq.~\eqref{eq:Adecomp}. The action is given by
\beqs
\label{eq:ActionModel1}
	\mathcal S &=&  \int \dd^5x \sqrt{-g} \, \bigg\{ \mathcal N_1 \bigg( \frac{R}{4} - \frac{\tilde\Lambda}{2} \bigg) - \mathcal N_2 \Tr \Big[ g^{MN} (D_M \Phi)^\dag D_N \Phi \Big] \nonumber \\ && \hspace {2.2cm} - \mathcal N_3 \Tr \Big[ \frac{1}{4} g^{MP} g^{NQ} \mathcal F_{MN} \mathcal F_{PQ} \Big] - \mathcal N_4 \mathcal V_\Phi(\Phi) \bigg\} \,,
\eeqs
where the cosmological constant is equal to $\tilde\Lambda = -6$, the potential $\mathcal V_\Phi(\Phi)$ is defined to be an $SU(2N_F)$ invariant, $\mathcal F_{MN}$ is the field strength associated with the gauge field $\mathcal A_M$, and the covariant derivative is given by $D_M \Phi = \partial_M \Phi + i g_5 \left(\mathcal A_M \Phi + \Phi \mathcal A_M^T\right)$. The overall normalisation factors in the action are chosen to be $\mathcal N_1 = \mathcal N_4 = N_C^2$ and $\mathcal N_2 = \mathcal N_3 = N_C$, so as to recover the expected large-$N_C$ counting of the dual field theory, as we shall now discuss.

First, as usual $\mathcal N_1$ is required to scale as $N_C^2$ in order to reproduce the correct large-$N_C$ scaling of correlation functions of the stress-energy tensor on the field theory side. Second, the large-$N_C$ scaling of $\mathcal N_2$ is fixed by the requirement that the pseudoscalar two-point functions scale as $N_C$. Third, the large-$N_C$ scaling of the decay constant is determined by $\mathcal N_3$ as $F_G^2 \sim \mathcal N_3 = N_C$, which is what is expected from field theory considerations.

The fourth feature that we would like to capture concerns the operator dual to $\sigma$, namely $\mathcal O_\sigma = \Tr (\Sigma_{ab} \psi^a \psi^b)$, which also will be responsible for symmetry breaking. In the Veneziano limit, its correlators scale as $N_C^2$. Note that with the above choice of $\mathcal N_2$, the coefficient in front of the kinetic term for $\sigma$ already scales as $N_F N_C \sim N_C^2$. In order to describe symmetry breaking, we need to ensure that $\sigma$ acquires a background profile, which leads us to also choose the factor $\mathcal N_4$ in front of the potential $\mathcal V_\Phi$ to scale as $N_C^2$. We will soon describe in more detail how we choose the form of the potential itself.

We anticipate that, with the choices explained in the previous paragraph, the action can be written as
\beqs
	\mathcal S = N_C^2 \int \dd^5x \sqrt{-g} \, \bigg\{ && \hspace{-0.4cm} \frac{R}{4} - \frac{x_F}{2} (\partial_M \sigma)^2 -  \frac{\tilde\Lambda}{2} - \mathcal V_\Phi(\Phi) \bigg\} \\ \nonumber
	+ N_C \int \dd^5x \sqrt{-g} \, \bigg\{ && \hspace{-0.4cm} - \frac{1}{2} (\partial_M S_{\hat A})^2 - \frac{1}{4} \sum_{i = A,V }(\mathcal F^{(i)}_{MN})^2
	- \frac{\sigma^2}{2} (\partial_M \pi_{\hat A} + g_5 A_{\hat A M})^2
	\bigg\} + \cdots \,,
\eeqs
where the field strengths are given by $\mathcal F^{(i)}_{MN} \equiv \partial_M \mathcal A^{(i)}_N - \partial_N \mathcal A^{(i)}_M$, and the dots contain interaction terms (we provide more details in section~\ref{sec:bosonicspectrum}). As we shall see, the part of the action on the first line, proportional to $N_C^2$, will be responsible for determining the background geometry.

\subsubsection{Equations of motion}

As in section~\ref{sec:warmup}, for the background solutions that we will consider, the gauge fields vanish, the metric takes the domain wall form given in Eq.~\eqref{eq:domainwall} with warp factor $A(r)$, and the scalar field $\Phi(r)$ only depends on the radial coordinate $r$. Furthermore, given the decomposition of $\Phi$ in Eq.~\eqref{eq:Phidecomp}, and the fact that $\mathcal V_\Phi$ is  $SU(2N_F)$ invariant and hence independent of $\Pi$, it is consistent to set $\Pi$ to zero in the ansatz for the background solutions. This leads to the background equations of motion for the scalars $\sigma$ and $S$, and the warp factor $A$:
\beqs
\label{eq:eomsModel1}
	\partial_r^2 \sigma + 4 \partial_r A \partial_r \sigma - x_F^{-1} \frac{\partial \mathcal V_\Phi}{\partial \sigma} &=& 0 \,, \nonumber \\
	\partial_r^2 S_{\hat A} + 4 \partial_r A \partial_r S_{\hat A} - N_C \frac{\partial \mathcal V_\Phi}{\partial S_{\hat A}} &=& 0 \,, \\ \nonumber
	6 (\partial_r A)^2 -  x_F (\partial_r \sigma)^2 -  N_C^{-1} (\partial_r S)^2 + \tilde\Lambda + 2 \mathcal V_\Phi &=& 0 \,.
\eeqs
The factor of $N_C$ in the second equation originates from the fact that $S$ and $\mathcal V_\Phi$ enter in the action with different factors of $N_C$. Yet, we anticipate that, because of our choice of potential $\mathcal V_\Phi$, the large $N_C$ limit will still lead to a well-defined equation of motion for $S$.

We will choose $\mathcal V_\Phi$ such that it is a function of an $SU(2N_F)$ invariant $\mathcal I$ built from $\Phi$, on which we impose two conditions,
\beq
\label{eq:conditionsI}
	\mathcal I \big|_{S=0} = \sigma \,, \hspace{1cm} \frac{\partial \mathcal I}{\partial S_{\hat A}} \bigg|_{S=0} = 0 \,.
\eeq
These conditions ensure that (i) it is consistent to put $S_{\hat A} = 0$ in Eq.~\eqref{eq:eomsModel1}, and (ii) the resulting equations of motion for $\sigma$ and $A$ do not depend on the choice of the invariant $\mathcal I$.
An example of an invariant satisfying the conditions in Eq.~\eqref{eq:conditionsI} is given by
\beq
\label{eq:I1}
	\mathcal I_1 \equiv \left[ \frac{2\Tr (\Phi^\dag \Phi)}{N_F} \right]^{1/2} = \sqrt{\sigma^2 + N_F^{-1} S^2} \,,
\eeq
We will comment on other choices of invariants in section~\ref{sec:bosonicspectrum}.

\subsubsection{Scalar potential}

We next follow a logic similar to that of the examples in section~\ref{sec:warmup}, chosing the potential $\mathcal V_\Phi$ such that it can be written in terms of a superpotential. Noting that both the cosmological constant $\tilde \Lambda$ and $\mathcal V_\Phi$ contribute to the full potential $\mathcal V$, we write
\beq
	\mathcal V(\mathcal I) \equiv \frac{\tilde\Lambda}{2} + \mathcal V_\Phi(\mathcal I) = \frac{1}{2x_F} \mathcal W'(\mathcal I)^2 - \frac{4}{3} \mathcal W(\mathcal I)^2 \,,
\eeq
where the factor $x_F^{-1}$ in front of the first term on the right hand side originates from that the sigma-model metric component associated with $\sigma$ is equal to $G_{\sigma\sigma} = x_F$. Making use of Eq.~\eqref{eq:conditionsI}, solutions to the background equations of motion \eqref{eq:eomsModel1} can then be obtained by putting $S_{\hat A} = 0$, and solving the first-order equations for the scalar $\sigma$ and the warp factor $A$,
\be
\label{eq:firstordereqsflavour}
	\partial_r \sigma = \frac{1}{x_F} \partial_\sigma\mathcal W(\sigma) \,, \qquad\qquad
	\partial_r A = - \frac{2}{3} \mathcal W(\sigma) \,.
\ee

Inspired by Example~B of section~\ref{sec:ExampleB}, we pick the superpotential to be
\beq
\label{eq:Model1W}
	\mathcal W(\mathcal I) = - \frac{3}{2} \left[ 1 + x_F \sinh^2\left( \sqrt{\frac{\Delta}{3}} \, \mathcal I \right) \right] \,,
\eeq
which for $x_F = 1$ coincides with the superpotential defined in Eq.~\eqref{eq:WExampleB}. For completeness, let us write explicitly the resulting potential
\beq
	\mathcal V_\Phi(\mathcal I;x_F) = \frac{3 x_F}{4} \sinh ^2\left( \sqrt{\frac{\Delta}{3}} \, \mathcal I \right)
   \left(\left(\Delta -2 x_F\right) \cosh \left( 2 \sqrt{\frac{\Delta}{3}} \, \mathcal I \right)+\Delta +2 x_F-8\right) \,.
\eeq
Let us make a few comments about the form of this potential. First, we observe that in the limit $x_F \rightarrow 0$, one obtains that $\mathcal V_\Phi$ vanishes, thus resulting in a model with only a cosmological constant and no backreacting flavours. Hence, in this limit our model resembles the composite-Higgs models of \cite{Contino:2003ve,Agashe:2004rs} in which the background geometry is a slice of AdS. Conversely, when $x_F \sim 1$, the backreaction of the flavour sector becomes significant. This is in agreement with expectations from field theory where, in the Veneziano limit, there is non-trivial (order $x_F$) mixing between the glue and matter sectors, and both contribute to the dynamics responsible for confinement. In this case, the geometry is close to that of Example~B of section~\ref{sec:ExampleB}, which in turn is a generalisation of the GPPZ model \cite{Girardello:1999bd}. The GPPZ model is dual to $\mathcal N = 1^*$ SYM, and the backreaction on the geometry occurs not because of the Veneziano limit, but because the fermion (and all other) fields of the dual field theory are in the adjoint representation (for instance, when $\Delta = 3$, the bulk scalar field encodes the dynamics leading to a non-trivial gaugino condensate). The reasons why we choose a GPPZ-inspired potential, despite these differences, are to capture the essential features of confinement, the link between a dynamically generated mass gap and the flavour symmetry-breaking scale, as well as to obtain a discrete mass spectrum. We also note that, while these features can be expected on general grounds, our models are not engineered to address detailed questions regarding the dynamics at the lower edge of the conformal window, in particular the critical values of $x_F$ and $\Delta$, which we hence keep as free parameters.

\subsubsection{Background solution}

The background solution that we will consider is obtained by solving the first order equations~\eqref{eq:firstordereqsflavour}, and is given by
\beqs
\label{eq:Model1solutions}
	\sigma(r) &=& \sqrt{\frac{3}{\Delta}} \arctanh \left(e^{-\Delta r} \right) \,, \\ \nonumber
	A(r) &=& r + \frac{x_F}{2\Delta} \log \left(1 - e^{-2\Delta r} \right) \,.
\eeqs
As can be seen, this is exactly the same solution as in Example~B, apart from the factor $x_F$ appearing in the warp factor. Notice that by choosing a small $x_F$, one can make the metric close to AdS except for an arbitrarily small region near $r=0$, thus resembling AdS with a hardwall cutoff. Conversely, for larger values of $x_F$ the backreaction of the flavours on the geometry becomes significant. We note that for the specific cases of $x_F=1$ and $\Delta = 1$ or $3$, the sectors corresponding to gravity and the scalar $\sigma$ coincide with two possible consistent truncations of the GPPZ model~\cite{Girardello:1999bd}.

Let us now comment on the asymptotic UV expansion of the background solution. In terms of the canonically normalized scalar field $\sigma^{(n)} \equiv x_F^{1/2} \sigma$, we have that
\beq
\label{eq:sigmanUV}
	\sigma^{(n)} = \sigma_c e^{-\Delta r} + \cdots \,, \hspace{1cm} \sigma_c \equiv \sqrt{\frac{3 x_F}{\Delta}}~,
\eeq
which implies that for $\Delta < 2$ the flavour symmetry is broken explicitly, while for $\Delta > 2$ it is broken spontaneously. 
Thus, the coefficient $\sigma_c$ is related to either the size of the source ($\Delta < 2$) or the VEV ($\Delta > 2$) of the operator $\mathcal O_\sigma /\sqrt{x_F}$ that is dual to $\sigma^{(n)}$. In the latter case, taking into account the $N_C^2$ prefactor in the action,
one expects $\langle \mathcal O_\sigma \rangle \sim \sqrt{x_F} \sigma_c N_C^2 = \sqrt{3/\Delta} \ N_F N_C$, which becomes enhanced at large number of flavours. We will come back to this observation in section~\ref{sec:bosonicspectrum} when we will study the spectrum.

\begin{figure}[t]
\begin{center}
\includegraphics[width=\figwidthsmall]{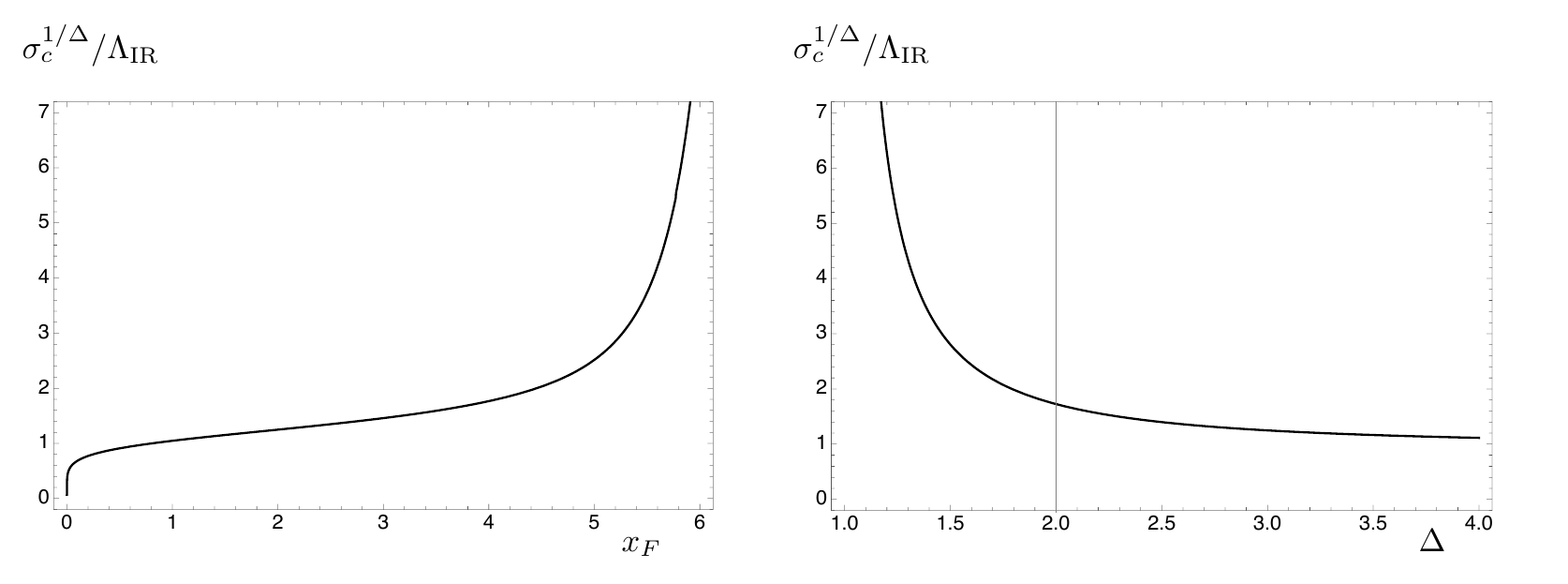}
\caption{Model I. The dimensionless ratio $\sigma_c^{1/\Delta} / \Lambda_{\rm IR}$, as a function of $x_F$ with $\Delta = 3$ (left panel) and as a function of $\Delta$ with $x_F = 2$ (right panel). The vertical gray line indicates the value of $\Delta = 2$ when the interpretation of $\sigma_c$ switches from being a source ($\Delta < 2$) or a VEV ($\Delta > 2$).}
\label{fig:Model1val}
\end{center}
\end{figure}

In Example~B of section~\ref{sec:ExampleB}, we showed that, as one approaches $\Delta \simeq \frac{1}{2}$, the IR scale $\Lambda_{\rm IR}$ vanishes, giving rise to multi-scale dynamics, which is exhibited in the spectrum approaching a gapped continuum. While these are interesting features, we aim to capture the dynamics of generic strongly-coupled field theories, and hence will require that $\Lambda_{\rm IR}$ remains finite. In the present case, this leads to an upper bound on the number of flavours, $x_F < 2\Delta$. In order to estimate whether more than one characteristic energy scale is present, we show the dependence of the dimensionless ratio $\sigma_c^{1/\Delta} / \Lambda_{\rm IR}$ on $x_F$ and $\Delta$ in Fig.~\ref{fig:Model1val}.

\subsection{Model II}
\label{sec:Model2}

In this model, we introduce an additional bulk scalar field $\phi$ that is a flavour singlet. The motivation for this is to allow for the possibility of explicit breaking of scale invariance without also explicitly breaking the flavour symmetry. The action of the model is given by
\beqs
\label{eq:ActionModel2}
	\mathcal S &=& \int \dd^5x \sqrt{-g} \, \bigg\{ N_C^2 \left[ \frac{R}{4}  - \frac{1}{2} g^{MN} \partial_M \phi \partial_N \phi - \mathcal V(\Phi, \phi) \right] \nonumber \\ && \hspace{1cm} - N_C \Tr \left[ g^{MN} (D_M \Phi)^\dag D_N \Phi + \frac{1}{4} g^{MP} g^{NQ} \mathcal F_{MN} \mathcal F_{PQ} \right] \bigg\} \,,
\eeqs
where we made the same choices for the factors of $N_C$ as in Model I, with the additional requirement that the kinetic term for $\phi$ scales as $N_C^2$. This latter choice is motivated by the requirement that the correlators of the operator $\mathcal O_\phi$ dual to $\phi$ be of order $N_C^2$, as would be the case if $\phi$ is the modulus of $Z$ (dual to $\chi \chi$) as suggested in the beginning of this section. It would also be the case if $\mathcal O_\phi$ were built from the glue of the field theory. In addition, requiring that $\phi$ is accompanied by the same $N_C$-scaling as that of gravity allows for $\phi$ to backreact on the geometry, such that it may play the role of explicitly breaking conformal invariance in the dual field theory and lift the mass of the dilaton. However, since we remain agnostic regarding the precise form of $\mathcal O_\phi$, we can only give general arguments for how $\phi$ enters into the bulk action. In particular, one could imagine that the factor in front of its kinetic term has a non-trivial dependence on $x_F$, whereas we have made the simplest possible choice in this regard.

We assume that the scalar potential $\mathcal V$ depends on $\phi$, as well as on an invariant $\mathcal I$ built from $\Phi$. As before, we require that $\mathcal I$ satisfies the conditions given in Eq.~\eqref{eq:conditionsI}. We next choose the scalar potential $\mathcal V(\mathcal I,\phi)$ such that it can be written in terms of a superpotential $\mathcal W(\mathcal I,\phi)$ as
\beq
	\mathcal V(\mathcal I,\phi) = \frac{1}{2x_F} \left( \frac{\partial \mathcal W}{\partial \mathcal I} \right)^2 + \frac{1}{2} \left( \frac{\partial \mathcal W}{\partial \phi} \right)^2 - \frac{4}{3} \mathcal W^2 \,,
\eeq
with $\mathcal W(\mathcal I,\phi)$ defined to be the sum of the superpotential of Model I given in Eq.~\eqref{eq:Model1W} and a function $w(\phi)$:
\beq
	\mathcal W(\mathcal I,\phi) =  - \frac{3}{2} \left[ 1 + x_F \sinh^2\left( \sqrt{\frac{\Delta}{3}} \mathcal I \right) \right] + w(\phi) \,,
\eeq
We consider two choices of $w(\phi)$, analogous to Example~A and B of section~\ref{sec:warmup}, namely
\beqs
\label{eq:wA}
	w_A(\phi) &=& - \frac{\Delta_\phi}{2} \phi^2 \,, \\
\label{eq:wB}
	w_B(\phi) &=& - \frac{3}{2} \sinh^2\left( \sqrt{\frac{\Delta_\phi}{3}} \phi \right) \,,
\eeqs
which we will refer to as Model IIA and Model IIB, respectively.

As for Model I, solutions to the equations of motion can be found by first setting $S_{\hat A} = 0$, and then solving the first order equations
\be
\label{eq:firstordereqsflavour2}
	\partial_r \sigma = \frac{1}{x_F} \mathcal \partial_\sigma \mathcal W(\sigma,\phi) \,, \qquad
	\partial_r \phi = \mathcal \partial_\phi \mathcal W(\sigma,\phi)  \,, \qquad
	\partial_r A = - \frac{2}{3} \mathcal W(\sigma,\phi) \,. 
\ee
We hence obtain the following solutions for the respective cases A and B:
\beqs
	\phi(r) &=& \phi_A \, e^{-\Delta_\phi r} \,,  \nonumber \\
	A(r) &=& r + \frac{x_F}{2\Delta} \log \left( 1 - e^{-2\Delta r} \right) - \frac{\phi_A^2}{6} e^{-2 \Delta_\phi r} \,,
\eeqs
and
\beqs
\label{eq:Model2Bsolutions}
	\phi(r) &=& \sqrt{\frac{3}{\Delta_\phi}} \arctanh \left( \phi_B \, e^{-\Delta_\phi r} \right) \,,  \nonumber \\
	A(r) &=& r + \frac{x_F}{2\Delta} \log \left( 1 - e^{-2\Delta r} \right) + \frac{1}{2\Delta_\phi} \log \left( 1 - \phi_B^2 e^{-2\Delta_\phi r} \right) \,.
\eeqs
In both cases, $\sigma(r)$ is the same as for Model I and given in Eq.~\eqref{eq:Model1solutions}. $\phi_A$ and $\phi_B$ are integration constants that for $\Delta_\phi < 2$ ($\Delta_\phi > 2$) govern the size of the source (VEV) of the operator $\mathcal O_\phi$ dual to $\phi$. We observe that, as for Model I, the requirement that the typical IR scale $\Lambda_{\rm IR}$ remains finite leads to the same bound $x_F < 2\Delta$. Furthermore, we note that for the specific case of $x_F = 1$ and $\Delta = 3$ and $\Delta_\phi = 1$, the sectors corresponding to gravity and the scalars $\sigma$, $\phi$ of Model IIB coincide with the GPPZ model~\cite{Girardello:1999bd}.

Consider Model IIB, and suppose that $\Delta > 2$ and $\Delta_\phi < 2$. In this case, as can be seen from the asymptotic UV behaviour of $\sigma$, the breaking of the global flavour symmetry of the dual field theory is spontaneous. At the same time, turning on a non-zero $\phi$ in the bulk corresponds to introducing explicit breaking of conformal invariance on the field theory side. If $0 \leq \phi_B < 1$, the end-of-space is dynamically generated due to the scalar $\sigma$ diverging at $r = 0$, and as in Example~B, the scales of the mass gap and the decay constant become linked. Conversely, when $\phi_B > 1$, it is rather the dynamics of $\phi$ that breaks the conformal invariance that also generates the mass gap, and there is no reason for the decay constant to be of the same order. For this reason, we will require that $0 \leq \phi_B < 1$ (for a discussion of the case $\phi_B > 1$ in a different context, see~\cite{Elander:2012fk}).

On the other hand, in Model IIA, it is always $\sigma$ that is responsible for the end-of-space of the geometry and the generation of the mass gap. However, as in Example~A, it can be argued that values of $\phi_A \gg 1$ do not capture the dynamics of generic strongly coupled field theories, thus leading us to a similar conclusion regarding the range of the integration constant $\phi_A$.

\section{Mass spectrum of composite states}
\label{sec:bosonicspectrum}

In this section, we compute the bosonic spectrum  of Models I and II introduced in section~\ref{sec:models}. The calculations proceed along lines similar to those of Examples~A and B of section~\ref{sec:warmup}, with a few key differences. First, we need to bring the actions of the models into a form that is amenable to the formalism presented in Appendices~\ref{sec:sigmamodel} and \ref{sec:AVandPS}. To this end we note that, as long as we are only interested in computing two-point functions, it is sufficient to retain in the action fluctuations to second order around a given background solution. We remind the Reader that, for the background solutions that we consider, the only non-zero fields are the $\sigma$-component of $\Phi$, the warp factor $A$, and (for Model II) $\phi$.

\subsection{Formalism} \label{Form}

Consider first the kinetic term for the gauge field $\mathcal A_M$ appearing in the actions of Eq.~\eqref{eq:ActionModel1} and Eq.~\eqref{eq:ActionModel2}. Expanding in $\mathcal A_M$, we have
\beq
	- N_C \int \dd^5x \sqrt{-g} \, \Tr \Big[ \frac{1}{4} g^{MP} g^{NQ} \mathcal F_{MN} \mathcal F_{PQ} \Big] = - N_C \int \dd^5x \sqrt{-g} \, \bigg\{ \frac{1}{4} \sum_{i = A,V }(\mathcal F^{(i)}_{MN})^2 + \cdots \bigg\} \,,
\eeq
where the field strengths are given by
\beq
	\mathcal F^{(i)}_{MN} \equiv \partial_M \mathcal A^{(i)}_N - \partial_N \mathcal A^{(i)}_M \,,
\eeq
we decomposed $\mathcal A_M$ along the broken and unbroken directions, writing $\mathcal A^{(i)}_M = (A_M,V_M)$, and the dots contain terms that are cubic or higher order in $\mathcal A_M$. As can be seen, the two-point functions are only sensitive to the Abelian part of the flavour group.

Next, let us work out the form of the kinetic term for $\Phi$. Using the decomposition of $\Phi$ in Eq.~\eqref{eq:Phidecomp}, we have
\beqs
	D_M \Phi &=& U \Big[ \partial_M \Phi_H \Sigma + \mathcal X_M \Phi_H \Sigma + \Phi_H \Sigma \mathcal X_M^T \Big] U^T \,, \nonumber \\
		(D_M \Phi)^\dag &=& U^* \Big[ - \Sigma \partial_M \Phi_H + \Sigma \Phi_H \mathcal X_M + \mathcal X_M^T \Sigma \Phi_H \Big] U^\dag \,, \\ \nonumber
	\mathcal X_M &\equiv& U^{-1} (\partial_M + i g_5 \mathcal A_M ) U \,.
\eeqs
The scalar kinetic term is hence given by
\beqs
\label{eq:kin}
	- \Tr \left( | D_M \Phi |^2 \right) &=& - \Tr \Big\{ (\partial_M \Phi_H)^2 + 2 \mathcal X^M [\Phi_H,\partial_M \Phi_H] - 2 \Phi_H^2 (\mathcal X_M)^2 + 2 \Phi_H \mathcal X^M \Phi_H \Sigma \mathcal X_M^T \Sigma \Big\} \nonumber \\ &=&
	- \frac{N_F}{2} (\partial_M \sigma)^2 - \frac{1}{2} (\partial_M S_{\hat A})^2 \nonumber \\ && - \Tr \Big\{ 2 \mathcal X^M [S,\partial_M S] - 2 \Phi_H^2 (\mathcal X_M)^2 + 2 \Phi_H \mathcal X^M \Phi_H \Sigma \mathcal X_M^T \Sigma \Big\} \\ \nonumber &=&
	- \frac{N_F}{2} (\partial_M \sigma)^2 - \frac{1}{2} (\partial_M S_{\hat A})^2 + \frac{\sigma^2}{2} \Tr \Big[ (\mathcal X_M)^2 - \mathcal X^M \Sigma \mathcal X_M^T \Sigma \Big] + \cdots
	\,,
\eeqs
where we used Eq.~\eqref{eq:genid}, and again the dots contain terms cubic or higher order in the fluctuations. 
At linear order in the fluctuations
\beq
	\mathcal X_M = i ( \partial_M \Pi + g_5 \mathcal A_M) + \cdots \,,
\eeq
thus using Eq.~\eqref{eq:genid} we obtain
\beq
	- \Tr \left( | D_M \Phi |^2 \right) = - \frac{N_F}{2} (\partial_M \sigma)^2 - \frac{1}{2} (\partial_M S_{\hat A})^2
	- \frac{\sigma^2}{2} (\partial_M \pi_{\hat A} + g_5 A_{\hat A M})^2 + \cdots \,.
\eeq

Putting everything together, we can write the  action of Model I defined in Eq.~\eqref{eq:ActionModel1} as
\beqs
\label{eq:S2Model1}
	\mathcal S = N_C^2 \int \dd^5x \sqrt{-g} \, \bigg\{ && \hspace{-0.4cm} \frac{R}{4} - \frac{x_F}{2} (\partial_M \sigma)^2 - \mathcal V\left[\mathcal I(\sigma,S)\right] \bigg\} \\ \nonumber
	+ N_C \int \dd^5x \sqrt{-g} \, \bigg\{ && \hspace{-0.4cm} - \frac{1}{2} (\partial_M S_{\hat A})^2 - \frac{1}{4} \sum_{i = A,V }(\mathcal F^{(i)}_{MN})^2
	- \frac{\sigma^2}{2} (\partial_M \pi_{\hat A} + g_5 A_{\hat A M})^2
	\bigg\} + \cdots \,.
\eeqs
Similarly, for the  action of Model II defined in Eq.~\eqref{eq:ActionModel2}, we obtain
\beqs
\label{eq:S2Model2}
	\mathcal S = N_C^2 \int \dd^5x \sqrt{-g} \, \bigg\{ && \hspace{-0.4cm} \frac{R}{4} - \frac{x_F}{2} (\partial_M \sigma)^2 - \frac{1}{2} (\partial_M \phi)^2 - \mathcal V\left[\mathcal I(\sigma,S), \phi\right] \bigg\} \\ \nonumber
	+ N_C \int \dd^5x \sqrt{-g} \, \bigg\{ && \hspace{-0.4cm} - \frac{1}{2} (\partial_M S_{\hat A})^2 - \frac{1}{4} \sum_{i = A,V }(\mathcal F^{(i)}_{MN})^2
	- \frac{\sigma^2}{2} (\partial_M \pi_{\hat A} + g_5 A_{\hat A M})^2
	\bigg\} +\cdots \,.
\eeqs
Neglecting the higher-order terms contained in the dots, these actions are on a form suitable for the formalism of Appendices~\ref{sec:sigmamodel} and \ref{sec:AVandPS}.

Let us now comment on how the computation of the spectrum and decay constant in Model I and Model II differs from the examples of section~\ref{sec:warmup}. First, the scalar fluctuations $\mathfrak a^a$, involve not only the fluctuation of $\sigma$, but also the fluctuations $S_{\hat A}$ in both Models I and II, and additionally the fluctuation of $\phi$ in Model II. Their equations of motion and boundary conditions are given in Eq.~\eqref{eq:fluceoms} and Eq.~\eqref{eq:flucbcs}. Second, the decay constant $f$ scales differently with $N_C$ 
with respect to \eq{eq:decayconstant}, and can be obtained as
\beq
	f^2 \equiv 2F_G^2 = \lim\limits_{r \rightarrow \infty} \bigg\{ 2N_C \frac{e^{2A}}{g_5^2} \frac{\partial_r a}{a} \Big|_{q^2 = 0} \bigg\} \,.
\eeq
We remind the Reader that $a(q,r)$  encodes the radial profile of the transverse part of the axial-vector, 
$P^{\mu\nu} A^{\hat A}_{\nu}(q,r) = \tilde A^{\hat A \mu}(q) a(q,r)$. We present the results in terms of a rescaled decay constant,
$\tilde f \equiv f/\sqrt{N_C}$. 
The remaining equations of motion and boundary conditions---for the tensor, vector, axial-vector, and pseudoscalar fluctuations---take the same form as in Eqs.~\eqref{eq:fluceoms22}--\eqref{eq:eomX2}.

Finally, let us derive a perturbative estimate for $g_5$, following~\cite{Erlich:2005qh}. The vector-current two-point function reads
\beqs
        q^2 \Pi_V(q^2) P^{\mu\nu} \delta^{{A}{B}} \equiv i \, \int \dd^4
x \, e^{i q_\sigma x^\sigma} \langle J^{{A}\mu}(x) J^{{B}\nu}(0) \rangle \,.
\eeqs
The asymptotic UV behaviour of the vector is determined by \eq{eq:eomV}. Writing $P^{\mu\nu} V^{A}_{\nu}(q,r) = \tilde V^{A \mu}(q) v(q,r)$, and expanding in powers of $e^{-r}$, we find that in all our models
\beq
	v(q,r) \propto 1 + \left[ v_2(q) + \frac{1}{4} q^2 \log\left(q^2 e^{-2r} \right) \right] e^{-2r} + \cdots \,,
\eeq
where $v_2(q)$ is an integration constant. Similar to the derivation leading to Eq.~\eqref{eq:JJT} for the transverse part of the axial-vector-current two-point function, this implies that the regularized expression for $\Pi_V$ is given by
\beq
	\Pi_V^{\rm reg}(q^2,r) = -N_C \frac{e^{2A}}{q^2 g_5^2} \frac{\partial_r v(q,r)}{v(q,r)} = \frac{N_C}{g_5^2} \left[ \frac{1}{2} + \frac{2v_2(q)}{q^2} + \frac{\log\left(q^2 e^{-2r} \right)}{2} \right] + \cdots \,.
\eeq
Comparing the coefficient in front of $\log(q^2)$ with the perturbative one-loop field-theory result in the large $q^2$ limit
\beq
	\Pi_V(q^2) \simeq \frac{d(R) T(R)}{24\pi^2} \log\left(\frac{q^2}{\mu^2}\right) = \frac{N_C}{24\pi^2} \log\left(\frac{q^2}{\mu^2}\right) \,,
\eeq
where $\mu$ is an arbitrary renormalization scale, and we used the values of $d(R)$ and $T(R)$ for fermions in the fundamental of $Sp(2N_C)$, given in Appendix \ref{beta},
one obtains the estimate $g_5 \simeq \sqrt{12} \pi \simeq 10.9$. While this gives some indication for the value of $g_5$, the estimate relies on a perturbative computation in a strongly-coupled field theory. Hence, in this section we rather take the approach of studying the dependence of the spectrum on $g_5$, keeping it as a free parameter. 
We will return to the choice of the $g_5$ value in section~\ref{sec:latticeNJL}, when we compare our results to those of lattice simulations.

\subsection{Model I}

Let us work out explicitly the linearised equations of motion and boundary conditions for the scalar fluctuations $\mathfrak a^a = \left(\mathfrak a^\sigma,\mathfrak a^S\right)$ starting from Eq.~\eqref{eq:fluceoms} and Eq.~\eqref{eq:flucbcs}. By making use of Eq.~\eqref{eq:conditionsI}, it can be shown that the equations for $\mathfrak a^\sigma$ and $\mathfrak a^S$ decouple, leading to
\beq
	\left[ \partial_r^2 + 4 A' \partial_r - \left( \frac{1}{x_F} \partial_{\mathcal I}^2 \mathcal V + \frac{8 \sigma'}{3A'} \partial_{\mathcal I} \mathcal V + x_F \frac{16\sigma'^2}{9A'^2} \mathcal V \right) - e^{-2A} q^2 \right] \mathfrak a^\sigma = 0 \,,
\eeq
with boundary conditions
\beq
	\partial_r \mathfrak a^\sigma \Big|_{r_i} = \frac{3A'}{2x_F\sigma'^2}  \left[ e^{-2A} q^2 - \frac{A'}{2} \partial_r \left( \frac{A''}{A'^2} \right) \right] \mathfrak a^\sigma \Big|_{r_i} \,,
\eeq
and
\beq
\label{eq:eomaS}
	\left[\partial_r^2 + 4A' \partial_r - N_C \partial_S^2 \mathcal I \, \partial_{\mathcal I} \mathcal V - e^{-2A} q^2 \right] \mathfrak a^S = 0 \,,
\eeq
with boundary conditions $\mathfrak a^S |_{r_i} = 0$. We observe that $\mathfrak a^S$ is the only fluctuation for which the choice 
of the invariant $\mathcal I$ matters.

We now have everything in place to compute the bosonic spectrum of Model I. As for Examples~A and~B of section~\ref{sec:warmup}, one needs to introduce IR and UV cutoffs that serve as regulators, with the physical results recovered as they are taken towards the end-of-space and the boundary, respectively. In order to ensure that the numerics captures the dynamics close to the end-of-space, we find it convenient to work with the radial coordinate $\rho$ defined as $r = \frac{1}{2} \big[ \rho + \log \left( 2 \cosh \rho \right) \big]$ in terms of which the IR ($r=0$) and UV ($r \rightarrow +\infty$) are located at $\rho = -\infty$ and $\rho = +\infty$, respectively, while $\rho \simeq r$ for large $r$. We report on the numerical results in terms of the IR and UV cutoffs $\rho_1$ and $\rho_2$, having checked that these are chosen such that the spectrum has converged sufficiently in order for cutoff effects to be negligible, unless explicitly stated.

\begin{figure}[t]
\begin{center}
\includegraphics[width=\figwidth]{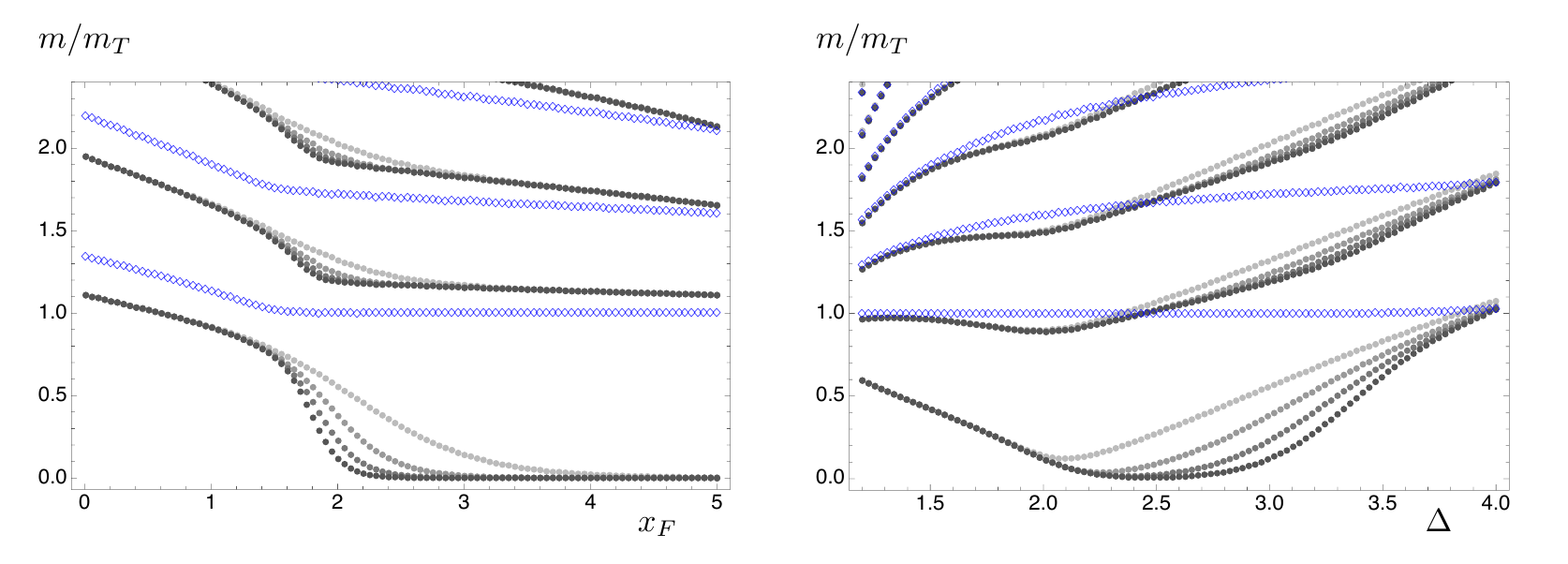}
\caption{Model I. Spectrum of $S$-resonances normalized to the mass $m_T$ of the lightest tensor, as a function of $x_F$ with $\Delta = 3$ (left panel) and as a function of $\Delta$ with $x_F = 2$ (right panel). Both panels have $\rho_2 = 10$. The result of using the scalar-potential invariant $\mathcal I_1$ is shown in shades of gray for different values of the IR cutoff $\rho_1 = -13, -10, -7, -4$ (darker to lighter), whereas the result of using the invariant $\mathcal I_2$ and $\rho_1 = -13$ is shown in blue diamonds.}
\label{fig:Model1S_1}
\end{center}
\end{figure}

\begin{figure}[t]
\begin{center}
\includegraphics[width=\figwidth]{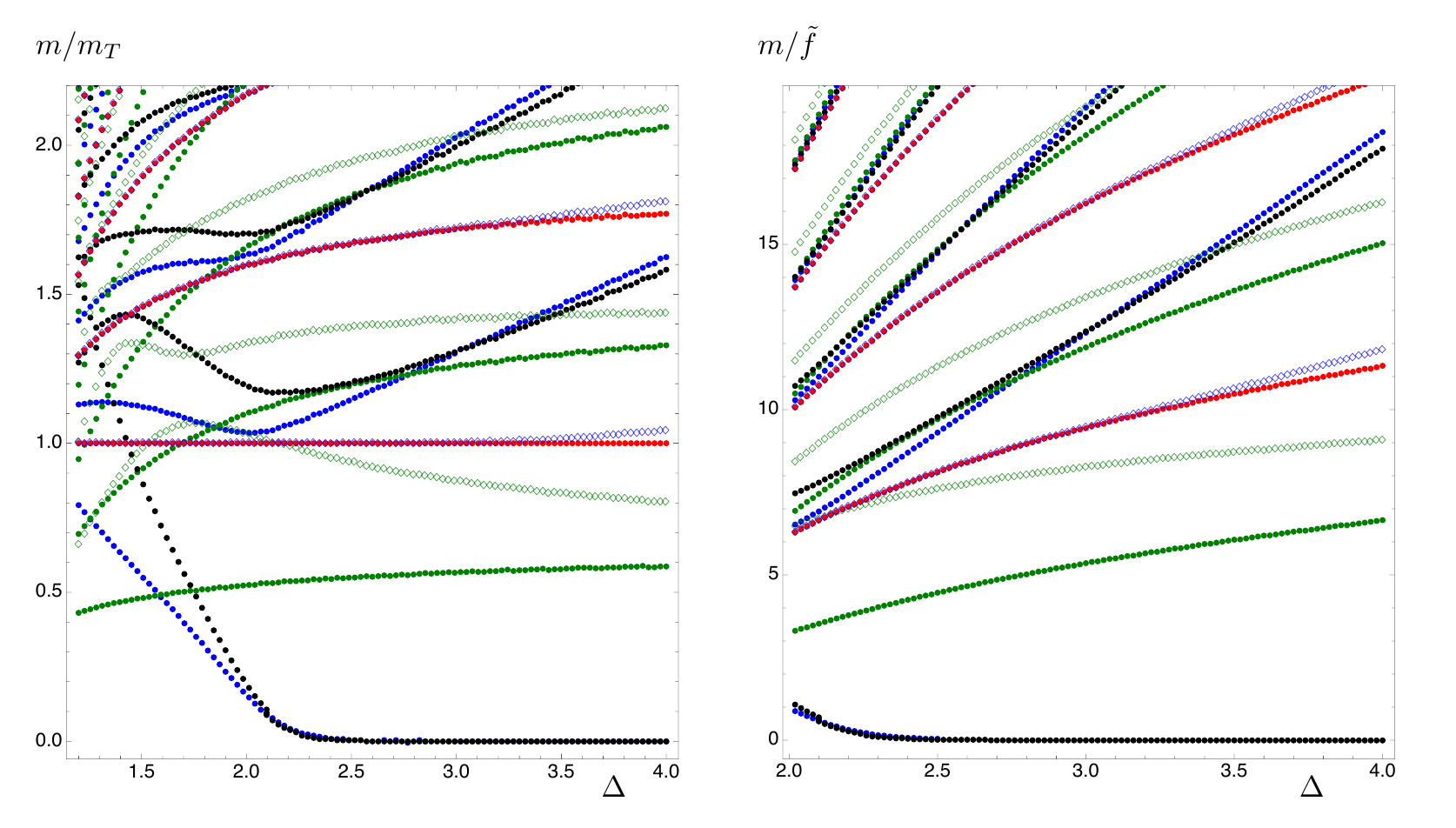}
\vspace{\ecap}
\caption{Model I. Spectrum as a function of $\Delta$ for $x_F = 2$, $g_5 = 5$, $\rho_1 = -7$, $\rho_2 = 10$. The left panel is normalized to the mass $m_T$ of the lightest tensor, while the right panel is normalized to the decay constant $\tilde f \equiv f / \sqrt{N_C}$. The colour coding for the spectrum is: singlet scalar (blue), non-singlet scalar (blue diamonds), tensor (red), pseudoscalar (black), vector (green), axial-vector (green diamonds).}
\label{fig:Model1_1}
\end{center}
\end{figure}

Let us first consider the spectrum of $S$-resonances, which as mentioned depends on the choice of invariant $\mathcal I$. The simplest non-trivial invariant $\mathcal I_1 = \left[ 2N_F^{-1}\Tr (\Phi^\dag \Phi) \right]^{1/2}$ was given in Eq.~\eqref{eq:I1}. The dependence of the resulting spectrum as a function of $x_F$ and $\Delta$ is represented by the black dots in Fig.~\ref{fig:Model1S_1}. Since the numerical results are rather sensitive to the IR cutoff, we have included a few different values of $\rho_1$, represented by the gray dots. As can be seen, for sufficiently large values of $x_F$, there is a parametrically light state. In order to investigate the origin of this light state, we computed the spectrum using a different invariant
\beq
	\mathcal I_2 = \left[ c_1 \Tr (\Phi^\dag \Phi \Phi^\dag \Phi) + c_2 \left( \Tr(\Phi^\dag \Phi)\right)^2 \right]^{1/4} \,,
\eeq
that also satisfies \eq{eq:conditionsI}, but is more generic in the sense that it contains more than one kind of single-trace term. For simplicity, we pick $c_1$ and $c_2$ such that $\mathcal I_2 = \sigma + \mathcal O(S^4)$,  in order for the third term of Eq.~\eqref{eq:eomaS} to vanish on the background, 
so that $\mathfrak a^S$ satisfies the equation of motion of a massless scalar field propagating on a fixed background geometry. The resulting spectrum of $S$ resonances is represented by the blue diamonds of Fig.~\ref{fig:Model1S_1}. While the qualitative features of the heavy states remain similar to the case of $\mathcal I = \mathcal I_1$, the light $S$ resonance is no longer present. This suggests that the choice $\mathcal I_1$ 
is too simplistic in order to result in a realistic model, since the HC gauge theory has no symmetry which could explain the lightness of $S$.

\begin{figure}[t]
\begin{center}
\includegraphics[width=\figwidth]{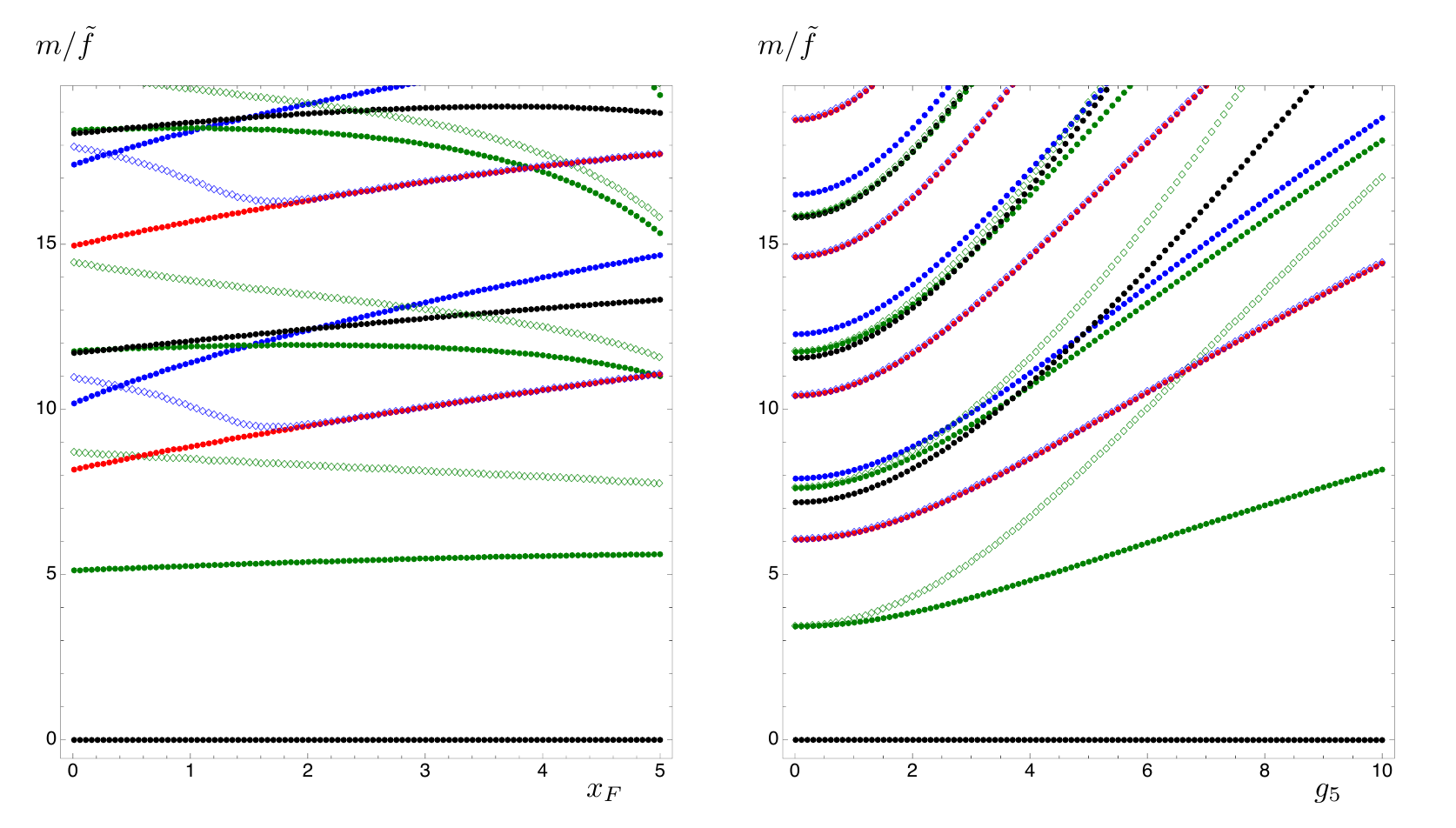}
\vspace{\ecap}
\caption{Model I. The left panel shows the spectrum as a function of $x_F$ for $\Delta = 3$, $g_5 = 5$, $\rho_1 = -7$, $\rho_2 = 8$. The right panel shows the spectrum as a function of $g_5$ for $x_F = 2$, $\Delta = 3$, $\rho_1 = -7$, $\rho_2 = 10$. Both panels are normalized to the decay constant $\tilde f \equiv f / \sqrt{N_C}$. The colour coding for the spectrum is: singlet scalar (blue), non-singlet scalar (blue diamonds), tensor (red), pseudoscalar (black), vector (green), axial-vector (green diamonds). Since $\Delta > 2$, there is also a massless dilaton in the spectrum, as can be seen from Fig.~\ref{fig:Model1_1}; due to the numerical resolution deployed only the Goldstone is shown in the present two plots.}
\label{fig:Model1_2}
\end{center}
\end{figure}

In Fig.~\ref{fig:Model1_1} we show the spectrum as a function of $\Delta$. For completeness, we included the spectrum of $S$-resonances, with the choice $\mathcal I = \mathcal I_2$, keeping in mind that such result may have a strong model dependence. For $\Delta > 2$, both $SU(2N_F)$ and scale invariance are broken spontaneously; however, the masses of both the dilaton and the NGBs are lifted, due to cutoff effects that are most pronounced around $\Delta \simeq 2$. We note that in Fig.~\ref{fig:Model1_1}, the dilaton always remains lighter than the Goldstone bosons associated with the breaking of $SU(2N_F)$. In Fig.~\ref{fig:Model1_2}, we show the dependence of the spectrum on $x_F$, and $g_5$, for fixed $\Delta = 3$. We remind the Reader that for $x_F \rightarrow 0$, the background geometry is close to AdS. The spectrum altogether has a mild dependence on the number of flavours, while the dependence on $g_5$ is stronger. We observe that the $S$-resonances coincide with the tensor for large $x_F$. Because of our choice of invariant $\mathcal I = \mathcal I_2$, these modes satisfy the same equations of motion, but their boundary conditions differ.

Let us stress again that the presence of a massless dilaton in the spectrum is an artefact of the minimality of Model I: the profile of the bulk scalar $\sigma$, given by \eq{eq:sigmanUV}, is chosen to realise a purely spontaneous breaking of scale invariance when $\Delta>2$. A realistic, composite-Higgs field theory, that we aim to describe, does contain explicit breaking sources as well. We move, therefore, to the analysis of the spectrum of Model II, where the dilaton mass will be lifted.

\subsection{Model II}

Consider the linearised equations of motion and boundary conditions for the scalar fluctuations $\mathfrak a^a = \left(\mathfrak a^\sigma,\mathfrak a^S,\mathfrak a^\phi\right)$ given in Eq.~\eqref{eq:fluceoms} and Eq.~\eqref{eq:flucbcs}. As in Model I, the fluctuations of $S$ decouple, leading to the same equation of motion~\eqref{eq:eomaS} and boundary conditions $\mathfrak a^S |_{r_i} = 0$. However, the fluctuations of $\sigma$ and $\phi$ mix, leading to the following linearised equations of motion:
\beqs
	\bigg[ \partial_r^2 + 4 A' \partial_r - \left( \frac{1}{x_F} \partial^2_{\mathcal I} \mathcal V + \frac{8 \sigma'}{3A'} \partial_{\mathcal I} \mathcal V + x_F \frac{16 \sigma'^2}{9 A'^2} \mathcal V \right) - e^{-2 A} q^2 \bigg] \mathfrak a^\sigma \hspace{0.2cm} && \\ \nonumber
	- \bigg[  \frac{1}{x_F} \partial_{\mathcal I} \partial_\phi \mathcal V + \frac{4 \phi'}{3 x_F A'} \partial_{\mathcal I} \mathcal V + \frac{4 \sigma'}{3A'} \partial_\phi \mathcal V + \frac{16 \sigma' \phi'}{9 A'^2} \mathcal V \bigg] \mathfrak a^\phi &=& 0 \,, \\
	\bigg[ \partial_r^2 + 4 A' \partial_r - \left(\partial^2_\phi \mathcal V + \frac{8 \phi'}{3 A'} \partial_\phi \mathcal V + \frac{16 \phi'^2}{9 A'^2} \mathcal V \right) - e^{-2 A} q^2 \bigg] \mathfrak a^\phi \hspace{0.2cm} && \\ \nonumber
	- \bigg[ \partial_{\mathcal I} \partial_\phi \mathcal V + \frac{4 \phi'}{3 A'} \partial_{\mathcal I} \mathcal V + x_F \frac{4 \sigma'}{3A'} \partial_\phi \mathcal V + x_F \frac{16 \sigma' \phi'}{9 A'^2} \mathcal V \bigg] \mathfrak a^\sigma &=& 0 \,.
\eeqs
with boundary conditions
\beqs
	\sigma' \left( x_F \sigma' \partial_r \mathfrak a^\sigma + \phi' \partial_r \mathfrak a^\phi \right) \Big|_{r_i} &=& \frac{3A'}{2}  \left[ e^{-2A} q^2 - \frac{A'}{2} \partial_r \left( \frac{A''}{A'^2} \right) \right] \mathfrak a^\sigma \Big|_{r_i} \,, \\
	\phi' \left( x_F \sigma' \partial_r \mathfrak a^\sigma + \phi' \partial_r \mathfrak a^\phi \right) \Big|_{r_i} &=& \frac{3A'}{2}  \left[ e^{-2A} q^2 - \frac{A'}{2} \partial_r \left( \frac{A''}{A'^2} \right) \right] \mathfrak a^\phi \Big|_{r_i} \,.
\eeqs

\begin{figure}[t]
\begin{center}
\includegraphics[width=\figwidth]{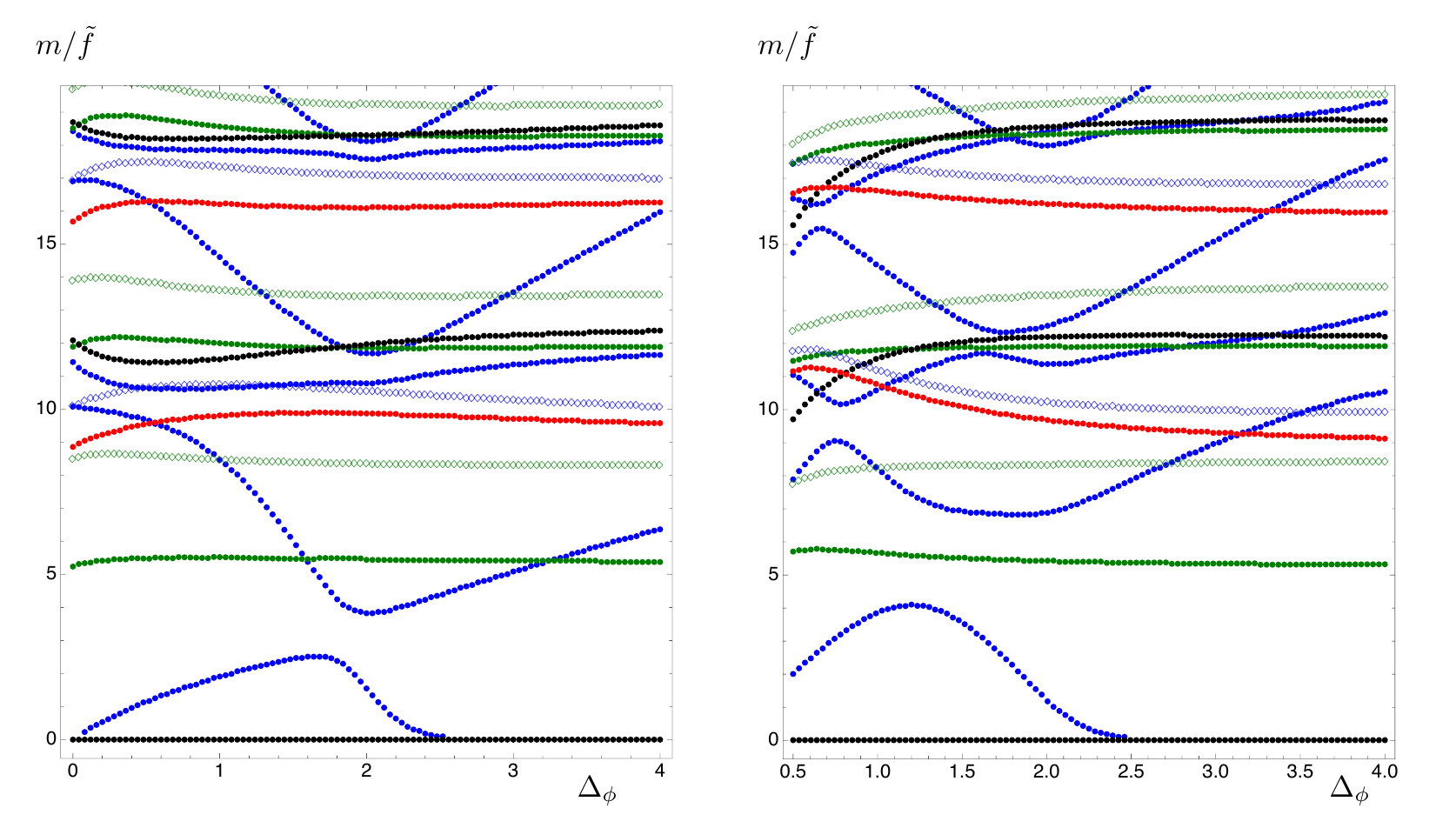}
\caption{Model IIA (left) and IIB (right). Spectrum as a function of $\Delta_\phi$ for $x_F = 1$, $\Delta = 3$, $g_5 = 5$, $\rho_1 = -13$, $\rho_2 = 8$, and $\phi_A = 1.5$ (Model IIA, left), $\phi_B = 0.9$ (Model IIB, right), normalized to the decay constant $\tilde f \equiv f / \sqrt{N_C}$. The colour coding for the spectrum is: singlet scalar (blue), non-singlet scalar (blue diamonds), tensor (red), pseudoscalar (black), vector (green), axial-vector (green diamonds).}
\label{fig:Model2AB_1}
\end{center}
\end{figure}

\begin{figure}[t]
\begin{center}
\includegraphics[width=\figwidth]{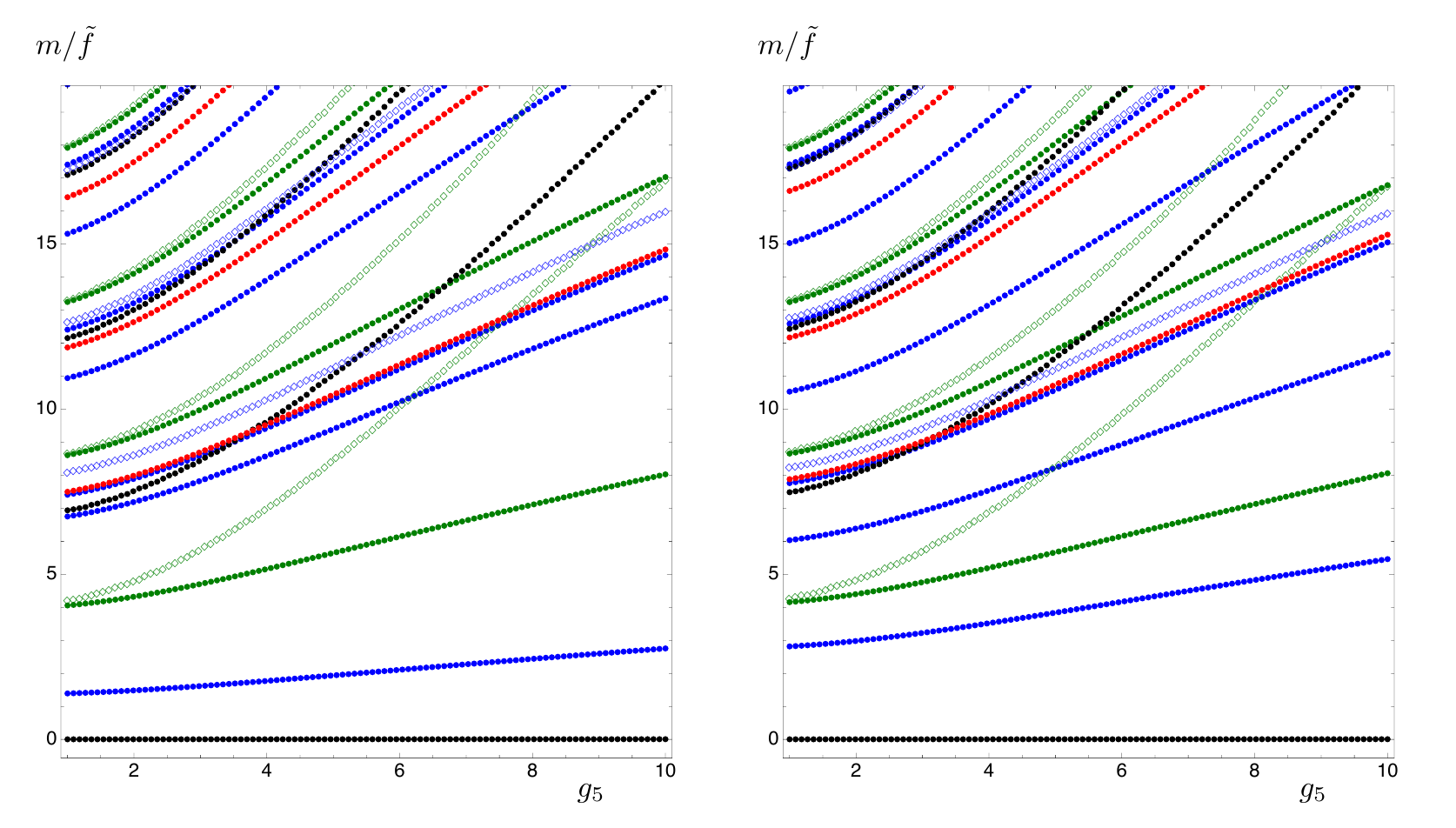}
\caption{Model IIA (left) and IIB (right). Spectrum as a function of $g_5$ for $x_F = 1$, $\Delta = 3$, $\Delta_\phi = 1$, $\rho_1 = -13$, $\rho_2 = 8$, and $\phi_A = 2$ (Model IIA, left), $\phi_B = 0.9$ (Model IIB, right), normalized to the decay constant $\tilde f \equiv f / \sqrt{N_C}$. The colour coding for the spectrum is: singlet scalar (blue), non-singlet scalar (blue diamonds), tensor (red), pseudoscalar (black), vector (green), axial-vector (green diamonds).}
\label{fig:Model2AB_3}
\end{center}
\end{figure}

\begin{figure}[t]
\begin{center}
\includegraphics[width=\figwidth]{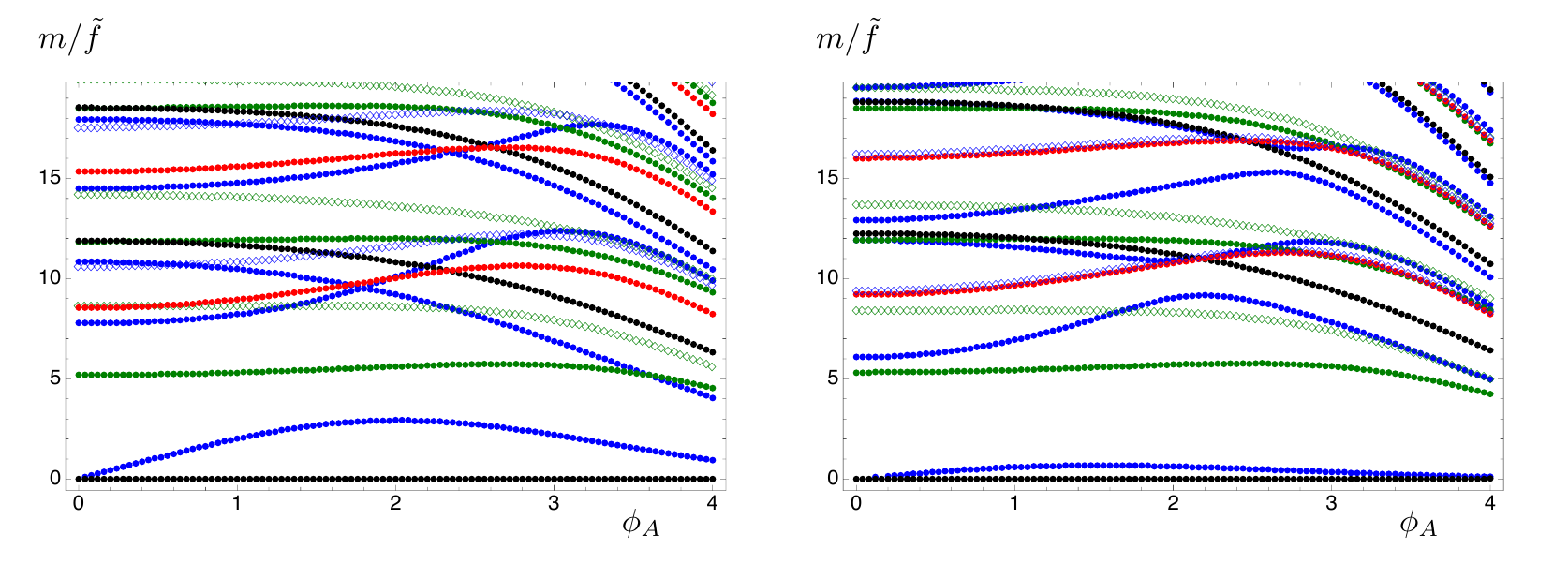}
\caption{Model IIA. Spectrum as a function of $\phi_A$ for $\Delta = 3$, $\Delta_\phi = 1$, $g_5 = 5$, $\rho_1 = -13$, $\rho_2 = 8$, and $x_F = 0.5$ (left), $x_F = 1.5$ (right), normalized to the decay constant $\tilde f \equiv f / \sqrt{N_C}$. The colour coding for the spectrum is: singlet scalar (blue), non-singlet scalar (blue diamonds), tensor (red), pseudoscalar (black), vector (green), axial-vector (green diamonds).}
\label{fig:Model2A_1}
\end{center}
\end{figure}

\begin{figure}[t]
\begin{center}
\includegraphics[width=\figwidth]{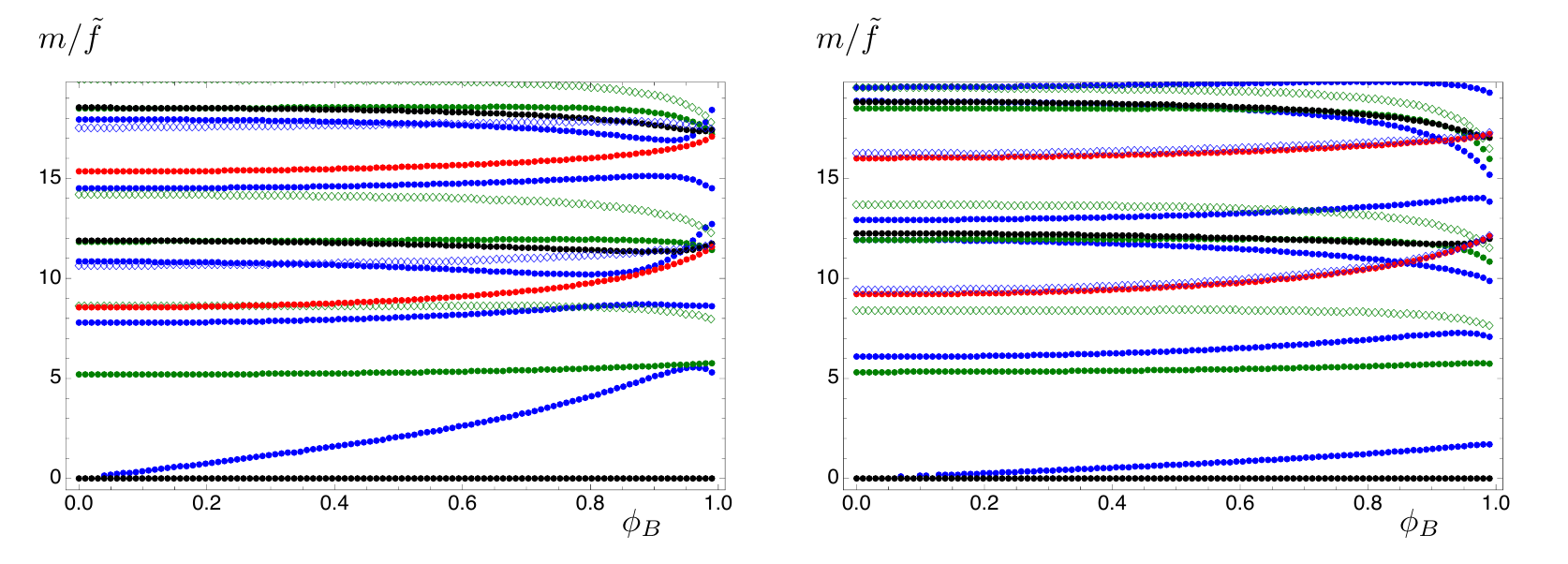}
\caption{Model IIB. Spectrum as a function of $\phi_B$ for $\Delta = 3$, $\Delta_\phi = 1$, $g_5 = 5$, $\rho_1 = -13$, $\rho_2 = 8$, and $x_F = 0.5$ (left), $x_F = 1.5$ (right), normalized to the decay constant $\tilde f \equiv f / \sqrt{N_C}$. The colour coding for the spectrum is: singlet scalar (blue), non-singlet scalar (blue diamonds), tensor (red), pseudoscalar (black), vector (green), axial-vector (green diamonds).}
\label{fig:Model2B_1}
\end{center}
\end{figure}

\begin{figure}[t]
\begin{center}
\includegraphics[width=\figwidth]{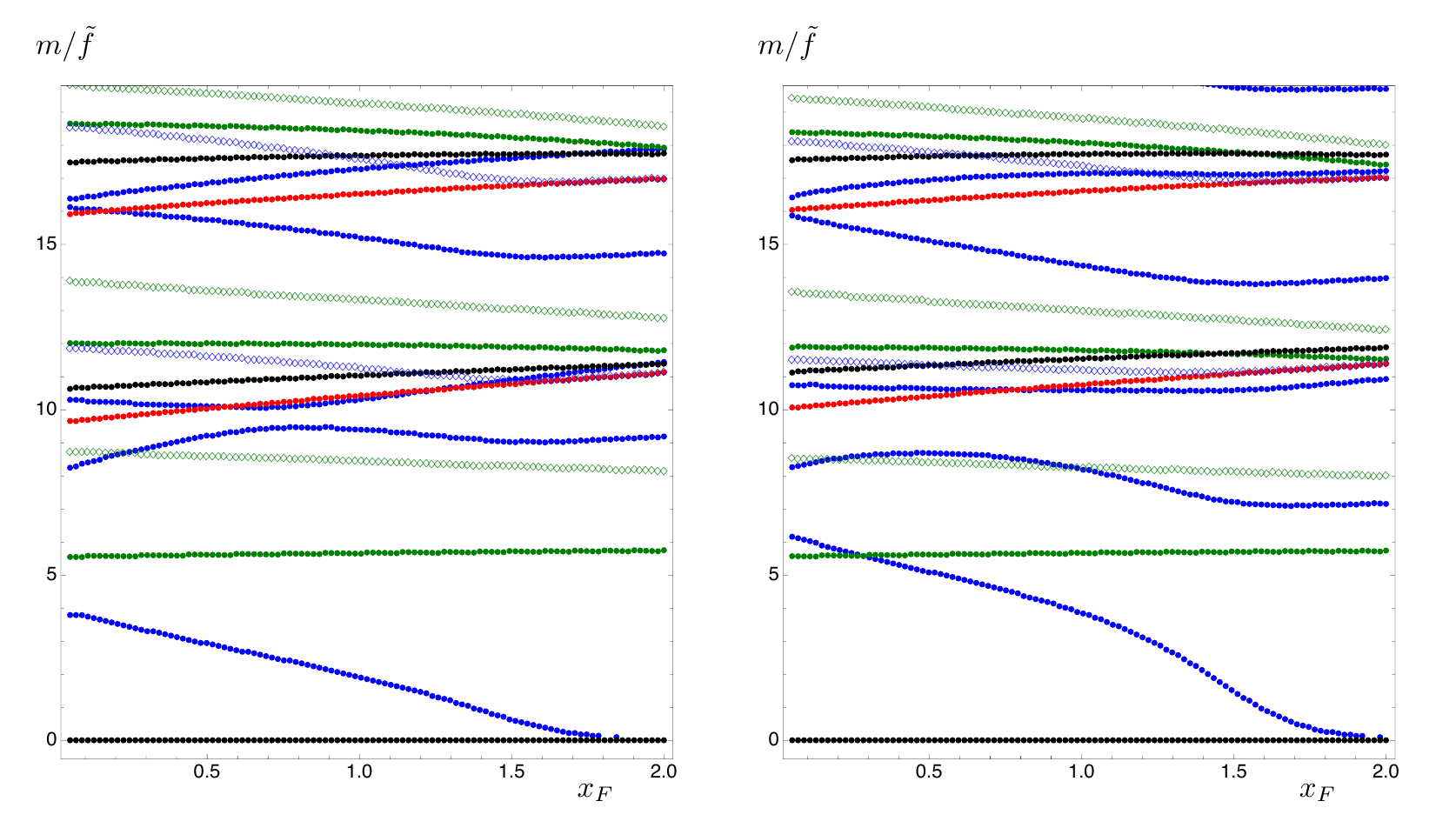}
\caption{Model IIA (left) and IIB (right). Spectrum as a function of $x_F$ for $\Delta = 3$, $\Delta_\phi = 1$, $g_5 = 5$, $\rho_1 = -13$, $\rho_2 = 8$, and $\phi_A = 2$ (Model IIA, left), $\phi_B = 0.9$ (Model IIB, right), normalized to the decay constant $\tilde f \equiv f / \sqrt{N_C}$. The colour coding for the spectrum is: singlet scalar (blue), non-singlet scalar (blue diamonds), tensor (red), pseudoscalar (black), vector (green), axial-vector (green diamonds).}
\label{fig:Model2AB_2}
\end{center}
\end{figure}

\begin{figure}[t]
\begin{center}
\hspace{-0.4cm}
\includegraphics[width=\figwidth]{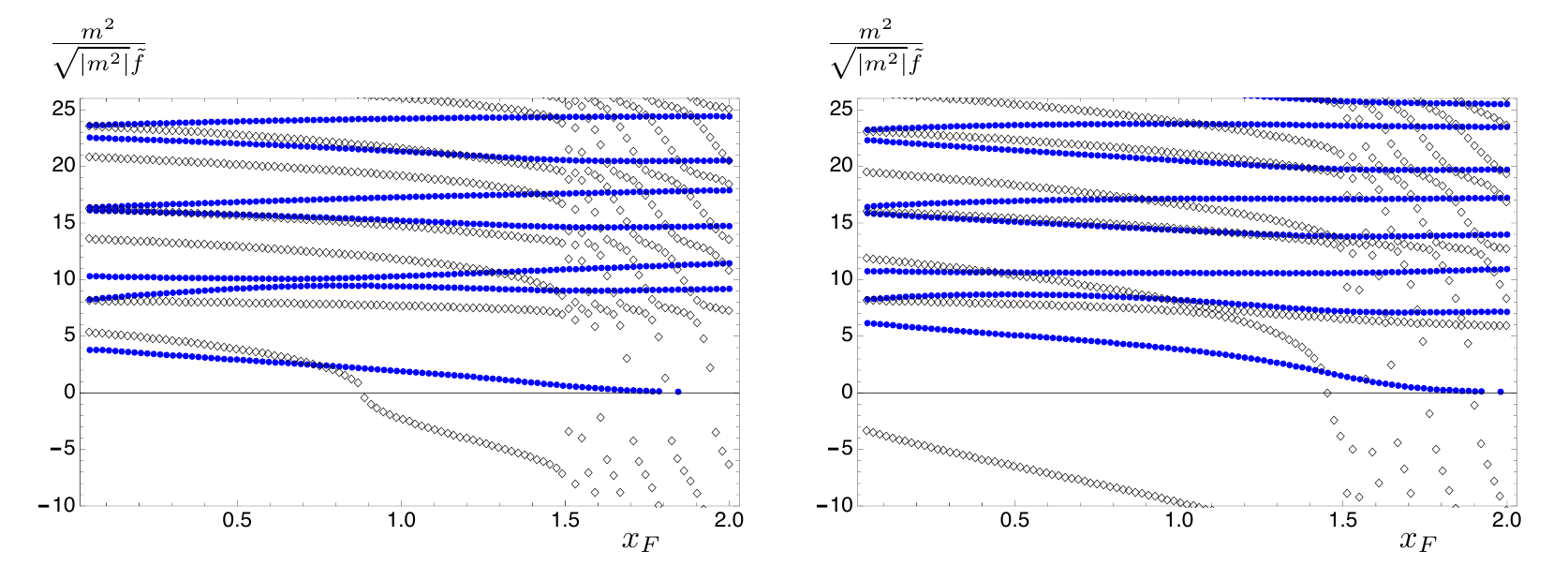}
\caption{Model IIA (left) and IIB (right). Spectrum of scalar resonances (blue) compared to the probe approximation (black diamonds) as a function of $x_F$ for $\Delta = 3$, $\Delta_\phi = 1$, $g_5 = 5$, $\rho_1 = -13$, $\rho_2 = 8$, and $\phi_A = 2$ (Model IIA, left), $\phi_B = 0.9$ (Model IIB, right), normalized to the decay constant $\tilde f \equiv f / \sqrt{N_C}$.}
\label{fig:Model2AB_4}
\end{center}
\end{figure}

The main motivation for introducing Model II is to study how explicit breaking of scale invariance can lift the mass of the dilaton while keeping the NGBs massless. In section \ref{sec:Model2} we introduced two possible forms of the superpotential $\mathcal W$, corresponding to Models IIA and IIB, leading to different background solutions for the scalar $\phi$.
In both models, we present the results obtained with the choice of invariant $\mathcal I = \mathcal I_2$, which only affects the $S$ resonances,
as for Model~I.

In Fig.~\ref{fig:Model2AB_1}, we show the spectrum as a function of $\Delta_\phi$, which encodes the scaling dimension of the operator dual 
to $\phi$. We have chosen a sizeable value (see later discussion) for the integration constant $\phi_{A,B}$, which determines the amount of 
explicit ($\Delta_\phi<2$) or spontaneous ($\Delta_\phi>2$) breaking. 
The resulting spacing among singlet-scalar resonances is roughly reduced by one half, with respect to theories with only one bulk scalar, such as Model I. Indeed, one can think of the spectrum as consisting of two towers of resonances, arising from the mixing of the fluctuations $\{ \mathfrak a^\sigma, \mathfrak a^\phi \}$.
Some features are similar to the spectrum of Example~A in Fig.~\ref{fig:ExampleA3}: 
a light dilaton when $\mathcal O_\phi$ is nearly marginal ($\Delta_\phi \simeq 0$), as well as when the scale-invariance breaking is spontaneous ($\Delta_\phi > 2$). 
However, contrary to Example A, the Goldstone bosons are always massless, as we fixed $\Delta = 3$. In Fig.~\ref{fig:Model2AB_3}, we show that the resonance masses moderately grow with $g_5$. From both Fig.~\ref{fig:Model2AB_1} and~\ref{fig:Model2AB_3}, we observe that the spectra of Model IIA and IIB are in close agreement for the heaviest resonances. Indeed, the UV asymptotics in both models is that of AdS, and the heavy modes are less sensitive to the differing dynamics in the deep IR.

Meanwhile, in Fig.~\ref{fig:Model2A_1}, we picked $\Delta = 3$ and $\Delta_\phi = 1$ such that conformal invariance is broken explicitly while the flavour symmetry is broken spontaneously, and computed the spectrum of Model IIA as a function of $\phi_A$ for $x_F = 0.5$ and $x_F = 1.5$.
As in Example~A, the model becomes problematic for large values of the integration constant $\phi_A$. The problem is manifest in the maximum value of the dilaton mass, around $\phi_A \sim 2$: we believe that the decrease of the dilaton mass at larger $\phi_A$ is unphysical. 
Interestingly, the maximum value of the dilaton mass is suppressed as $x_F$ grows. In order to confirm the model-independence of this result,
we carried out the same study for Model IIB, see Fig.~\ref{fig:Model2B_1}. 
The dilaton mass grows in the range $0 \leq \phi_B \lesssim 1$, and its maximum value decreases as the number of flavours increases.

This phenomenon is further illustrated in Fig.~\ref{fig:Model2AB_2}, which shows the spectrum as a function of $x_F$ for both Models IIA and IIB,
with $\phi_{A,B}$ chosen such that the mass of the dilaton is near its maximum value. We interpret the presence of a parametrically light dilaton, suppressed by the number of flavours, as follows: the VEV $\langle \mathcal O_\sigma \rangle$, that breaks spontaneously the flavour symmetry as well as scale invariance, is of order $N_C N_F \simeq x_F N_C^2$, becoming enhanced at large $x_F$. In contrast, no such enhancement is present for the explicit breaking of scale invariance, encoded by the profile of the scalar $\phi$.

Finally, in order to confirm that the light scalar state seen in Fig.~\ref{fig:Model2AB_2} for large $x_F$ indeed is a dilaton, we perform a test suggested in \cite{Elander:2020csd}, namely to compare the correct calculation of the scalar spectrum to the result in the probe approximation. The former calculation, detailed in appendix \ref{sec:sigmamodel}, involves the gauge-invariant variables $\mathfrak a^a = \varphi^a - 
\frac{\Phi'^a}{6A'} h$, where $\varphi^a$ contains the fluctuations of the scalar fields $\sigma$ or $\phi$, 
while $h$ originates from the trace of the metric and is associated with the dilatation operator on the dual field theory side. The probe approximation consists of neglecting the contribution of the metric fluctuation $h$ (for details we refer to \cite{Elander:2020csd}). If the resulting spectrum agrees with the correct gauge-invariant calculation, we may conclude that the light scalar state is not a dilaton. Conversely, if the probe approximation fails to capture the light state, we may conclude that it has significant overlap with the dilaton. 
As can be seen from Fig.~\ref{fig:Model2AB_4}, in both Model IIA and IIB, the probe approximation not only fails to capture the light scalar state, but also some of the heavy resonances (the failure is particularly pronounced at large $x_F$ for which the probe approximation results in several tachyonic states). Hence, we conclude that the light scalar state is indeed a dilaton.

\section{Comparing different approaches to non-perturbative dynamics}
\label{sec:latticeNJL}

Let us compare, as far as possible, our present results
with existing ones for the bosonic spectra in closely related models, 
obtained from non-perturbative calculations alternative to the present holographic framework.
We mainly compare with rather recent lattice simulation
results~\cite{latt_sp4,latt_sp4_dyn}
of an $Sp(4)$ gauge theory.
We also compare with a previous analysis of some of us~\cite{Bizot:2016zyu} in the  
framework of the NJL
model~\cite{NJL,njlrev}. Before comparing concrete meson spectra results, it is useful to 
recapitulate the main features of these alternative calculations, stressing their differences, as well as some of their limitations. 

\subsection{Lattice simulations}

We first consider the analysis of \cite{latt_sp4} for a $Sp(2N_C)$ gauge symmetry, more precisely performed for $Sp(4)$, with fermions in two different representations, the fundamental and a two-index representation. The meson masses and decay constants have been extracted as usual from 
Euclidean two-point correlation functions with appropriate quantum numbers. 
Note most importantly that the analysis in \cite{latt_sp4} is performed
in the quenched approximation, i.e. with no dynamical fermions involved. Among the limitations 
that this implies, it means that the resulting 
meson masses and decay constant are insensitive to the number of fermion flavors,
either for the fermions of the fundamental 
or two-index representations (respectively $N_F$ and $n_F$ in the notation of appendix \ref{BL}). 
While the quenched approximation is exact for the $\psi$-fermions (in the fundamental representation) when $N_C\to\infty$ at fixed value of $N_F$, the same is not true for the $\chi$-fermions (in the anti-symmetric representation), even when $n_F$ remains small as compared to $N_C$. Another source of difference is that the holographic models require $N_C \gg 1$ in order for the gravity description to be weakly coupled. We can also compare our results with available lattice simulations with dynamical fermions, 
that have been performed for $Sp(4)$~\cite{latt_sp4_dyn}. Note, however, that this study only considers the fermions in the fundamental representation.

Another limitation of \cite{latt_sp4,latt_sp4_dyn} is that the singlet mesons
were not considered, the corresponding
disconnected  contributions to the correlators being computationally very challenging on the lattice. 
Concerning the important chiral and continuum extrapolation limits, these were performed using the
appropriate Wilson chiral perturbation theory~\cite{WChPT} 
(i.e. the double expansion in both small fermion mass and lattice spacing), 
at the so-called tree-level NLO (i.e. not including
chiral logarithms).  
Finally one shoud keep in mind that the analyses in \cite{latt_sp4,latt_sp4_dyn}
do not address any dynamical considerations (this is obviously true for \cite{latt_sp4} where the quenched
approximation is considered)
about possibly large anomalous dimensions and near-conformal dynamics.

Since our holographic study focused on the $Sp(2N_F)$ sector, we compare our results with the
lattice results for the (non-singlet) mesons from the fundamental-fermion sector.
In order to compare with the holographic results, we extract the ratios $m_i/F_G$, in the chiral and continum limits, from Table V of \cite{latt_sp4}, that provide the (lattice-normalized)  
equivalents of $F^2_G$ and $m^2_i$, $i=V,A,S$ in our notation. The meson mass hierarchy 
obtained in this way is\footnote{Statistical and systematic uncertainties are combined from 
the fit values given 
in Table V of \cite{latt_sp4}.}
\beq
\frac{m_S}{F_G} : \frac{m_A}{F_G} : \frac{m_V}{F_G} 
= (14.1 \pm 0.5) : (12.2 \pm 0.9) : (7.7 \pm 0.2) 
\label{sp4_fund} \,,
\eeq
where $S$ stands for the non-singlet scalar and $A, V$ for the (axial-)vectors, respectively. We next consider the unquenched lattice simulations for $Sp(4)$ with dynamical 
fermions in the fundamental representation~\cite{latt_sp4_dyn}. The results in the continuum
and chiral limits, explicitly given in their Table 11, give slightly larger $V, A$ masses 
(although with slightly larger uncertainties) compared to the above quenched results:
\beq
\frac{m_S}{F_G} : \frac{m_A}{F_G} : \frac{m_V}{F_G} 
= (14.2 \pm 1.7) : (13.4 \pm 1.5) : (8.1 \pm 0.3) \,.
\label{sp4dyn}
\eeq

Let us also quote some independent and complementary lattice results for 
a closely related model~\cite{latt_su4}. This study considered a more standard $SU(4)$ gauge
theory and, for technical reasons, a somewhat simplified
model with two Dirac fermions respectively in the fundamental and two-index antisymmetric
representations, with a symmetry breaking $SU(4)/SO(4)$ coset
that does not accomodate the SM Higgs. 
Despite this simplification the main qualitative features of other important aspects of 
a composite Higgs scenario
are assumed \cite{latt_su4} to be essentially captured. This study also illustrates the reasonably good agreement 
with large-$N_C$ expectations in the two fermion sectors, concerning the behavior of masses and decay constants. A qualitative addition of~\cite{latt_su4} 
with respect to~\cite{latt_sp4} and \cite{latt_sp4_dyn} is that 
the two fermion species are dynamical and simultaneously simulated.
The chiral and continuum limits were also performed from a more elaborate 
NLO Wilson chiral perturbation theory
(in particular~\cite{latt_su4} includes
chiral logarithms and a generalized chiral perturbation accounting for the non-trivial
mixing of the two fermion species~\cite{chpt_2f}).
For the fundamental-fermions sector, we extract the ratio $m_V/F_G$ in the continuum limit by taking the chiral 
extrapolation of the bands shown in Fig.~12 (top) and Fig.~5 (bottom) of 
\cite{latt_su4}.\footnote{The decay constant $F_P$ 
in \cite{latt_su4} is differently normalized than in \cite{latt_sp4}:
$F_P = f = \sqrt{2} F_G$. In QCD, $F_G\simeq 92$ MeV.} 
One obtains roughly
\be
\frac{m_V}{F_G} \simeq 8.8 \pm 0.6 \,,
\label{su4_fund}
\ee
which is somewhat larger compared to the quenched result in \eq{sp4_fund} and the unquenched one in \eq{sp4dyn}. For completeness,
we mention another independent lattice analysis~\cite{latt_su4_other} of the very same model
with a different lattice setup, giving 
complementary results. Since this paper does not address the chiral and continuum extrapolations, we do not elaborate further.

\begin{figure}[t]
\begin{center}
\includegraphics[width=\figwidth]{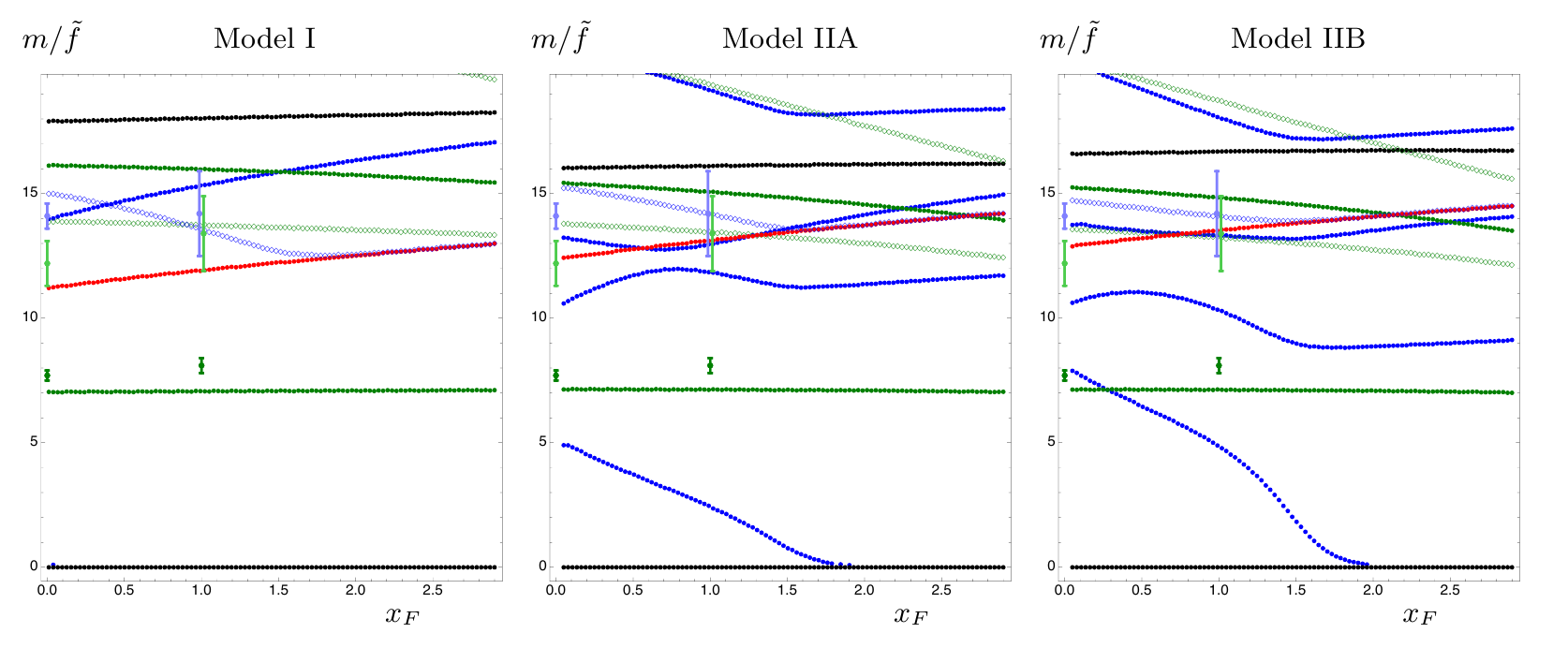}
\caption{Spectrum as a function of $x_F$ for Models I (left), IIA (center) and IIB (right) compared to lattice results taken from Eq.~\eqref{sp4_fund} (quenched, at $x_F = 0$) and Eq.~\eqref{sp4dyn} (unquenched, at $x_F = 1$), with the vector in green, axial-vector in light green, and the scalar $S$ in light blue. In Model I: $\Delta = 3$, $g_5 = 8$, $\rho_1 = -7$, $\rho_2 = 8$. In Models IIA and IIB: $\Delta = 3$, $\Delta_\phi = 1$, $g_5 = 8$, $\rho_1 = -13$, $\rho_2 = 8$, and $\phi_A = 2$ (Model IIA), $\phi_B = 0.9$ (Model IIB). All masses are normalized to the decay constant $\tilde f \equiv f / \sqrt{N_C}$. The colour coding for the spectrum (computed in the holographic models) is: singlet scalar (blue), non-singlet scalar (blue diamonds), tensor (red), pseudoscalar (black), vector (green), axial-vector (green diamonds).}
\label{fig:CompareToLattice}
\end{center}
\end{figure}

We now compare these lattice results with the spectra obtained in the holographic models of section~\ref{sec:models}. Although this comparison is most suitable for Models~IIA and~IIB, which incorporate the explicit breaking of conformal invariance, we also include the results from Model~I in our analysis.
The results of Ref.~\cite{latt_sp4}, displayed in Eq.~\eqref{sp4_fund}, were obtained in the quenched approximation, and hence we make the comparison for small $x_F \rightarrow 0$. Conversely, the results of Ref.~\cite{latt_sp4_dyn}, displayed in Eq.~\eqref{sp4dyn}, were obtained with dynamical fermions, and we make the comparison at $x_F\equiv (N_F=2)/(N_C=2) =1$.
As can be seen from Figs.~\ref{fig:Model1_2} and~\ref{fig:Model2AB_3}, the spectra computed holographically have a strong dependence on the 5D gauge coupling $g_5$. Several studies on holographic QCD~\cite{Erlich:2005qh} (and also very recently in similar composite Higgs models holography analysis~\cite{Erdmenger:2020lvq}) extract $g_5$ by matching the large-$q^2$ behaviour of the (axial-)vector current two-point function computed holographically to the perturbative field theory result. In our models, this would give $g_5=\sqrt{12}\pi \simeq 10.9$. However, while this value gives some indication, it requires trusting the perturbative field theory result in the strongly-coupled regime for which the holographic models are applicable.
In order to estimate the value of $g_5$, we rather extract masses from Figs.~\ref{fig:Model1_2} and~\ref{fig:Model2AB_3} in a reasonable range $7 \lesssim g_5\lesssim 9$. For the first low-lying resonances, this gives approximately
\beq
\frac{m_S}{\tilde f} : \frac{m_A}{\tilde f} : \frac{m_V}{\tilde f} \simeq
  (11.5-15.1) : (11.5-15.3) : (6.5-7.7)
\label{fig17}
\eeq
for Models I, IIA, and IIB, in units of $\tilde f \equiv F_G \sqrt{2/N_C} $. 
Note that the numbers in Eqs.~\eqref{fig17} can be directly compared 
with the $Sp(4)$ lattice results in Eqs.~\eqref{sp4_fund} and
\eqref{sp4dyn}, since for $N_C=2$, $\tilde f\equiv F_G$. 
Within both lattice and holographic uncertainty ranges as above quoted, the previous numbers appears 
in reasonably good agreement
with lattice results from \cite{latt_sp4,latt_sp4_dyn,latt_su4} in Eqs.~(\ref{sp4_fund}), (\ref{sp4dyn}), (\ref{su4_fund}). In Fig.~\ref{fig:CompareToLattice}, we show the spectrum in all three holographic models (I, IIA, and IIB) as a function of $x_F$, having fixed $g_5 = 8$. We also include in this figure the lattice results from Eqs.~\eqref{sp4_fund} and~\eqref{sp4dyn}. As can be seen, the spectra obtained from holography, while predicting a slightly lower value of $m_V$, agree well with the unquenched lattice results of Ref.~\cite{latt_sp4_dyn}. We also observe that, when restricting ourselves to the (axial-)vector and scalar $S$ resonances, the predictions of all three holographic models are very similar, while their differences manifest primarily in the singlet-scalar and pseudoscalar sectors.

Let us make a somewhat digressive comment regarding the value of bulk gauge coupling $g_5$. In a top-down model, the action of the gravity theory is constrained to take a particular form. In contrast, since the holographic models of section~\ref{sec:models} were built from the bottom-up, various simplifying assumptions regarding their actions were made. In particular, if we were to allow for the rescaling of the action by an overall factor, this would not affect the location of poles of two-point functions, and hence the spectrum, but it would rescale their residues. Hence, in units of the decay constant $\tilde f$, the entire spectrum would be rescaled. With this extra freedom, it is possible to get good agreement with lattice results also for smaller values of $g_5 \simeq 5$.

We close our comparison with lattice  by considering very recent
results~\cite{Bennett:2020hqd,Bennett:2020qtj} for the scalar and tensor glueball masses in the $Sp(2N_C)$ model,
for different $N_C$ values, where extrapolation to large $N_C$ is also provided. 
The glueball masses in \cite{Bennett:2020hqd,Bennett:2020qtj} are obtained
in units of the string tension of the pure Yang-Mills theory. In order to compare with 
our holographic results, we rather quote the lattice result for the ratio of glueball masses. 
For the lightest scalar ($A^+_1$) and spin-2 glueball ($E^+$),\footnote{Strictly speaking, $A^+_1$ and $E^+$
correspond to irreducible representations of the lattice {\it octahedral} group, but in the continuum limit
their masses correspond respectively to  
a spin-0 and spin-2 representation of the Poincar\'e group.} the lattice finds
\beq
 \frac{m_{A^+_1}}{m_{E^+}}  
= 0.711\pm 0.021 ~~ (0.678\pm 0.032) ~,
\label{sp4_glueball}
\eeq
where the first (second) value corresponds to $N_C=2$ (the large $N_C$ extrapolation)~\cite{Bennett:2020qtj}. Note, however, that these are results for pure Yang-Mills glueballs, 
thus far from a possible conformal window. Accordingly, the explicit breaking of scale invariance is large, and one does not expect a light dilaton to be present in the spectrum. Our results reproduce Eq.~\eqref{sp4_glueball} to the extent that i) our models match the GPPZ model~\cite{Girardello:1999bd} for some specific choices of the parameters, as explained below Eqs.~\eqref{eq:Model1solutions} and~\eqref{eq:Model2Bsolutions}, and ii) it has been observed~\cite{Bennett:2020hqd} that the lattice results agree well with GPPZ, which predicts the ratio corresponding to Eq.~\eqref{sp4_glueball} to be $1/\sqrt{2} \simeq 0.71$~\cite{Mueck:2004qg}.

\subsection{Nambu--Jona-Lasinio model}

The NJL model is based on four-fermion 
interactions, that provide an effective low-energy  
approximation of the underlying strongly-coupled gauge dynamics.
Restricting ourselves to the HC fermions $\psi^a$, with coset $SU(2N_F)/Sp(2N_F)$,
the NJL Lagrangian reads
\be
{\cal L}_{NJL} = 
\frac{\kappa_A}{2N_C}(\psi^a \psi^b) (\bar\psi_a \bar\psi_b)+ \dots~,
\label{NJLbasic}
\ee
where brackets indicate HC and Lorentz singlets, and the dimensionful coupling $\kappa_A$ is $N_C$-independent in the large $N_C$ limit. 
The dots designate other possible fermion effective interactions invariant under $SU(2N_F)$ \cite{Barnard:2013zea,Bizot:2016zyu},
whose role is briefly discussed below. 
Eq.~(\ref{NJLbasic}) is sufficient to capture 
the main features of the $SU(2N_F)\to Sp(2N_F)$ SSB, 
that occurs above a certain critical value for $\kappa_A$, 
resulting in a mass gap and massless Goldstone bosons. 
The meson masses  and  decay constants are 
obtained respectively as the poles and residues of large-$N_C$ resummed correlators in the 
appropriate meson channels. 
For vector mesons, the relevant four-fermion interactions can be obtained from \eq{NJLbasic}
by a Fierz transformation, an approximation justified in the large-$N_C$ limit
(the analogous assumption gives good predictions in the QCD case).

The NJL model being a large-$N_C$ approximation, valid for large $N_F$ as well, 
it is appropriate to compare its predictions with our holographic approach in the Veneziano limit,
while lattice simulations are typically limited to small $N_C$ and $N_F$.
Among the NJL limitations, however, one should remark
that the model does not immediately incorporate dynamical effects from large anomalous dimensions, 
and it does not describe specific properties of gauge theories close to the conformal window,
at least in the simplest NJL realisation.\footnote{For an extended ``gauged-NJL'' 
framework, possibly addressing large anomalous dimensions and near-conformal dynamics, mostly in the
context of technicolour models, see e.g. \cite{Yamawaki:1996vr,Miransky:1996pd}.}

One can define the NJL mass gap by $x\equiv M^2_\psi/\Lambda^2$, where $M_\psi$ is the dynamical fermion mass induced by the strong dynamics,
and $\Lambda$ is the UV cutoff of the four-fermion interactions.
The mass gap is determined by the equation
\begin{equation}
 1-x \ln\left(1+\frac{1}{x}\right) = \frac{1}{\xi} ~,\qquad \xi\equiv  \frac{N_F}{2} \frac{(\kappa_A+\kappa_B) \Lambda^2}{4\pi^2}  ~,
\label{mgap2}
\end{equation}
where $\kappa_B$ parametrises the potential contribution of other operators in \eq{NJLbasic}.
The effective, dimensionless coupling $\xi$ should lie in the range $1\le \xi \lesssim 3$,
where the lower bound comes from the requirement of SSB with a non-vanishing mass gap, while the upper bound
follows from the condition $M_\psi \lesssim \Lambda$,
as the NJL predictions are no longer reliable for $M_\psi \sim \Lambda$.
One can relate $\xi$ to the underlying HC gauge coupling $g_{HC}$, by assuming that \eq{NJLbasic} is generated, through a $Sp(2N_C)$ Fierz identity, by the current-current
operator induced by single-gluon exchange.  
In this approximation, and neglecting $\kappa_B$, one finds \cite{Bizot:2016zyu}
\be
\xi \simeq \frac{N_FN_Cg_{HC}^2}{8\pi^2} = \frac{N_F\,\lambda}{8\pi^2} ~.
\label{xitolambda}
\ee
According to this rough estimate, $N_F$ should not be taken too large to trust the NJL results, say $N_F^{max} \simeq  16\pi^2/\lambda$.
This bound can be easily satisfied for $N_F\lesssim 10$, even for a large 't Hooft coupling $\lambda\sim 4\pi$, as required
in order for our holographic models to be applicable.

Once a value for $\xi$ is chosen, the only other independent parameter that determines the NJL meson spectrum is $\kappa_B/\kappa_A$.
It turns out \cite{Bizot:2016zyu} that
$\kappa_B$ effectively parametrizes the axial $U(1)_A$ breaking by the HC anomaly,
thus providing a mass to the HC meson $\eta^\prime$, in analogy with QCD.
This effect can be described in the IR by a four-fermion operator when $N_F=2$, and by a larger operator when $N_F>2$ \cite{Bizot:2016zyu}.
The bottom line is that $\kappa_B/\kappa_A$ should be sizeable, 
if one wants the $\eta'$ mass,  $m^2_{\eta^\prime}\sim (\kappa_B/\kappa_A) N_F/N_C$, to be of the same order 
as the other (non-Goldstone) meson masses.
As already discussed at the beginning of section \ref{sec:models}, in our holographic framework we did
not study the $\eta'$ sector.
However, $\kappa_B/\kappa_A$ has some influence on the other meson masses, in particular it breaks the degeneracy between
the scalar non-singlet $S$ and the scalar singlet $\sigma$.

\begin{figure}[!]
    \centering
    \hspace{-0.5cm}
    \includegraphics[width=0.48\textwidth]{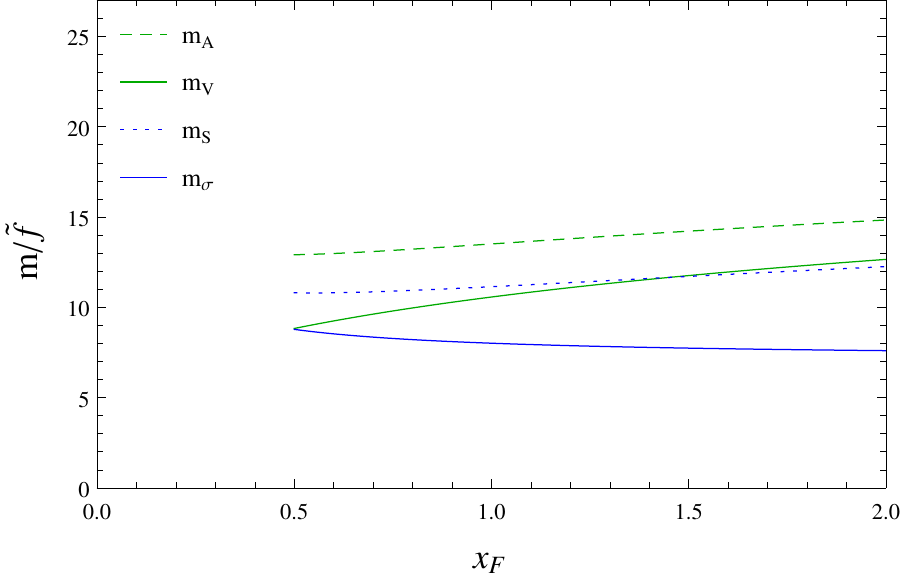}\hspace{0.2cm}
    \includegraphics[width=0.48\textwidth]{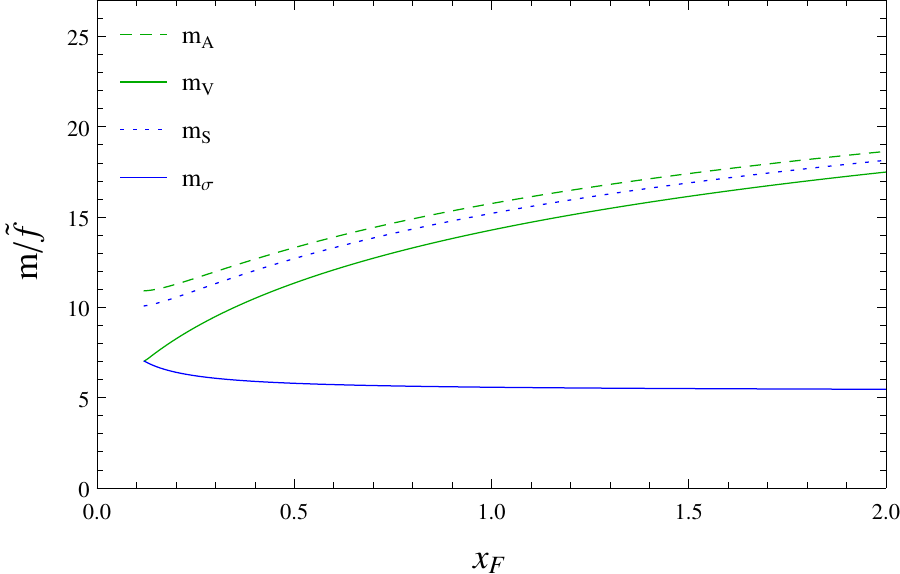}
      \caption{NJL mass spectrum $m_i/\tilde f$ as a function of $x_F=N_F/N_C$, for  $\xi=1.3$ and $\kappa_B/\kappa_A=0.5$ (left panel), as well as for $\xi=1.1$ and $\kappa_B/\kappa_A=0.7$ (right panel).}
\label{NJLfig}    
\end{figure}

With the above summary of the NJL framework in mind, we can compare the NJL meson spectrum with our holography results. 
In the limit $\xi\to 1$,
$m_\sigma$ and  $F_G \propto M_\psi$ rapidly vanish, restoring the chiral symmetry \cite{Bizot:2016zyu}. 
Thus, the scalar singlet $\sigma$ is substantially lighter than the other (non-Goldstone) mesons when $\xi$ is close to one,
potentially matching holographic scenarios with a light dilaton. 
It is not easy to match all other meson masses to our holography results.
For example, choosing the NJL parameters $x_F\simeq 1$, $\xi \simeq 1.3$ and $\kappa_B/\kappa_A\simeq 0.5$, we find
\beq
\frac{m_A}{\tilde f} : \frac{m_S}{\tilde f} : \frac{m_V}{\tilde f} :  \frac{m_\sigma}{\tilde f} 
\simeq  
13.5 : 11.1 : 10.5 : 8.0 ~,
\label{njlmassn2}
\eeq
(with $\sim 10\%$ variations in $m_V, m_A$ in the window $\xi \sim 1.2-1.6$ and $\kappa_B/\kappa_A\sim 0.3-0.7$), 
where $\tilde f\equiv F_G \sqrt{2/N_C}$ is the same unit adopted in our results from holography.
Comparing with Fig.~\ref{fig:Model2AB_3}, one observes that the NJL masses are typically closer to each other 
with respect to the holography ones.
Varying $g_5$, one can roughly reproduce the NJL values for $A$ and $S$, however $V$ is typically lighter 
in our holographic models.

The behaviour of the NJL meson masses as a function of $x_F$ is illustrated in Fig. \ref{NJLfig},
for two representative values of $\xi$ and 
$\kappa_B/\kappa_A$.\footnote{In Fig.~\ref{NJLfig} the curves start at a value $x_F^{min}$, chosen in order for  the full NJL  calculation
(involving form factors and pole-mass dependences) to remain reliable \cite{Bizot:2016zyu}. }
The increase of the NJL masses with $x_F$ is simple to understand: neglecting subleading form-factor
dependences \cite{Bizot:2016zyu}, the meson squared masses receive
a contribution $\propto 1/\kappa_{A} +{\cal O}(M_\psi^2) $,  
except for $m^2_\sigma \simeq 4 M_\psi^2$. 
Thus, for a fixed $\xi$,
the masses increase with $N_F$
while $m_\sigma$ remains constant.
This behaviour is partly damped by the normalisation to
$\tilde f$, which also moderately increases with $N_F$ due to form-factor effects.
One finds $m_A > m_S$, except
for very large $\kappa_B/\kappa_A \gtrsim 0.8$, that may be problematic, as some of the NJL masses are singular for $\kappa_B/\kappa_A\to 1$ \cite{Bizot:2016zyu}. 
Note that in holography $m_A >m_S$ can be
obtained for sufficiently large $g_5$, see Fig.~\ref{fig:Model2AB_3}, since $m_A$ grows faster than $m_S$ with $g_5$.
The growth at large $x_F$, not accessible in quenched lattice simulations,
is not present in holography, where 
the meson masses (except for the singlet scalar) are almost $x_F$ independent,
see Fig.~\ref{fig:Model2AB_2}. 
This is hardly surprising, as 
the set of parameters which has been fixed, as $x_F$ varies, is not necessarily equivalent in the NJL and holographic approaches.

Finally let us compare the NJL prediction for  $m_\sigma$, in Eq.~(\ref{njlmassn2}) and in 
Fig.~\ref{NJLfig}, with the relatively light dilaton obtained in holography. According to 
Fig.~\ref{fig:Model2AB_3},  for $x_F\simeq 1$ and $g_5\simeq 8$ the lightest singlet scalar has mass 
\be
\frac{m_{\rm dilaton}}{\tilde f} \simeq (2.5 -5)~, 
\ee
depending on whether Model IIA or IIB is considered.
 As explained above, 
in the NJL model one can lower $m_\sigma$ 
by taking $\xi\to 1$, as apparent from the comparison of the two panels in  Fig.~\ref{NJLfig}: agreement with Model IIB
can be reached at $\xi= 1.1$.
Note that, for $\xi \lesssim 1.1$ (depending also on $\kappa_B/\kappa_A$ and $x_F$) 
the complete NJL calculation (accounting for form factors and pole-mass dependences) 
is no longer reliable for $m_V$ and $m_A$, as the latter develop a large
unphysical imaginary part \cite{Bizot:2016zyu}.
Besides, the NJL model behavior very close to $\xi=1$ is, in any case, conceptually
problematic, calling for extensions of the simplest NJL framework to evade  
triviality \cite{Yamawaki:1996vr} \cite{Barnard:2013zea}.
We conservatively conclude that, within the parameter range
where the NJL approximation is more reliable, it is difficult to correctly describe 
a near-conformal regime.

\section{Conclusions}
\label{sec:Conclusions}

We showed that 
strongly-coupled gauge theories with a large number of flavours provide several attractive features to UV-complete composite Higgs models. 
They allow for protection of the SM accidental symmetries, while keeping the evolution of the gauge couplings under control.
Moreover, gauge theories in which the number of flavours and colours are of the same order may sit close to the conformal window, leading to walking dynamics that provides a mechanism to generate the SM Yukawa couplings while suppressing flavour violation.

The Veneziano limit, in which the ratio $x_F = N_F/N_C$ is kept fixed while $N_C \rightarrow \infty$, allows making use of large-$N_C$ arguments, while going beyond the quenched approximation. In holographic models, large $N_C$ ensures that the gravity theory is weakly coupled, while $x_F \sim 1$ implies that the flavour sector backreacts on the geometry.
Within the bottom-up approach to holography, we presented models in which the flavour symmetry of the dual field theory is broken due to a bulk scalar acquiring a non-trivial profile along the extra radial dimension. At the same time, the dynamics of this scalar field generates an end-of-space for the geometry in the IR, leading to a mass gap in the dual field theory. This implies that the decay constant, associated with the spontaneous breaking of the flavour symmetry, becomes dynamically related to the mass gap.

In Model I, consisting of gravity coupled to a single scalar field charged under an $SU(2N_F)$ symmetry, together with the $SU(2N_F)$ gauge field, we choose a scalar potential such that the backreaction on the geometry grows with $x_F$. The resulting solution of the equations of motion
interpolates between an AdS background for $x_F \rightarrow 0$, and geometries with a significant backreaction on the metric for $x_F \sim 1$. An additional parameter $\Delta$ controls the dimension of the operator $\mathcal O_\sigma$ responsible for the breaking of the flavour symmetry,  $[\mathcal O_\sigma] = 2 + |\Delta - 2|$. We chose a special form of the bulk scalar profile, such that for $\Delta > 2$ the breaking is purely spontaneous, while for $\Delta < 2$ explicit breaking is present.

We computed the spectrum of scalar, pseudoscalar, vector, axial-vector, and tensor resonances, as a function of $x_F$ and $\Delta$. In addition to states of the order of the mass gap and heavier, for $\Delta > 2$ the spectrum contains massless NGBs, associated with the flavour SSB, as well as a massless dilaton, present due to the spontaneous breaking of scale invariance. For $\Delta < 2$, the explicit breaking causes both the dilaton and the Goldstone bosons to acquire masses. We found that the dilaton is typically lighter than the NGBs. In addition, we found that the spectrum has a rather mild dependence on the number of flavours.

In Model II, we introduced an additional scalar in the bulk, as a way to capture a more realistic, non-conformal dynamics on the field theory side. We chose this scalar field to be a singlet under the flavour symmetry, allowing for the possibility of breaking scale invariance explicitly while breaking flavour symmetry spontaneously. We proposed two variations, Model IIA and IIB, corresponding to a different choice of the scalar potential. The resulting spectra of (pseudo-) scalar, (axial-)vector, and tensor resonances show similar features in the two models, testifying to the genericness of our predictions. Thanks to the independent source of scale-invariance breaking, it is possible to lift the mass of the dilaton, while keeping massless Goldstone bosons. Still, the dilaton may become light when the operator dual to the singlet scalar, $[O_\phi]$, is close to marginal.

Even more interestingly, we found that, as the number of flavours is increased, it becomes progressively more difficult to lift the mass of the dilaton. In our models the IR end-of-space, and hence the mass gap, is generated by the scalar associated with the breaking of the flavour symmetry.
In such setting the maximum possible mass of the dilaton (as a function of the remaining parameters of the model) is suppressed as $x_F$ grows:
already for $x_F \sim 1.5$ the dilaton is much lighter than the mass gap.
We interpret this effect 
to be due to the enhancement of the flavour condensate, responsible for SSB of scale invariance and flavour symmetry, which scales as $N_F N_C \sim x_F N_C^2$ in the Veneziano limit. 
The question of whether this mechanism for a light dilaton can be accomplished in a strongly-coupled field theory is subtle.
The lower edge of the conformal window may well lie at large $x_F$, however it is far from obvious that one can approach this edge
keeping the SSB of scale invariance parametrically larger than its explicit breaking. While our bottom-up approach to holography can well describe such a hierarchy of scales, a convincing proof of existence requires to rigorously study top-down models obtained from supergravity, for which a field-theory dual is known to exist. We leave these interesting questions for future studies.

We attempted a comparison of our holographic spectra with those extracted from lattice simulations of similar gauge theories. It is remarkable that---despite several, complementary limitations of the two approaches---the spectra of the first resonances are qualitatively similar. We also compared with the NJL model, suitable to describe non-perturbative flavour SSB at large $N_C$, finding some useful correspondence between the holography and NJL parameters.

As a concluding remark, we note that the large number of flavours in the Veneziano limit need not all be treated equally. In general, one could conceive of splitting the flavours into different subsets, e.g. adding an explicit mass only for some of the hyper-fermions, implying that large $Sp(2N_F)$ multiplets are divided into smaller multiplets of composite states. In this case, the shape of the background solutions in the dual gravity theory would be affected, to account for a flavour-dependent symmetry breaking. The Higgs would be accompanied by fewer light composite states. Besides, the light flavours would not be subject to the issue of the Veneziano limit, where 5D loops become non-perturbative because of the large flavour multiplets. We leave the study of such generalised scenarios for the future.

\section*{Acknowledgements}
We thank Kaustubh Agashe, William A. Bardeen, David Lin, Maurizio Piai, Alex Pomarol, and Alberto Zaffaroni for useful discussions.
The project leading to this publication has received funding from the Excellence Initiative of Aix-Marseille University - A*MIDEX, a French ``Investissement d'Avenir" programme (ANR-11-LABX-0060 - OCEVU and AMX-19-IET-008 - IPhU). This project has received support from the European Union's Horizon 2020 research and innovation programme under the Marie Sk\l odowska-Curie grant agreement No 860881-HIDDeN.

\appendix

\section{How to preserve baryon number}\label{BL}

As baryon number is extremely well conserved, the HC sector  should not induce dangerous baryon-number-violating operators suppressed by the IR scale $m_*$, as we wish to keep $m_*$ close to the electroweak scale. Some hyper-fermions need to carry baryon number, in order to form operators mixing linearly with the quarks, especially the top. Thus, both the SM and the HC sector have to transform non-trivially under  $U(1)_B$.
In previous HC models for composite Higgs, the baryon number is necessarily present, either implicitly  
(see e.g. \cite{Barnard:2013zea,Ferretti:2013kya})
or explicitly  (see e.g. \cite{Gertov:2019yqo}). 
In this section we present a general discussion of the possible ways to implement this symmetry, as well as the specific implementation
that we adopt. At the end, we will comment on generalisations including lepton number as well.

The HC fermion kinetic terms have a global symmetry group $G_F\times U(1)_A$, where $G_F$ is given in \eq{GF} and $U(1)_A$ is the
independent linear combination of fermion numbers, anomalous with respect to hypercolour.
In principle, there are three qualitatively different possibilities to embed baryon number within this group: $U(1)_B$ could be identified with 
$U(1)_A$, or with a generator in the coset $G_F/H_F$, or with a generator in $H_F$, where the latter is defined as the subgroup that cannot undergo SSB. Let us discuss these three options in turn.

The $U(1)_A - G_{HC} - G_{HC}$ anomaly implies that $U(1)_A$ is broken non-perturbatively. One might hope that baryon number violation could be exponentially suppressed, roughly proportionally to $\exp(-8\pi^2/g_{HC}^2)=\exp(-8\pi^2N_C/\lambda)$. 
This estimate, based
on the instanton dilute-gas approximation, may provide a strong suppression in the large-$N_C$ limit,
even if the 't Hooft coupling $\lambda$ is large, but fixed. However, experience from QCD indicates that $U(1)_A$ breaking is actually stronger,
as the $\eta'$ mass is not exponentially suppressed in the large-$N_c$ limit, rather $m_{\eta'}^2\sim 1/N_c$.
As we consider HC in the strongly-coupled regime, given the tight constraints on baryon number violation, the possibility $U(1)_B=U(1)_A$ must be discarded.

If $U(1)_B$ is embedded in the coset $G_F/H_F$, 
it can be spontaneously broken by the VEV of any Lorentz- and hypercolour-invariant operator ${\cal O}$ which carries a non-zero baryon number.
For any factor group $G_F^i\subset G_F$, the subgroup $H_F^i$ defined in \eq{RPC} is maximal. As a consequence, either $G_F^i$ is entirely preserved, or all generators in 
$G_F^i/H_F^i$ undergo spontaneous breaking \cite{Vafa:1983tf}. 
Therefore, baryon number conservation requires to preserve each $G_F^i$ that has 
an intersection with $U(1)_B$,
otherwise spontaneous $U(1)_B$ breaking at scale $m_*$ would typically induce too large baryon-number violating operators.
In principle, one could conceive that the HC dynamics sets to zero the VEV of all operators with $B\ne 0$. 
However this is a very non-trivial assumption, that is hard to justify. 
In addition, SSB may even be unavoidable, to match the $G_F$ global anomalies \cite{tHooft:1979rat}: 
the only other possibility to match the UV anomalies
requires massless composite fermions in the IR, an option non-trivial to realise in practice \cite{Bizot:2016zyu}.

Thus, we are left with the unique option to embed $U(1)_B$ in the vector subgroup $H_F$, as anticipated in \eq{HFparent}.
In this case baryon number is protected from SSB and anomaly breaking. Still, one should worry about other potential sources of explicit breaking of baryon number. 

Firstly, consider a non-zero hyper-fermion mass matrix, $M$, with eigenvalues $m_1,\dots,m_{N_F}$. 
If these eigenvalues are not all equal, $H_F$ is explicitly broken to a smaller subgroup $H_M$.
When $m_1,\dots,m_{N_F}$ are all different, the three types of flavour symmetries listed in \eq{RPC} break
according to $SO(N_F)\to Z_2^{N_F}$ (real), $Sp(2N_F)\to Sp(2)^{N_F}$ (pseudoreal), and $SU(N_F)\times U(1) \to U(1)^{N_F}$ (complex).
For a given embedding of $U(1)_B$ within $H_F$, one should restrict fermion masses to preserve it.
In the pseudoreal and complex cases, this is not a significant restriction, since one can always consider the basis where each fermion flavour has a definite $B$-charge, and then choose $M$ to be diagonal and $B$-preserving in this basis.

Secondly, consider the interactions between the HC sector and the SM.
The (linear) mixing of composite operators with the SM fermions may generically violate $U(1)_B$. 
Therefore, one is forced to forbid (or sufficiently suppress) by hand all those couplings incompatible with the chosen assignment of $B$-charges.
Concerning the mixing with the SM gauge bosons, since the SM gauge group and $U(1)_B$ are separately embedded into $H_F$, the weak gauging of the SM does not break baryon number perturbatively. The electroweak anomaly of $U(1)_B$ 
represents a negligible, exponentially-suppressed violation of baryon number, as in the SM.

We are now in the position to provide an explicit embedding of $U(1)_B$ in models with $H_F=Sp(2N_F)$.
The minimal number of flavours is $N_F=5$, with
\be
Sp(10)\supset SU(3)_c \times SU(2)_L \times SU(2)_R \times U(1)_B~, \qquad Y=\pm T^R_3+\dfrac B2 ~.
\label{app3221}\ee
The embedding of the $SU_{3221}$ subgroup within $Sp(10)$ is unique, and we indicated the two inequivalent ways to embed hypercharge.
The 10 Weyl fermions $\psi^a$ in the fundamental of HC transform as
\be 
10_{SU(10)}=10_{Sp(10)}=\left[(3,1,1)_{1/3}+(\bar 3, 1,1)_{-1/3}+(1,2,1)_0+(1,1,2)_0\right]_{SU_{3221}}~.
\label{10Sp}\ee
In general, we are interested in fermion-bilinear operators with flavour structure $(\psi^a\psi^b)$ or $(\overline\psi_a \psi^b)$, as well as fermion-trilinear operators with flavour structure
$(\psi^a\psi^b\chi)$ or $(\overline\psi_a \psi^b \chi)$. Upper and lower $SU(2N_F)$ indexes are indistinguishable from the perspective of the $Sp(2N_F)$ subgroup, in particular $\overline\psi_a$ transform in the same way as $\psi^a$, since $\overline{10}_{SU(10)}=10_{Sp(10)}$.
Therefore, all relevant operators transform as 
\be
(10\times 10)_{Sp(10)} = (1_A + 44_A + 55_S)_{Sp(10)}~.
\ee
Note that each specific operator is either symmetric ($55_S$) or antisymmetric ($1_A+44_A$) in the flavour indexes $a,b$, depending on the symmetries of the Lorentz and HC contractions among the anti-commuting spinors, see appendix \ref{zoo}.
The SM charges can be read from the decomposition under the subgroup of \eq{app3221}, which reads
\be\ba{rcl}
(1_A)_{Sp(10)} &=& \left[(1,1,1)_0\right]_{SU_{3221}}~,\\
(44_A)_{Sp(10)} &=& \left[2\times (1,1,1)_0 + (1,2,2)_0 + (8,1,1)_0 + (3,1,1)_{-2/3} + (\overline{3},1,1)_{2/3} \right.\\
&&+\left.(3,2,1)_{1/3} + (\overline{3},2,1)_{-1/3}+(3,1,2)_{1/3}+(\overline{3},1,2)_{-1/3}\right]_{SU_{3221}}~,\\
(55_S)_{Sp(10)} &=& \left[(1,1,1)_0 + (1,2,2)_0 + (1,1,3)_0 + (1,3,1)_0 \right.\\
&&+\left.(8,1,1)_0 + (6,1,1)_{2/3} + (\overline{6},1,1)_{-2/3} + (3,2,1)_{1/3} \right.\\
&&+\left.(\overline{3},2,1)_{-1/3}+(3,1,2)_{1/3}+(\overline{3},1,2)_{-1/3}\right]_{SU_{3221}}~.
\ea
\label{decompo}\ee
The Higgs doublet is identified with the antisymmetric $(1,2,2)_0$ component, as it is embedded in the operator $(\psi^a\psi^b)=-(\psi^b\psi^a)$.
The embedding of top and bottom quark multiplets were given in \eq{topP}.
Depending on the sign in the definition of hypercharge in \eq{app3221}, the top and bottom components are embedded in a flipped way into the $SU(2)_L\times SU(2)_R$ doublets.

By inspection of \eq{decompo}, one observes that  $(10\times 10)_{Sp(10)}$ also contains components with the charges of the SM lepton multiplets. However, these potential lepton partners do not preserve lepton number, that is, they interact with the other composite states in a generic way.
The simple reason is that there is no room, within $Sp(10)$, for an additional conserved $U(1)_L$ factor. To avoid lepton number violation, one has to forbid any SM operator carrying non-zero lepton number to couple to the HC sector, that is, one assumes the HC  sector to be neutral under $U(1)_L$.

In the absence of composite states carrying lepton number, one cannot implement partial compositeness for leptons. However, the latter would be useful to explain the hierarchy (and the size) of the charged lepton Yukawa couplings, as well as the suppression of lepton flavour and CP violation,
see \cite{Frigerio:2018uwx} for a recent analysis.
To implement lepton partial compositeness, the rank of $Sp(2N_F)$ should be increased to include $U(1)_L$.
This generalisation is straightforward for e.g. $N_F=6$, but we do not need to display it explicitly here.
Finally, the UV origin of (tiny) baryon and lepton number violations and their interplay are important, of course, for the analysis of e.g. Majorana neutrino masses and proton decay channels. These developments go beyond our present purposes.

\section{The hypercolour and colour $\beta$-functions}\label{beta}

Let us fix our conventions for the Renormalisation Group Equation (RGE) of a gauge coupling $g$. We define the $\beta$-function by
\be
\frac{d\alpha}{d\ln\mu}= \beta(\alpha)~,\qquad \beta(\alpha)=-\frac{\alpha^2}{2\pi}\left(b_0 + \frac{\alpha}{4\pi} b_1 + \dots\right)~,
\ee
where $\alpha\equiv g^2/(4\pi)$ and $\mu$ is the renormalisation scale. 
For a gauge theory of fermions, the first two coefficients of the $\beta$-function are given by
\be\ba{l}
\displaystyle b_0=\dfrac{11}{3}C(R_A) -\dfrac{2}{3} \sum_{f} T(R_f)~,\\
\displaystyle b_1=\dfrac{34}{3}C(R_A)^2 - 2 \sum_f C(R_f)T(R_f) -\dfrac{10}{3} C(R_A)\sum_f T(R_f)~,
\ea\ee
where $R_A$ is the adjoint representation, the sums run over all Weyl fermions in various representations $R_f$, $C(R)$ is the quadratic Casimir
and $T(R)$ the Dynkin index, normalised to $1/2$ for the fundamental.

The condition for asymptotic freedom is $b_0>0$, such that $\alpha$ and $\beta(\alpha)$ vanish in the UV. We are interested in theories with an approximate IR fixed point, where $\beta(\alpha)$ goes back to zero for some finite value $\alpha=\bar\alpha$.
We are also interested in the limit of large number of colours, $N_C\gg 1$. In this limit each additional loop is proportional to 
$\alpha N_C/(4\pi) \equiv \lambda/(16\pi^2)$, where $\lambda$ is the 't Hooft coupling, so that the $\beta$-function coefficients 
scale as $b_i \sim N_C^{i+1}$. A perturbative IR fixed point requires
\be 
-\frac{b_0 N_C}{b_1} \simeq \frac{\bar\alpha N_C}{4\pi} \ll 1~.
\ee
This occurs for $b_1$ negative and $b_0/b_1$ sufficiently small. As $b_0/b_1$ increases, one enters the non-perturbative regime, and the existence of the fixed point becomes speculative. The dual description of the gauge theory in terms of classical gravity is expected to be more reliable in 
the regime $N_C\gg 1$ (small quantum gravity corrections) and $\lambda\gg 1$ (small string corrections). 
For orientation, we will take $b_1\sim - 4\pi N_C b_0$ as a suggestive value for the existence of an IR fixed-point at $\bar\lambda \sim 4\pi$.

For definiteness, let us concentrate on a $Sp(2N_C)$ gauge theory, where the tensor product of two fundamentals reads
\begin{equation}
\Yvcentermath1
{\small\yng(1)}\times {\small \yng(1)}= \left.\bullet\right._A ~+~ \left.{\small\yng(2)}\right._S ~+~ \left.{\small\yng(1,1)}\right._A ~.
\label{product-rep1}
\end{equation}
Here the bullet stands for a singlet and the subscripts indicate the (anti)symmetry under the exchange of the two factors.
The two-index symmetric representation is actually the adjoint of the group $Sp(2N)$. 
Some properties of the relevant representations are collected in table \ref{DDC}
(see e.g.~\cite{Feger:2012bs,Yamatsu:2015npn}).

\begin{table}[t]
\begin{center}\begin{tabular}{|c|c|c|c|}
\hline
$Sp(2N)$ & ${\small \yng(1)}$ & ${\small\yng(2)}$ & $\Yvcentermath1{\small\yng(1,1)}$ \\[.1cm]
\hline
$d(R)$ & $2N$ & $N(2N+1)$ & $2N^2-N-1$ \\
\hline
$T(R)$ & $\dfrac 12$ & $N+1$ & $N-1$ \\
\hline
$C(R)$ & $\dfrac{2N+1}{4}$ & $N+1$ & $N$ \\
\hline
\end{tabular}\end{center}
\caption{The dimension, Dynkin and Casimir of the smallest $Sp(2N)$ representations.}
\label{DDC}\end{table}

If the gauge theory contains the Weyl fermions $\psi^a\sim{\tiny\yng(1)}$ for $a=1,\dots,2N_F$,  and 
$\Yvcentermath1\chi^b\sim{\tiny\yng(1,1)}$ for $b=1,\dots,n_F$, one obtains
\be\ba{l}
\displaystyle b_0=\dfrac{1}{3}(11-2n_F)N_C -\dfrac{2}{3}N_F + \frac{1}{3} (11+2n_F)~,\\
\displaystyle b_1=\dfrac{2}{3}(17-8n_F)N_C^2 -\frac{13}{3} N_F N_C + \frac{2}{3} (34+3n_F)N_C -\frac{23}{6}N_F + \frac 23 (17+5n_F)~.
\ea\ee
Note that asymptotic freedom sets an upper bound on the number of fermion flavours, $2[N_F+(N_C-1)n_F]< 11(N_C+1)$. 
On the other hand, one expects a lower bound on the number of flavours to generate an IR fixed point, and enter the so-called conformal window.
Perturbatively, the criterion would be $b_1<0$, however the lower boundary of the conformal window may well correspond to a strongly non-perturbative regime for the gauge coupling. 

Let us study the large $N_C$ limit, keeping $x_F\equiv N_F/N_C$ finite while $n_F/N_C \ll 1$. 
In the case $n_F=1$, 
the two-loops estimate for the IR fixed point reads
\be
\dfrac{\bar\lambda(n_F=1)}{16\pi^2} \simeq \dfrac{9-2x_F}{13x_F-18} ~,\qquad  {\rm e.g.}~x_F = 2.5 ~~\Rightarrow~~ \frac{\bar\lambda(n_F=1)}{4\pi} \simeq 3.4~.
\label{IRFP}\ee
Note that we
neglected subleading terms in $1/N_C$, even though they may be
quantitatively important for $N_C= {\cal O}(10)$.
From this perturbative estimate,  one could infer that $x_F$ of order one (a few) is preferable to realise a strongly-coupled IRFP.

Taking an increasing $n_F=2,\dots,5$, a given $\bar\lambda$ is obtained for a decreasing value of $x_F$.  For $n_F\ge 6$, the limit $N_C\to \infty$ becomes incompatible with asymptotic freedom.
In the case $n_F=6$ \cite{Barnard:2013zea,Bizot:2016zyu}, asymptotic freedom requires $N_C < 23 - 2N_F$. The perturbative estimate for the IR fixed point reads 
\be
\dfrac{\bar\lambda(n_F=6)}{16\pi^2} \simeq \dfrac{23-N_C-2N_F}{62 N_C + 13 N_F-104} ~,\qquad  {\rm e.g.}~N_C=4,~x_F=0.5 ~~\Rightarrow~~ \frac{\bar\lambda(n_F=6)}{4\pi} \simeq 1.1  ~.
\ee
Note that only moderate values of $N_C$ are compatible with a large 't Hooft coupling, and gauge-gravity duality is expected to provide less
accurate results for small $N_C$. In other words, HC  models with $n_F\ge 6$ appear disfavoured to realise a strongly-coupled, walking regime.

If we modify the theory by replacing $\chi^b$ with $\Yvcentermath1\chi'^b\sim{\small\yng(2)}$ for $b=1,\dots,n_F$, then $3b_0=(11-2n_F)(N_C+1)-2N_F$. Therefore, if
$n_F\ge 6$ asymptotic freedom is lost for any value of $N_C$. On the other hand, for $1\le n_F \le 5$ the behaviour is qualitatively the same as before, 
in particular \eq{IRFP} is unchanged.

Let us now consider the RGE of the SM gauge couplings $\alpha_{1,2,3}$,  where we define $\alpha_1\equiv(5/3)\alpha_{U(1)_Y}$, $\alpha_2\equiv\alpha_{SU(2)_L}$ and $\alpha_3\equiv\alpha_{SU(3)_c}$. 
As the HC  sector carries SM charges, it modifies the SM evolution
above the scale $m_*$. Too many additional SM-charged states may drive $\alpha_i(\mu)$ to a Landau pole at some scale $\Lambda_{LP}$ close above $m_*$, limiting the range of validity of the model.
 As the HC  sector is supposed to be in the strongly-coupled over a large walking region, $m_*<\mu<\Lambda_{UV}$,
the HC  contribution to the running should be extracted by summing over composite states. More precisely, one would need to evaluate the two-point correlators of SM gauge currents, integrated over squared momenta $p^2>m_*^2$.
Such correlators could be estimated e.g. via gauge-gravity duality, but this relies on a detailed knowledge of the composite spectrum over a large range of scales.
For a naive estimate, one can compute the contribution to the one-loop $\beta$-function generated by the constituent fermions $\psi^a$, and in the following we will take it as a rough approximation of the non-perturbative result. 

In the minimal model the HC  fermion content is  defined by \eq{10Sp}, which implies 
\be
b_0^{1,HC}=-\dfrac{8}{15}N_C~,\qquad b^{2,HC}_0=-\dfrac{2}{3}N_C~,\qquad b^{3,HC}_0=-\dfrac{4}{3}N_C~.
\ee
 At scale $m_*$, these coefficients must be added to the usual SM contributions,
$b_0^{1,SM}=-41/10$, $b_0^{2,SM}=19/6$ and $b_0^{3,SM}=7$.
From the UV perspective, one observes that the asymptotic freedom of $SU(2)_L$ ($SU(3)_c$) is preserved for $N_C<5$ ($N_C<6$), while for a larger number of hypercolours a Landau pole should be reached at some finite scale $\Lambda_{LP}$.
Taking the experimental values of $\alpha_i(m_Z)$, evolving them with $b_0^i=b_0^{i,SM}$ up to $m_* = 10$ TeV, and then using $b_0^i=b_0^{i,SM}+b_0^{i,HC}$, we find $1/\alpha_3(\Lambda_{LP})\to 0$ for $\Lambda_{LP}\simeq 7\cdot 10^6$ $(7\cdot 10^3)$ TeV if $N_C=10~(15)$, 
while $\alpha_{1,2}$ remain perturbative longer.
Thus, our scenario remains reliable up to scales much larger than $m_*$, even for  a moderately large number of hypercolours, $N_C\sim 10$. 

The situation would badly deteriorate if the hyperfermions transforming as $SU(3)_c$-triplet and antitriplet were six copies of $\chi^a$  rather than  $\psi^a$, because in this case $b_0^{3,SM}$ would be proportional to their HC  dimension, $(2N_C^2-N_C-1)$ rather than $(2N_C)$.
In this case asymptotic freedom is lost already for $N_C=3$, and the Landau pole in $\alpha_3(\mu)$ occurs already at $\Lambda_{LP}\simeq 20$ TeV if $N_C=10$.  
This strongly disfavours models with several flavours in HC  representations other than the fundamental.

\section{The inventory of composite bosonic operators}
\label{zoo}

We consider a HC gauge theory $Sp(2N_C)$, with $2N_F$ Weyl fermions $\psi$ in the fundamental representation,
and one additional  Weyl fermion, either $\chi$ in the two-index antisymmetric, traceless representation, or $\chi'$ in the two-index symmetric
representation.
Once the HC  theory confines, the constituent degrees of freedom, HC gauge bosons
and fermions,
are replaced by composite, HC-singlet states. They are associated to operators constructed out of the constituents, and they transform in given Lorentz and flavour representations.
The transformation properties of the constituents are collected in table \ref{consti}. Here we classify all possible bosonic operators, while fermionic ones are classified in \cite{part2}.

\begin{table}[tb]
\renewcommand{\arraystretch}{1.2}
\begin{center}
\begin{tabular}{|c|c|c|c|c|}
\hline
& Lorentz & $Sp(2N_C)$ & $SU(2N_F)$ & $U(1)$  \\
\hline \hline
$A^{\mu}_{ij}$ & $(1/2,1/2)^\mu$ & ${\Yvcentermath1 \tiny \yng(2)}_{\, ij}$ & $\bullet$ & $0$ \\ 
\hline \hline
 $\psi^{\alpha a}_i$ & $(1/2,0)^\alpha$ & ${\Yvcentermath1 \tiny \yng(1)}_{\, i}$ 
 & ${\Yvcentermath1 \tiny \yng(1)}^{\, a}$ & $q_\psi$ \\
\hline
 $\overline{\psi}^{\dot\alpha}_{ai} \equiv \psi^{\dagger\dot\alpha}_{aj} \Omega_{ji}$ & $(0,1/2)^{\dot\alpha}$ & 
 ${\Yvcentermath1 \tiny \yng(1)}_{\, i}$ &
$\overline{\Yvcentermath1 \tiny \yng(1)}_{\, a}$  & $-q_\psi$ \\
\hline\hline
 $\chi^\alpha_{ij}$ ~($\chi'^\alpha_{ij}$) & $(1/2,0)^\alpha$ & ${\Yvcentermath1 \tiny \yng(1,1)}_{\, ij}$ ~(${\Yvcentermath1 \tiny \yng(2)}_{\, ij}$)     
 & $\bullet$ & $q_\chi$ ~($q_{\chi'}$)\\
\hline
 $\overline{\chi}^{\dot\alpha}_{ij}\equiv \Omega_{ik}\chi^{\dag\dot\alpha}_{kl}\Omega_{lj}$ ~($\overline{\chi}'^\alpha_{ij}$) 
 & $(0,1/2)^{\dot\alpha}$ & ${\Yvcentermath1 \tiny \yng(1,1)}_{\, ij}$ ~(${\Yvcentermath1 \tiny \yng(2)}_{\, ij}$)     
 & $\bullet$ & $-q_\chi$ ~($-q_{\chi'}$)\\
\hline
\end{tabular}\end{center}
\caption{The transformation properties of the HC  gauge bosons $A$ and fermions $\psi$ and $\chi$ ($\chi'$).
The $Sp(2N_C)$ indexes are denoted by $i,j,\dots$,  and $\Omega_{ij}$ is the antisymmetric invariant tensor of $Sp(2N_C)$.
Lorentz vector and spinor indexes are denoted by $\mu,\nu,\dots$, and $\alpha,\beta\dots$, $\dot\alpha,\dot\beta,\dots$, respectively. 
The flavour $SU(2N_F)$ indexes are denoted by $a,b,\dots$. 
The bullet stands for the singlet representation.
The ratio $q_\psi/q_\chi$ ($q_\psi/q_{\chi'}$) is fixed by requiring the 
$U(1)-Sp(2N_C)-Sp(2N_C)$ anomaly to vanish.}
\label{consti}
\end{table}

\subsection{Gluon bilinears}\label{glue}

Let us consider the HC  field strength $F_A^{\mu\nu}$,
transforming in the adjoint, and its contraction with the $Sp(2N_C)$ generators,
$(F^{\mu\nu})^i_{\, j}\equiv F_A^{\mu\nu} (T^A)^i_{\, j}$.
The hypercolour-invariant operator with the smallest possible canonical dimension (four) is
\be
(F^{\mu\nu})^i_{\, j} (F^{\rho\sigma})^j_{\, i} = \frac 12 F^{\mu\nu}_{A} F^{\rho\sigma}_{A}~.
\ee
Note the adjoint representation of $Sp(2N_C)$ coincides with the two-index symmetric representation.
One can indeed define a two-index symmetric field strength,
\be
F^{\mu\nu}_{ij}\equiv \Omega_{ik} (F^{\mu\nu})^k_{\, j} = (\Omega T^A)_{ij} F_A^{\mu\nu}~, 
\label{Fadjoint}\ee
where $\Omega$ is the 
$Sp(2N_C)$ invariant tensor, satisfying $\Omega_{ij}=-\Omega_{ji}$, $\Omega_{ik}\Omega_{kj}=-\delta_{ij}$, and $(T^A)^T=\Omega T^A \Omega$.

Let us decompose $F^{\mu\nu}F^{\rho\sigma}$ in irreducible Lorentz representations (dropping HC  indexes from now on).
The field strength $F^{\mu\nu}=\frac ig [D_\mu,D_\nu]$ transforms as $(1,0)+(0,1)$ under Lorentz.
One can project out the two irreducible components as
\be 
F_\alpha^{\ \beta} \equiv F_{\mu\nu}(\sigma^{\mu\nu})_\alpha^{\ \beta}~,\qquad 
F^{\dot\alpha}_{\ \dot\beta}\equiv F_{\mu\nu} (\overline\sigma^{\mu\nu})^{\dot\alpha}_{\ \dot\beta}~,
\ee
where $\alpha,\beta,\dots$, $\dot\alpha,\dot\beta,\dots$ are spinor indexes, 
$4\sigma^{\mu\nu}\equiv i(\sigma^\mu\overline{\sigma}^\nu-\sigma^\nu\overline{\sigma}^\mu)$, and
$4\overline{\sigma}^{\mu\nu}\equiv i(\overline{\sigma}^\mu\sigma^\nu-\overline{\sigma}^\nu\sigma^\mu)$.
Note that $F_{\alpha\beta}=F_\alpha^{\ \gamma}\epsilon_{\gamma\beta}$ is symmetric, as required for a $(1,0)$ representation, and analogously for the dotted counterpart. The product $F_{\alpha\beta}F_{\gamma\delta}$ transforms as $(1,0)\times (1,0)= (0,0)_s + (1,0)_a + (2,0)_s$.
The spin-1 component vanishes by antisymmetry, while the spin-0 and spin-2 components can be written as
\be 
F_{\alpha}^{\ \beta}F_{\beta}^{\ \alpha} ~,\qquad F_{\alpha\beta}F_{\gamma\delta}+F_{\alpha\gamma}F_{\beta\delta}+F_{\alpha\delta}F_{\beta\gamma}~. 
\label{1010}\ee
Note the latter, fully symmetric tensor is automatically traceless, that is, it vanishes when contracted with $\epsilon$ tensors.
Analogous considerations hold for the product $F_{\dot\alpha\dot\beta}F_{\dot\gamma\dot\delta}$. The two scalar components can be rewritten in the familiar form
\be
F_{\mu\nu}F^{\mu\nu} = \frac 12 \left(F_{\alpha}^{\ \beta}F_{\beta}^{\ \alpha}+F^{\dot\alpha}_{\ \dot\beta}F^{\dot\beta}_{\ \dot\alpha}\right)
~,\qquad  
F_{\mu\nu}\tilde{F}^{\mu\nu} = \frac i2 \left(F_{\alpha}^{\ \beta}F_{\beta}^{\ \alpha}-F^{\dot\alpha}_{\ \dot\beta}F^{\dot\beta}_{\ \dot\alpha}\right)~,
\label{scalars}\ee
where $\tilde F^{\mu\nu}\equiv \epsilon^{\mu\nu\rho\sigma}F_{\rho\sigma}/2$.
They correspond to scalar and pseudoscalar glueballs, while the second operator in \eq{1010} and its dotted counterpart correspond to spin-2 glueballs.
Finally, the product $F_\alpha^{\ \beta} F^{\dot\gamma}_{\ \dot\delta}$ transforms in the Lorentz representation $(1,0)\times(0,1)=(1,1)$. This operator corresponds, in tensor notation, to the traceless energy-momentum tensor shown in \eq{tmunu}, that is associated to additional spin-2 glueballs.

\subsection{Fermion bilinears}\label{Sbili}

Let us build hypercolour-invariant fermion bilinears. The available fermion constituents are $\psi^a_i$ for $a=1,\dots,2N_F$, and 
$\chi_{ij}=-\chi_{ji}$ with $\chi_{ij}\Omega_{ij}=0$ (or alternatively $\chi'_{ij}=\chi'_{ji}$), all defined as left-handed Weyl (anticommuting) spinors.

We list all possible hypercolour-singlet bilinears in table \ref{bili}, together with their Lorentz and flavour representations.
The $\psi\psi$ operators obviously have $\overline{\psi}\,\overline{\psi}$ conjugate operators, and similarly for the $\chi\chi$ ones.
The Young tableaux for $SU(2N_F)$ representations involve upper and lower indexes, associated to the fundamental and anti-fundamental representations, ${\Yvcentermath1 \tiny \yng(1)}^{\, a}$ and $\overline{\Yvcentermath1 \tiny \yng(1)}_{\, a}$, where the bar stands for a column of $2N_F-1$ boxes. As $Sp(2N_F)$ does not have complex representations, in the associated Young tableaux we do not distinguish upper and lower indexes.
Note that the operator $T_{\mu\nu} = \chi \sigma_{\mu\nu} \chi$ vanishes, because the contraction of HC  and Lorentz indexes are both symmetric, under the exchange of the two anticommuting fermions.

When $\chi$ is replaced by $\chi'$, the non-vanishing bilinears $s'$ and $\mathcal J'_\mu$ have exactly the same structure 
as $s$ and $\mathcal J_\mu$, basically
because the different symmetry of HC  indexes reduces to $(-1)^2=1$.

\begin{table}[tb]
\renewcommand{\arraystretch}{1.2}
\begin{center}
\begin{tabular}{|c|c|c|c|c|}
\hline
& Lorentz & $SU(2N_F)$ & $U(1)$ & $Sp(2N_F)$  \\
\hline \hline
 $S^{ab} = \psi^{\alpha a}_i \psi^{\beta b}_j \Omega_{ij} \epsilon_{\alpha\beta}$ 
 & $(0,0)$ & ${\Yvcentermath1 \tiny \yng(1,1)}^{\, ab}$ & $2q_\psi$ & ${\Yvcentermath1 \tiny \yng(1,1)}_{\, ab} + \bullet_{aa}$  \\
\hline
 $T^{ab}_{\mu\nu} = \psi^{\alpha a}_i \psi^{b}_{\beta j} \Omega_{ij} (\sigma_{\mu\nu})_\alpha^{\ \beta}$ 
 & $(1,0)_{\mu\nu}$ & ${\Yvcentermath1 \tiny \yng(2)}^{\, ab}$ & $2q_\psi$ & ${\Yvcentermath1 \tiny \yng(2)}_{\, ab}$  \\ 
\hline
$\mathcal J^{~~b}_{\mu a} = \overline{\psi}_{\dot\alpha i a} \psi^b_{\beta j} \Omega_{ij} \overline{\sigma}^{\dot\alpha\beta}_\mu $ &
$(1/2,1/2)_\mu$ & ${\Yvcentermath1 \tiny \overline{\yng(1)}\yng(1)}_{\, a}^{\, b} + \bullet_a^a$ & $0$ 
& ${\Yvcentermath1 \tiny \yng(2)}_{\, ab} + {\Yvcentermath1 \tiny \yng(1,1)}_{\, ab} + \bullet_{aa}$ \\
\hline\hline
 $s = \chi^{\alpha}_{ij} \chi^{\beta}_{kl} \Omega_{jk}\Omega_{li} \epsilon_{\alpha\beta}$ 
 & $(0,0)$ & $\bullet$ & $2q_\chi$ & $\bullet$  \\
\hline
$\mathcal J_{\mu} = \overline{\chi}_{\dot\alpha ij} \chi_{\beta kl} \Omega_{jk}\Omega_{li} \overline{\sigma}^{\dot\alpha\beta}_\mu $ &
$(1/2,1/2)_\mu$ & $\bullet$ & $0$ 
& $\bullet$ \\
\hline
\end{tabular}\end{center}
\caption{The fermion bilinear operators, and their transformation properties with respect to Lorentz and to the flavour symmetry, before and after SSB. One can straightforwardly replace $\chi$ with $\chi'$ everywhere, if desired.}
\label{bili}
\end{table}

These bilinear operators excite spin-zero and spin-one composite states, that are organised in multiplets of the flavour symmetry,
either $SU(2N_F)$ if SSB did not occur, or $Sp(2N_F)$ if SSB took place. We assume the latter possibility is realised, in order to obtain
the Higgs as a composite pNGB. 
According to \eq{HFparent}, the SM symmetries are embedded into $Sp(2N_F)$, therefore the
flavour multiplets, ${\Yvcentermath1 \tiny \yng(2)}_{\, ab}$ and ${\Yvcentermath1 \tiny \yng(1,1)}_{\, ab}$, decompose into SM multiplets, 
while the flavour singlets carry no SM charges.
In the minimal case $N_F=5$, the meson assignments under $SU(3)_C\times SU(2)_L\times SU(2)_R \times U(1)_B$ can be found in \eq{decompo}.

\subsection{Operators with more than two constituents}

Let us classify operators involving more than two constituents, beginning from operators involving HC fermions only. 
The unique $Sp(2N_C)$-invariant tensor is the antisymmetric matrix $\Omega_{ij}$, therefore it is easy to enumerate all possible HC-invariant, single-trace operators:
\be  
\psi^T\Omega(\chi\Omega)^n\psi~,\qquad \Tr\left[\chi\Omega\chi\Omega(\chi\Omega)^n\right]~,\qquad n=0,1,2,\dots~.
\label{master}\ee 
These operators involve $2+n$ fermion constituents, therefore they are bosonic (fermionic) for $n$ even (odd). 
It is understood that each $\psi$ ($\chi$) can be replaced by $\overline\psi$ ($\overline\chi$, $\chi'$, or $\overline\chi'$), as they carry the same $Sp(2N_C)$ indexes.
Since the canonical scaling dimension is $3+3/2\,n$, one naively expects the operators to become more irrelevant as $n$ grows, and the associated composite states to become heavier.
These fermion chains have some analogies but also qualitative differences with the concept of baryon
in $SU(N_C)$ theories, which relies on the invariant tensor $\epsilon_{i_1\dots i_{N_C}}$. In this paper we need only consider the minimal baryonic operators, with  $n=0$,
which we already analysed above. In \cite{part2} we will analyse the minimal fermionic operators, with $n=1$.

Coming to operators involving also the HC field strength $F^{\mu\nu}$, it is sufficient to notice that it transforms in the HC two-index symmetric representation, according to \eq{Fadjoint}. Therefore, one can replace $\chi$'s in \eq{master} with $F$'s. Beside the minimal $F^2$ operators already studied in section \ref{glue}, the next-to-minimal bosonic operators are made of three constituents: $F^3$ (canonical dimension 6), $\psi^2 F$ and $\chi^2 F$ (canonical dimension 5). Note also the minimal fermionic operator $\chi'F$, with canonical dimension 7/2, that will be discussed in \cite{part2}.

For completeness, we mention that each operator in the above classification can be dressed with (extra) derivatives.
Each additional derivative increases the canonical scaling dimension by one unit and, of course, it changes the Lorentz representation of the operator.

\section{Sigma-model coupled to gravity}
\label{sec:sigmamodel}

In this appendix, we summarise the formalism~\cite{Bianchi:2003ug,Berg:2005pd,Berg:2006xy,Elander:2009bm,Elander:2010wd} that we use to study a sigma-model consisting of a number of scalars coupled to gravity. For further details, the Reader is referred to \cite{Elander:2010wd}. We start with the 5D action given by
\beqs
\label{eq:action}
	\mathcal S = \frac{1}{4\pi G_5} \int \dd^4x \dd r \Bigg\{ && \hspace{-0.4cm} \sqrt{-g} \, \bigg[
	\frac{R}{4} - \frac{1}{2} G_{ab}(\Phi) g^{MN} \partial_M \Phi^a \partial_N \Phi^b - \mathcal V(\Phi) \bigg] \nonumber\\
	 && + \sum_{i=1,2} \delta(r-r_i) (-)^i \sqrt{-g} \bigg[ \frac{K}{2} + \mathcal L_i \bigg] \Bigg\} \,,
\eeqs
where $\Phi^a$ ($a = 1, \cdots, n$) are scalars with sigma-model metric $G_{ab}$ and potential $\mathcal V$. The signature of $g_{MN}$ is mostly plus, and $M = 0,1,2,3,5$ while $\mu = 0,1,2,3$. Here $r_1$ and $r_2$ are regulators introduced in order to perform the numerical calculations. The physical results are obtained taking the limits $r_1 \rightarrow r_o$, where $r_o$ denotes the end-of-space corresponding to the IR, together with $r_2 \rightarrow +\infty$ approaching the UV boundary. The second line contains the boundary terms necessary to make the variational problem well-defined, including the Gibbons-Hawking term containing the extrinsic curvature $K$. Our conventions, as well as the detailed form of $\mathcal L_i$ can be found in \cite{Elander:2010wd}.

We will study backgrounds for which the scalars $\Phi^a$ and the metric only depend on the radial coordinate $r$. The background metric is taken to be of the domain-wall form of \eq{eq:domainwall},
\beq
\label{eq:ds2dw}
g_{MN} ={\rm diag}(-e^{2A},e^{2A},e^{2A},e^{2A},1)_{MN} ~,
\eeq
where $A(r)$ is the warp factor. The background equations of motion are
\beqs
\label{eq:eoms}
	\partial_r^2 \Phi^a + 4 \partial_r A \, \partial_r \Phi^a + \mathcal G^a{}_{bc} \partial_r \Phi^b \partial_r \Phi^c - \mathcal V^a &=& 0 \,, \nonumber \\
	6 (\partial_r A)^2 - G_{ab} \partial_r \Phi^a \partial_r \Phi^b + 2 \mathcal V &=& 0 \,,
\eeqs
where $\mathcal V^a = G^{ab} \mathcal V_b = G^{ab} \partial_b \mathcal V = G^{ab} \frac{\partial \mathcal V}{\partial \Phi^b}$ and $\mathcal G_{abc} = \frac{1}{2} (\partial_b G_{ca} + \partial_c G_{ab} - \partial_a G_{bc})$.
In the case where the potential $\mathcal V$ can be obtained from a superpotential $\mathcal W$ as
\beq
	\mathcal V = \frac{1}{2} \mathcal W_a \mathcal W^a - \frac{4}{3} \mathcal W^2 \,,
\eeq
one can obtain solutions to the equations of motion~\eqref{eq:eoms} by solving the first order equations
\beq
\label{eq:eomW}
	\partial_r \Phi^a = \mathcal W^a \,, \qquad\qquad
	\partial_r A = - \frac{2}{3} \mathcal W \,. 
\eeq

The spectrum of spin-0 and spin-2 states can be found by studying the fluctuations of the scalar fields and gravity around a given background solution. One Fourier transforms along Minkowski directions, writing the fluctuations as functions of the four-momentum $q^\mu$ and the radial coordinate $r$. The linearised equations of motion admit solutions satisfying the appropriate boundary conditions in the IR and the UV only for certain values of $q^2 = -m^2$.  
These boundary conditions are chosen such that, as long as $m^2 > 0$, they select the poles of the two-point functions of the dual
field-theory operators, thus providing a method for calculating the spectrum.

More precisely, given a background solution, defined by $\bar \Phi^a(r)$ and $A(r)$, we expand around it in the fluctuations  $\{ \varphi^a, \nu, \nu^\mu, \mathfrak e^\mu{}_\nu, h, H, \epsilon^\mu \}$ as
\beqs
\label{eq:ADM}
	\Phi^a &=& \bar \Phi^a + \varphi^a, \nonumber\\
	\dd s^2 &=& (1 + 2\nu + \nu_\mu \nu^\mu) \dd r^2 + 2 \nu_\mu \dd x^\mu \dd r + e^{2A} ( \eta_{\mu\nu} + h_{\mu\nu} ) \dd x^\mu \dd x^\nu, \nonumber \\
	h^\mu_{\ \nu} &=& \mathfrak e^\mu{}_\nu + i q^\mu \epsilon_\nu + i q_\nu \epsilon^\mu + \frac{q^\mu q_\nu}{q^2} H + \frac{1}{3} 
	\delta^\mu{}_\nu h~.
\eeqs
Here $\mathfrak e^\mu{}_\nu$ is transverse and traceless, $\epsilon^\mu$ is transverse, and the four-dimensional indices $\mu$, $\nu$ are raised and lowered by the boundary metric $\eta$.

The spin-2 fluctuation $\mathfrak e^\mu{}_\nu$ satisfies the linearised equation of motion
\beq
\label{eq:fluceoms2}
	\left[\partial_r^2 + 4A' \partial_r - e^{-2A} q^2 \right] \mathfrak e^\mu{}_\nu = 0 \,,
\eeq
with boundary conditions given by $\partial_r \mathfrak e^\mu{}_\nu |_{r_i} = 0$. We denote differentiation with respect to $r$ with a prime, e.g.~$A' \equiv \partial_r A$. After forming the gauge-invariant combination (invariant under diffeomorphisms) \cite{Berg:2005pd,Elander:2009bm}
\beq
	\mathfrak a^a = \varphi^a - \frac{\bar\Phi'^a}{6A'} h \, ,
\eeq
it can be shown that the linearised equation of motion for the fluctuations in the spin-0 sector can be written as $n$ second-order differential equations,
\beq
\label{eq:fluceoms}
	\Big[ \mathcal D_r^2 + 4A' \mathcal D_r - e^{-2A} q^2] \mathfrak a^a - \Big[ \mathcal V^a{}_{|c} - \mathcal R^a{}_{bcd} \bar \Phi'^b \bar \Phi'^d + \frac{4 (\bar \Phi'^a \mathcal V_c + \mathcal V^a \bar \Phi'_c )}{3A'} + \frac{16 \mathcal V \bar \Phi'^a \bar \Phi'_c}{9A'^2} \Big] \mathfrak a^c = 0 \,,
\eeq
while the boundary conditions are given by \cite{Elander:2014ola}
\beq
\label{eq:flucbcs}
	\bar \Phi'^a \bar \Phi'_b \mathcal D_r \mathfrak a^b \Big|_{r_i} = \frac{3A'}{2}  \left[ e^{-2A} q^2 - \frac{A'}{2} \partial_r \left( \frac{A''}{A'^2} \right) \right] \mathfrak a^a \Big|_{r_i} \,.
\eeq
The different quantities involved in these expressions are
\be
\begin{array}{rclcrcl}
\mathcal D_r \mathfrak a^a &=& \partial_r \mathfrak a^a + \mathcal G^a{}_{bc} \bar \Phi'^b \mathfrak a^c \,,
&\qquad\qquad&
\bar\Phi'_a &= &G_{ab} \bar \Phi'^b \,,\\ 
\mathcal R^a{}_{bcd}& =& \partial_c \mathcal G^a{}_{bd} - \partial_d \mathcal G^a{}_{bc} + \mathcal G^a{}_{ce} \mathcal G^e{}_{bd} - \mathcal G^a{}_{de} \mathcal G^e{}_{bc}\,,
&\qquad\qquad&
\mathcal V^a{}_{|b}& =& \dfrac{\partial \mathcal V^a}{\partial \Phi^b} + \mathcal G^a_{\ bc} \mathcal V^c \,.
\end{array}
\ee
As can be seen, $\mathcal R^a{}_{bcd}$ is the Riemann tensor corresponding to the sigma-model metric.

\section{Axial-vector and pseudoscalar sector}
\label{sec:AVandPS}

In this appendix, we work out in detail how to compute two-point functions in the axial-current and pseudoscalar sector of the dual field theory. For further details on the formalism of holographic renormalization, see \cite{Bianchi:2001kw,Bianchi:2001de}.

Consider a $U(1)$ gauge field $A_M$ and a pseudoscalar $\pi$ with action
\beq
	\mathcal S = \int_{r_1}^{r_2} \dd r \int \dd^4x \sqrt{-g} \, \bigg\{ \hspace{-0.1cm}
	- \frac{1}{4} H(r) F_{MN}^2 
	- \frac{1}{2} G(r) (\partial_M \pi + g_5 A_M)^2
	 \bigg\} \,.
\eeq
We assume that the background metric takes the domain-wall form given in Eq.~\eqref{eq:ds2dw}. The functions $G$ and $H$ depend on the radial coordinate $r$, and their form is model-dependent.

The boundary-localized counter-term action needed to cancel divergences in this sector is given by
\beq
	\mathcal S_{\rm ct} = \int \dd^4x \sqrt{- \tilde g} \, \bigg\{ \hspace{-0.1cm}
	- \frac{\mathcal C}{2} (\partial_\mu \pi + g_5 A_\mu)^2
	- \frac{\mathcal D}{4} F_{\mu\nu}^2
	- \frac{\mathcal E}{2} \pi^2
	 \bigg\} \bigg|_{r=r_2} \,,
\eeq
where $\tilde g_{\mu\nu}$ is the induced metric on the boundary, and $\mathcal C$, $\mathcal D$, $\mathcal E$ are required to be local. Note that the term containing $\mathcal E$ breaks gauge invariance on the boundary (as explained in \cite{Bianchi:2001kw}, it is necessary for the case when there is a Goldstone).

We will work in the gauge $A_r = 0$. The equations of motion for the gauge field and pseudoscalar written in Fourier space are (our conventions for Fourier transforms are explained in footnote~\ref{FourierConvention})
\beqs
\label{eq:eomA}
	\Bigg[ \partial_r^2 +\left( 2\partial_r A + \frac{\partial_r H}{H} \right) \partial_r - \left( q^2 e^{-2A} + g_5^2 \frac{G}{H} \right) \Bigg] P^{\mu\nu} A_\nu(q,r) &=& 0 \,, \\
	\Bigg[ \partial_r^2 +\left( 2\partial_r A + \frac{\partial_r H}{H} \right) \partial_r - g_5^2 \frac{G}{H} \Bigg] \frac{q^\mu q^\nu}{q^2} A_\nu(q,r) - g_5 \frac{G}{H} i q^\mu \pi(q,r) &=& 0 \,, \\
\label{eq:eomAL}
	\Bigg[ \partial_r^2 +\left( 4\partial_r A + \frac{\partial_r G}{G} \right) \partial_r - q^2 e^{-2A} \Bigg] \pi(q,r) + g_5 e^{-2A} i q^\mu A_\mu(q,r) &=& 0 \,, \\
\label{eq:eompi}
	g_5 e^{2A} \frac{G}{H} \partial_r \pi(q,r) + i q^\mu \partial_r A_\mu(q,r) &=& 0 \,,
\eeqs
where indices are raised with $\eta^{\mu\nu}$ so that $q^2 = \eta^{\mu\nu} q_\mu q_\nu$ and the projector is given by 
$P^{\mu\nu} = \eta^{\mu\nu} - \frac{q^\mu q^\nu}{q^2}$.

The variational problem demands that we impose the IR boundary conditions
\beq
\label{eq:BCs}
	\partial_r A^\mu(q,r) \big|_{r=r_1} = 0, \ \ \ \partial_r \pi(q,r) \big|_{r=r_1} = 0 \,,
\eeq
after which the action $\mathcal S + \mathcal S_{\rm ct}$ evaluated on-shell becomes
\beqs
\label{eq:Ssub}
	\mathcal S_{\rm sub} = \int \dd^4 q \bigg\{ && \hspace{-0.3cm}
	- \frac{e^{2A}}{2} A_\mu(-q,r) \left( g_5^2 \mathcal C + q^2 e^{-2A} \mathcal D + H \partial_r \right) P^{\mu\nu} A_\nu(q,r) \nonumber \\ && \hspace{-0.4cm} \nonumber
	- \frac{e^{2A}}{2} A_\mu(-q,r) \left( g_5^2 \mathcal C + H \partial_r \right) \frac{q^\mu q^\nu}{q^2} A_\nu(q,r) \\ && \nonumber \hspace{-0.4cm}
	- \frac{e^{4A}}{2} \pi(-q,r) \left( q^2 e^{-2A} \mathcal C + \mathcal E + G \partial_r \right) \pi(q,r) \\ &&  \hspace{-0.4cm}
	- e^{2A} g_5 \mathcal C i q^\mu A_\mu(-q,r) \pi(q,r)
	\bigg\} \Bigg|_{r=r_2} \,.
\eeqs

Correlators in the dual field theory are obtained by differentiating $\mathcal S_{\rm sub}$ with respect to the boundary values of the fields. The gauge field $A_\mu$ couples to the current $J_\mu$ as $\int \dd^4 x \, g_5 A_\mu J^\mu$, while the pseudoscalar field $\pi$ couples to an operator which we denote $\mathcal O_\pi$ as $\int \dd^4 x \, N_\pi^{-1} \pi \mathcal O_\pi$ where $N_\pi(r_2)$ is a normalization factor that is included so that $\lim_{r_2 \rightarrow \infty} N_\pi^{-1}(r_2) \, \pi(r_2)$ is finite. The precise form of $N_\pi(r_2)$ is model-dependent and can be read off from the UV expansion of $\pi$. We obtain the following two-point functions
\beqs
\label{eq:twopointfunctions}
	\langle J^\mu(q) J^\nu(-q) \rangle &=&
	- \frac{i}{g_5^2} \lim\limits_{r_2 \rightarrow \infty} \left\{ \frac{\delta^2 \mathcal S_{\rm sub}}{\delta A_\mu(-q,r_2) \delta A_\nu(q,r_2)} \right\} \,, \nonumber \\
	\langle J^\mu(q) \mathcal O_\pi(-q) \rangle &=&
	- \frac{i}{g_5} \lim\limits_{r_2 \rightarrow \infty} \left\{ N_\pi(r_2) \frac{\delta^2 \mathcal S_{\rm sub}}{\delta A_\mu(-q,r_2) \delta \pi(q,r_2)} \right\} \,, \\ \nonumber
	\langle \mathcal O_\pi(q) \mathcal O_\pi(-q) \rangle &=&
	(- i) \lim\limits_{r_2 \rightarrow \infty} \left\{ N_\pi^2(r_2) \frac{\delta^2 \mathcal S_{\rm sub}}{\delta \pi(-q,r_2) \delta \pi(q,r_2)} \right\} \,.
\eeqs

This whole procedure is straightforward for the transverse part of the vector. After writing $P^{\mu\nu} A_\nu(q,r) = \tilde A_\nu(q) a(q,r)$,\footnote{This decomposition is unique up to a relative rescaling of $\tilde A_\nu(q)$ and $a(q,r)$ that does not affect the final results for the correlation functions.} and differentiating $\mathcal S_{\rm sub}$ with respect to $\tilde A_\nu(q)$, we obtain
\beqs
\label{eq:JJT}
	P_{\mu\sigma} P_{\nu\rho} \, \langle J^\sigma(q) J^\rho(-q) \rangle &=&
	- \frac{i}{g_5^2} \lim\limits_{r_2 \rightarrow \infty} \left\{ P_{\mu\sigma} P_{\nu\rho} \frac{\delta^2 \mathcal S_{\rm sub}}{\delta A_\mu(-q,r_2) \delta A_\nu(q,r_2)} \right\}
	\\ \nonumber
	 &=& \lim\limits_{r_2 \rightarrow \infty} \left\{ i \, e^{2A} \left( \mathcal C + q^2 e^{-2A} \frac{\mathcal D}{g_5^2} + \frac{H}{g_5^2} \frac{\partial_r a}{a} \right) P_{\mu\nu} \, \bigg|_{r=r_2} \right\} \,.
\eeqs
On the other hand, since the longitudinal part of $A_\mu$ couples to the pseudoscalar, we need to be careful about how to vary their boundary values independently. The general solution to the equations of motion of the longitudinal part of the axial-vector and the pseudoscalar, given in Eqs.~\eqref{eq:eomAL} and~\eqref{eq:eompi}, satisfying the boundary conditions Eq.~\eqref{eq:BCs} can be written as
\beqs
\label{eq:piAL}
	\pi(q,r) &=& g_5 \left[ c_1(q) \int_{r_1}^r \dd \tilde r \, \frac{X(q,\tilde r)}{e^{4A} G} + c_2(q) \right] \,, \\ \nonumber
	i q^\mu A_\mu(q,r) &=& c_1(q) \left[ q^2 \int_{r_1}^r \dd \tilde r \, \frac{X(q,\tilde r)}{e^{4A} G} - \frac{\partial_r X(q,r)}{e^{2A} G} \right] + q^2 c_2(q) \,.
\eeqs
where $c_1(q)$ and $c_2(q)$ are integration constants, and $X(q,r)$ is a solution to the second order differential equation\footnote{Note the agreement with \cite{Elander:2018aub} where a different approach of using $R_\chi$-gauge was implemented.}
\beq
\label{eq:eomX}
	\Bigg[ \partial_r^2 - \left( 2\partial_r A + \frac{\partial_r G}{G} \right) \partial_r - \left( q^2 e^{-2A} + g_5^2 \frac{G}{H} \right) \Bigg] X(q,r) = 0 \,,
\eeq
with boundary condition
\beq
	X(q,r) \big|_{r=r_1} = 0 \,.
\eeq
Hence, if we vary the action with respect to the integration constants $c_1(q)$ and $c_2(q)$, we are ensured to stay within the space of solutions.

Defining $\chi(q,r) = (i q^\mu A_\mu(q,r),\pi(q,r))^T$, $C(q,r) = (c_1(q,r), c_2(q,r))^T$, we have that
\beqs
	\chi(q,r) &=& B(q,r) C(q) \,, \\
	B(q,r) &=& \left(
\begin{array}{cc}
q^2 \int_{r_1}^r \dd \tilde r \, \frac{X(q,\tilde r)}{e^{4A} G} - \frac{\partial_r X(q,r)}{e^{2A} G} & q^2 \\
 g_5 \int_{r_1}^r \dd \tilde r \, \frac{X(q,\tilde r)}{e^{4A} G} & g_5  \\
\end{array}
\right) \,.
\eeqs
After writing the relevant part of the action $\mathcal S_{\rm sub}$---the last three lines of Eq.~\eqref{eq:Ssub}---on the form
\beqs
	&& \int \dd^4 q \left\{ - \frac{1}{2} \chi(-q,r)^T \left( \mathcal M \partial_r + \mathcal N \right) \chi(q,r) \right\} \bigg |_{r=r_2} \nonumber \\ && = \int \dd^4 q \left\{ - \frac{1}{2} \chi(-q,r)^T \left( \mathcal M \left(\partial_r B\right) B^{-1} + \mathcal N \right) \chi(q,r) \right\} \bigg |_{r=r_2} \,,
\eeqs
where
\beqs
\mathcal M &=& {\rm diag} \left(e^{2A} q^{-2} H, e^{4A} G \right) \,, \\
\mathcal N &=&
\left(
\begin{array}{cc}
\frac{e^{2A}}{q^2} g_5^2 \mathcal C  & - e^{2A} g_5 \mathcal C \\
- e^{2A} g_5 \mathcal C & e^{4A} \left( q^2 e^{-2A} \mathcal C + \mathcal E \right)  \\
\end{array}
\right) \,,
\eeqs
we have that
\beq
	- \frac{\delta^2 \mathcal S_{\rm sub}}{\delta \chi_i(-q,r_2) \delta \chi_j(q,r_2)} = \left[ \mathcal M \left(\partial_r B\right) B^{-1} + \mathcal N \right]_{ij} \Big |_{r=r_2}  \,.
\eeq

Putting everything together, we obtain
\beqs
\label{eq:d2S}
	- \frac{1}{g_5^2} \frac{\delta^2 \mathcal S_{\rm sub}}{\delta A_\mu(-q,r_2) \delta A_\nu(q,r_2)} && \nonumber \\ && \hspace{-3cm} = e^{2A} \left[ \left( \mathcal C + q^2 e^{-2A} \frac{\mathcal D}{g_5^2} + \frac{H}{g_5^2} \frac{\partial_r a}{a} \right) P^{\mu\nu} + \left( \mathcal C + G \frac{X}{\partial_r X} \right) \frac{q^\mu q^\nu}{q^2} \right] \, \bigg|_{r=r_2} \,, \nonumber \\
	- \frac{1}{g_5} \frac{\delta^2 \mathcal S_{\rm sub}}{\delta A_\mu(-q,r_2) \delta \pi(q,r_2)}  &=& e^{2A} \left( \mathcal C + G \frac{X}{\partial_r X} \right) i q^\mu \, \bigg|_{r=r_2} \,, \nonumber \\ 
	- \frac{\delta^2 \mathcal S_{\rm sub}}{\delta \pi(-q,r_2) \delta \pi(q,r_2)}  &=& e^{4A} \left[ \mathcal E + q^2 e^{-2A} \left(\mathcal C + G \frac{X}{\partial_r X} \right) \right] \, \bigg|_{r=r_2} \,,
\eeqs
which we can plug into Eq.~\eqref{eq:twopointfunctions} in order to obtain two-point functions of the dual field theory. Note that these expressions do not rely on knowing the solutions to the equations of motion analytically as in \cite{Bianchi:2001kw}. The precise form of $\mathcal C$, $\mathcal D$, and $\mathcal E$ appearing in the counter-terms depends on the specific model and which divergencies need to be cancelled. We will describe this for a particular example in the next section.

The behaviour of the two-point functions around $q^2 = 0$ is important in order to identify Goldstone bosons and their decay constants. One can show that the residue of the pole at $q^2 = 0$ in the first equation of~\eqref{eq:d2S} is given by
\beq
	\mathcal Y(r) \equiv e^{2A} \left( G \frac{X}{\partial_r X} - \frac{H}{g_5^2} \frac{\partial_r a}{a} \right) \bigg|_{q^2 =0} \,.
\eeq
By differentiating with respect to $r$ and using the equations of motion for $a$ and $X$, one can show that $\mathcal Y$ satisfies the first-order differential equation
\beq
	\partial_r \mathcal Y(r) = - \mathcal F(r) \mathcal Y(r) \,, \hspace{1cm} \mathcal F(r) \equiv \frac{g_5^2}{H} \left[ G \frac{X}{\partial_r X} + \frac{H}{g_5^2} \frac{\partial_r a}{a} \right] \,,
\eeq
with general solution $\mathcal Y(r) = C \exp \left( - \int_{r_1}^r \dd \tilde r \mathcal F(\tilde r)\right)$, where $C$ is an integration constant. However, by using the IR boundary conditions $\partial_r a |_{r_1} = 0$ and $X |_{r_1} = 0$, one has that $\mathcal Y(r_1) = 0$, so that
\beq
\label{eq:q2zero}
	\mathcal Y(r) = 0
\eeq	
for all values of $r$. Now, this may constitute a welcome property for the case where the symmetry is explicitly broken and no massless pole develops in the axial channel. There is, however, a subtle point to be made about the order of limits. Naively, the preceding argument would imply that a massless pole is also absent when the symmetry is spontaneously broken. However, in the computation of the correlation functions, the correct order of limits is to first take $r_2 \rightarrow \infty$ in Eq.~\eqref{eq:d2S}, and, only after the result has been obtained, take the limit of zero momentum. In this way one will generate a massless Goldstone pole when the symmetry is spontaneously broken. In the example of the next subsection, Eq.~\eqref{eq:q2zero} will prove useful in deriving various relations that serve as consistency checks of the model.

\subsection{Explicit computation of correlators}

As an application of the formalism developed in the previous section, let us study Example~B, for which we have that $G = \sigma^2$ and $H = 1$ (for the purposes of this appendix, we suppress the overall factor $N_C^2$ appearing in the action). The background solutions for the scalar $\sigma$ and the warp factor $A$ are given in Eq.~\eqref{eq:ExampleBbackground}. We focus on the two cases $\Delta = 3$ and $\Delta = 1$. In the former case, the global $U(1)$ symmetry of the model is spontaneously broken, leading to a Goldstone boson, which is not present in the latter case that describes explicit breaking.

\begin{center}{$a.~~~\Delta = 3$}\end{center}

For $\Delta = 3$, the UV expansions of $a$ and $X$ are given conveniently in terms of the coordinate $z \equiv e^{-r}$ as
\beqs
\label{eq:UVexpDelta3}
	a &=& a_0(q) \left[ 1 + \left( a_2(q) + \frac{q^2}{2} \log(z) \right) z^2 \right] + \mathcal O(z^4) \,, \nonumber \\
	X &=& X_0(q) \bigg[ 1 - \frac{q^2}{4} z^2 + \left( X_4(q) - \frac{q^4}{16} \log(z) \right) z^4 \nonumber \\ && \hspace{1.2cm} + \left( \frac{1}{288} \left( 24 g_5^2 + q^6 + 24 q^2 X_4(q) \right) - \frac{q^6}{192} \log(z) \right) z^6 \bigg] + \mathcal O(z^8) \,,
\eeqs
where $a_0$, $a_2$, $X_0$, and $X_4$ are integration constants.\footnote{In a slight abuse of notation, we denote by $\mathcal O(z^n)$ terms that are of order $z^n$ up to possible logarithmic factors.} Note that $a_0$ and $X_0$ are overall normalizations and hence will not appear in the two-point functions, while $a_2$ and $X_4$ are to be determined by solving the equations of motion for $a$ and $X$ with the appropriate boundary conditions in the IR.

Using Eq.~\eqref{eq:UVexpDelta3} and Eq.~\eqref{eq:piAL}, we have that
\beqs
	\pi &=& \pi_0(q) z^{-2} + \frac{q^2}{2} \pi_0(q) \log(z) + \frac{1}{8q^2} \Big[ 8g_5 \alpha_0(q) + \pi_0(q) \left( q^4 - 64 X_4(q) \right) \Big] + \mathcal O(z^2) \,, \nonumber \\
	i q^\mu A_\mu &=& \alpha_0(q) + \mathcal O(z^2) \,,
\eeqs
where $\alpha_0(q)$ is an integration constant and $\pi_0 = \frac{g_5 c_1 X_0}{2}$. This shows that the normalization factor in the coupling of $\pi$ to the operator $\mathcal O_\pi$ should be chosen to be $N_\pi = z^{-2}$. As anticipated, no such normalization factor is needed for the coupling of $A_\mu$ to the current $J^\mu$.

In order to cancel the divergencies in the on-shell action, the counter-terms need to be chosen to be
\beq
	\mathcal C = \sigma^2 \left( \tilde c + \log(z) \right) \,, \ \ \
	\mathcal D = \tilde d + \frac{1}{2} + \log(z) \,, \ \ \
	\mathcal E = - 2 \sigma^2 \,,
\eeq
where $\tilde c$ and $\tilde d$ give finite contributions and correspond to the choice of regularization scheme. Plugging Eq.~\eqref{eq:d2S} into Eq.~\eqref{eq:twopointfunctions}, we obtain the two-point functions\footnote{Note that an alternative way to obtain the two-point functions of Eq.~\eqref{eq:twopointfunctionsDelta3} is to plug the UV expansions of $a$, $\pi$, and $i q^\mu A_\mu$ into the on-shell action $\mathcal S_{\rm sub}$ given in Eq.~\eqref{eq:Ssub} and differentiate with respect to $a_0$, $\pi_0$, and $\alpha_0$ (note that these parameters can all be varied independently while staying in the space of solutions). The final result of this exercise agrees with the above analysis as it should.}
\beqs
\label{eq:twopointfunctionsDelta3}
	i \langle J^\mu(q) J^\nu(-q) \rangle &=& q^2 \Pi_A(q) P^{\mu\nu} = \frac{1}{g_5^2} \left( 2 a_2(q) - \tilde d \, q^2 \right) P^{\mu\nu} \,, \nonumber \\
	\langle J^\mu(q) \mathcal O_\pi(-q) \rangle &=& - \frac{2 q^\mu}{q^2} \,, \\ \nonumber
	i \langle \mathcal O_\pi(q) \mathcal O_\pi(-q) \rangle &=& \left( \frac{3}{4} - \tilde c \right) q^2 - \frac{16 X_4(q)}{q^2} \,,
\eeqs
which leads to the presence of a massless pole. Moreover, the decay constant $F_G$ is defined as
\beq
	F_G^2 = \lim\limits_{q^2 \rightarrow 0} \Big\{ -q^2 \Pi_A(q) \Big\} = - \frac{2 a_2(0)}{g_5^2} \,,
\eeq
and is regularization scheme independent.

We can define another decay constant $G_G$ as
\beq
	G_G^2 = \lim\limits_{q^2 \rightarrow 0} \Big\{ q^2 \, i \langle \mathcal O_\pi(q) \mathcal O_\pi(-q) \rangle \Big\} = -16 X_4(0) \,.
\eeq
As a consistency check, one needs to satisfy
\beq
	F_G G_G = - q_\mu \langle J^\mu(q) \mathcal O_\pi(-q) \rangle = 2 \,.
\eeq
This can be derived using Eq.~\eqref{eq:q2zero} from which it follows that $a_2(0) X_4(0) = \frac{g_5^2}{8}$.

\newpage
\begin{center}{$b.~~~\Delta = 1$}\end{center}

For $\Delta = 1$, the UV expansions of $a$ and $X$ are given by
\beqs
\label{eq:UVexpDelta1}
	a &=& a_0(q) \left[ 1 + \left( a_2(q) + \frac{1}{2} (3g_5^2 + q^2) \log(z) \right) z^2 \right] + \mathcal O(z^4) \,, \nonumber \\
	X &=& X_0(q) \bigg[ \log(z) + \tilde X_0(q) \\ && \hspace{1cm} + \left(\frac{1}{4} \big(3 g_5^2 + q^2\big) \big(\tilde X_0(q) - 1\big) - \frac{1}{6} + \frac{1}{4} (3g_5^2 + q^2) \log(z) \right) z^2 \bigg] + \mathcal O(z^4) \,, \nonumber
\eeqs
where $a_0$, $a_2$, $X_0$, and $\tilde X_0$ are integration constants. Note that $a_0$ and $X_0$ are overall normalizations and hence will not appear in the two-point functions, while $a_2$ and $\tilde X_0$ are to be determined by solving the equations of motion for $a$ and $X$ with the appropriate boundary conditions in the IR.

Using Eq.~\eqref{eq:UVexpDelta1} and Eq.~\eqref{eq:piAL}, we have that
\beqs
	\pi &=& \pi_0(q) + \frac{1}{4} \big( 1 - 2 \tilde X_0(q) - 2 \log(z) \big) \big( g_5 \alpha_0(q) - q^2 \pi_0(q) \big) z^2 + \mathcal O(z^4) \,, \nonumber \\
	i q^\mu A_\mu &=& \alpha_0(q) + \mathcal O(z^2) \,,
\eeqs
where $\alpha_0(q)$ is an integration constant and $\pi_0 = \frac{g_5}{3 q^2} \left( 3 \alpha_0 - c_1 X_0 \right)$. This shows that the normalization factor in the coupling of $\pi$ to the operator $\mathcal O_\pi$ should be chosen to be $N_\pi = 1$.

In order to cancel the divergencies in the on-shell action, the counter-terms need to be chosen to be
\beq
	\mathcal C = \sigma^2 \left( \tilde c + \log(z) \right) \,, \ \ \
	\mathcal D = \tilde d + \frac{1}{2} + \log(z) \,, \ \ \
	\mathcal E = 0 \,,
\eeq
where $\tilde c$ and $\tilde d$ again correspond to the choice of regularization scheme. Plugging Eq.~\eqref{eq:d2S} into Eq.~\eqref{eq:twopointfunctions}, we obtain the two-point functions
\beqs
\label{eq:twopointfunctionsDelta1}
	i \langle J^\mu(q) J^\nu(-q) \rangle &=& \left( \frac{3}{2} - 3 \tilde c - \frac{\tilde d \, q^2}{g_5^2} + \frac{2a_2(q)}{g_5^2} \right) P^{\mu\nu} + 3 \left( \tilde X_0(q) - \tilde c \right) \frac{q^\mu q^\nu}{q^2} \,, \nonumber \\
	\langle J^\mu(q) \mathcal O_\pi(-q) \rangle &=& 3 \left( \tilde X_0(q) - \tilde c \right) q^\mu \,, \\ \nonumber
	i \langle \mathcal O_\pi(q) \mathcal O_\pi(-q) \rangle &=& 3 \left( \tilde X_0(q) - \tilde c \right) q^2 \,.
\eeqs
The expression for $\langle J^\mu(q) J^\nu(-q) \rangle$ might at first look worrisome because of the appearance of a pole at $q^2=0$. Again, we can use Eq.~\eqref{eq:q2zero} to show that $\tilde X_0(0) = \frac{2a_2(0)}{3g_5^2} + \frac{1}{2}$, so that the poles in the longitudinal and transverse parts cancel.

\end{document}